\documentclass{aa} 
\usepackage[varg]{txfonts}

\usepackage{amsmath}
\usepackage{amsfonts}
\usepackage{amssymb}
\usepackage{graphicx}
\usepackage{fourier}

\usepackage{natbib}
\bibpunct{(}{)}{;}{a}{}{,} 
\usepackage{booktabs} 

\usepackage{makecell}
\usepackage{txfonts}
\usepackage{graphicx,natbib,url,twoopt}
\usepackage{lipsum}
\usepackage{multicol}
\usepackage{float}
\usepackage{stackengine}
\usepackage{xcolor}

\usepackage{subcaption}
\usepackage{threeparttable}
\usepackage{adjustbox} 

\usepackage{placeins}
\def\logg{\hbox{$\log{g}$}}         
\def\rhk{\hbox{$\log{(R^{'}_{HK})}$}}         
\def\ms{\hbox{\,m\,s$^{-1}$}}         
\def\m2s2{\hbox{\,m$^{2}$\,s$^{-2}$}} 
\def\kms{\hbox{\,km\,s$^{-1}$}}       
\def\vsini{\hbox{$v$\,sin\,$i$}}      
\def\sini{\hbox{sin\,$i$}}      
\def\Msun{\hbox{$\mathrm{M}_{\odot}$}}             

\def\teff{\hbox{$T_{\rm eff}$}}
\def\feh{\hbox{[Fe/H]}}

\begin{document}
\title{A review of planetary systems around HD\,99492, HD\,147379 and HD\,190007 with HARPS-N\thanks{The HARPS-N data of these stars is available at the CDS via anonymous ftp to cdsarc.cds.unistra.fr (130.79.128.5).}}

\author{M. Stalport\inst{\ref{i:liege}, \ref{i:geneva}}
\and M.\,Cretignier\inst{\ref{i:oxford}, \ref{i:geneva}}
\and S.\,Udry\inst{\ref{i:geneva}}
\and A.\,Anna John\inst{\ref{i:StAndrews}} 
\and T.\,G.\,Wilson\inst{\ref{i:StAndrews}, \ref{i:Warwick}}
\and J.-B.\,Delisle\inst{\ref{i:geneva}}
\and A. S. Bonomo\inst{\ref{i:Torino}}
\and L.\,A.\,Buchhave\inst{\ref{i:Lyngby}}
\and D.\,Charbonneau\inst{\ref{i:Harvard}} 
\and S.\,Dalal\inst{\ref{i:Exeter}}
\and M.\,Damasso\inst{\ref{i:Torino}}
\and L.\,Di Fabrizio\inst{\ref{i:INAFSpain}} 
\and X.\,Dumusque\inst{\ref{i:geneva}}
\and A.\,Fiorenzano\inst{\ref{i:INAFSpain}}
\and A.\,Harutyunyan\inst{\ref{i:INAFSpain}}
\and R.\,D.\,Haywood\inst{\ref{i:Exeter}, \ref{f:Ernest}} 
\and D.\,W.\,Latham\inst{\ref{i:Harvard}}
\and M.\,López-Morales\inst{\ref{i:Harvard}}
\and V.\,Lorenzi\inst{\ref{i:INAFSpain}, \ref{i:Laguna}}
\and C.\,Lovis\inst{\ref{i:geneva}} 
\and L.\,Malavolta\inst{\ref{i:Padova}, \ref{i:INAFPadova}} 
\and E.\,Molinari\inst{\ref{i:Selargius}}
\and A.\,Mortier\inst{\ref{i:Birming}}
\and M.\,Pedani\inst{\ref{i:INAFSpain}}
\and F.\,Pepe\inst{\ref{i:geneva}}
\and M.\,Pinamonti\inst{\ref{i:Torino}}
\and E.\,Poretti\inst{\ref{i:INAFSpain}} 
\and K.\,Rice\inst{\ref{i:Edinb_SUPA}} 
\and A.\,Sozzetti\inst{\ref{i:Torino}}
}

\institute{Space sciences, Technologies and Astrophysics Research (STAR) Institute, Université de Liège, Allée du 6 Août 19C, 4000 Liège, Belgium \label{i:liege} 
\and D\'epartement d'Astronomie, Universit\'e de Gen\`eve, Chemin Pegasi 51b, CH-1290 Versoix, Suisse \label{i:geneva} 
\and Department of Physics, University of Oxford, Oxford, UK \label{i:oxford} 
\and Centre for Exoplanet Science, SUPA School of Physics and Astronomy, University of St Andrews, North Haugh, St Andrews KY16 9SS, UK \label{i:StAndrews} 
\and Department of Physics, University of Warwick, Gibbet Hill Road, Coventry, CV4 7AL, UK \label{i:Warwick} 
\and INAF - Osservatorio Astrofisico di Torino, Via Osservatorio 20, I10025 Pino Torinese, Italy \label{i:Torino}
\and DTU Space, National Space Institute, Technical University of Denmark, Elektrovej 328, DK-2800 Kgs. Lyngby, Denmark \label{i:Lyngby}
\and Center for Astrophysics ${\rm \mid}$ Harvard {\rm \&} Smithsonian, 60 Garden Street, Cambridge, MA 02138, USA  \label{i:Harvard} 
\and Astrophysics Group, University of Exeter, Exeter EX4 2QL, UK \label{i:Exeter} 
\and STFC Ernest Rutherford Fellow \label{f:Ernest}
\and Fundaci\'{o}n Galileo Galilei -- INAF, Rambla Jos\'{e} Ana Fernandez P\'{e}rez 7, 38712 -- Bre\~{n}a Baja, Spain \label{i:INAFSpain}
\and Instituto de Astrofísica de Canarias, C/Vía Láctea sn, 38205 La Laguna, Spain \label{i:Laguna}
\and Dipartimento di Fisica e Astronomia "Galileo Galilei", Universitá di Padova, Vicolo del l'Osservatorio 3, I-35122 Padova, Italy \label{i:Padova}
\and INAF - Osservatorio Astronomico di Padova, Vicolo dell'Osservatorio 5, Padova, 35122, Italy \label{i:INAFPadova}
\and INAF - Osservatorio Astronomico di Cagliari, Via della Scienza 5, I-09047, Selargius, Italy \label{i:Selargius}
\and School of Physics $\&$ Astronomy, University of Birmingham, Edgbaston, Birmingham, B15 2TT, UK \label{i:Birming}
\and SUPA, Institute for Astronomy, University of Edinburgh, Blackford Hill, Edinburgh EH9 3HJ, Scotland, UK \label{i:Edinb_SUPA}} 
 
\date{Received ?? / Accepted ??} 
 
\abstract
{The Rocky Planet Search (RPS) program is dedicated to a blind radial velocity (RV) search of planets around bright stars in the Northern hemisphere, using the high-resolution echelle spectrograph HARPS-N installed on the Telescopio Nazionale Galileo (TNG).}
{The goal of this work is to revise and update the properties of three planetary systems by analysing the HARPS-N data with state-of-the-art stellar activity mitigation tools. The stars considered are HD\,99492 (83Leo\,B), HD\,147379 (Gl617\,A) and HD\,190007.}
{We employ a systematic process of data modelling, that we selected from the comparison of different approaches. We use YARARA to remove instrumental systematics from the RV, and then use SPLEAF to further mitigate the stellar noise with a multidimensional correlated noise model. We also search for transit features in the Transiting Exoplanets Survey Satellite (TESS) data of these stars.} 
{We report on the discovery of a new planet around HD\,99492, namely HD\,99492 c, with an orbital period of 95.2 days and a minimum mass of $m~\sin i =$ 17.9 $M_{\oplus}$, and refine the parameters of HD\,99492 b. We also update and refine the Keplerian solutions for the planets around HD\,147379 and HD\,190007, but do not detect additional planetary signals. We discard the transiting geometry for the planets, but stress that TESS did not exhaustively cover all the orbital phases.} 
{The addition of the HARPS-N data, and the use of advanced data analysis tools, has allowed us to present a more precise view of these three planetary systems. It demonstrates once again the importance of long observational efforts such as the RPS program. Added to the RV exoplanet sample, these planets populate two apparently distinct populations revealed by a bimodality in the planets' minimum mass distribution. The separation is located between 30 and 50\,$M_{\oplus}$.} 
{}


\maketitle
\section{Introduction} \label{Intro} 
Since the early 2000s, the HARPS spectrograph \citep{Pepe2000, Mayor2003}, installed on the ESO-3.6m telescope at La Silla observatory, Chile, has seen numerous breakthroughs in the detection and characterisation of new worlds. Initially with a Guaranteed Time Observation (GTO) programme, and then an ESO Large Programme, the instrument dedicated its time to observe a large sample of stars. This blind planet search was performed on a well-chosen sample of main sequence and quiet stars. Among the principal results of this HARPS survey was the discovery of a new category of exoplanets, the sub-Neptune and Super-Earth planets on close-in orbits \citep{Mayor2008, Lovis2009}. A significant number of detections later suggested that these planets are among the most abundant in the exoplanet population \citep[e.g.][]{Mayor2011, Lovis2011, LoCurto2013, Bonfils2013, Astudillo2017b, Delisle2018, Unger2021}. Their over-abundance was strengthened by the Kepler space telescope, which scrutinised a small area of the North sky for 3.5 years in search for transiting planets \citep{Latham2011, Borucki2011, Fabrycky2014}. 

The successful HARPS story motivated the development of a 'HARPS twin' in the Northern hemisphere, HARPS-N \citep{Cosentino2012, Cosentino2014}. This high resolution spectrograph is mounted on the 3.6m Telescopio Nazionale Galileo (TNG), located at the Roque de Los Muchachos observatory on the La Palma Island, Spain. The instrument is embedded in a vacuum-controlled tank, ensuring a high level of stability in temperature and air pressure. It covers the wavelength range 383 - 693 nm, and reaches a spectral resolution of R = 115\,000. The instrument was built with two primary science goals. 1) To ensure a synergy with the Kepler \citep{Borucki2010}, then K2 \citep{Howell2014}, TESS \citep{Ricker2015} and CHEOPS \citep{Benz2021} missions. This is achieved via a precise RV follow-up of promising transiting candidates. Ultimately, such a strategy provides the community with bulk density measurements of planets, which are essential to precisely characterise the planet population and constrain formation and evolution processes \citep[e.g.][]{Rajpaul2017, Malavolta2018, Bonomo2019, Cloutier2020, Mortier2020, Lacedelli2021}. 2) To undertake a blind RV search for low-mass exoplanets, similar to the historical program of HARPS but in the Northern sky. To achieve this, we started the Rocky Planet Search (RPS) program with HARPS-N. It consists in a sample of 58 bright stars widely-spread in right ascension, many of which presenting low levels of activity and hence amenable for precise planet characterisation \citep[see][for a presentation of the RPS program]{Motalebi2015}. Unlike the HARPS sample however, we highlight that some targets with moderate activity are also part of this sample, to serve as test cases for the development of tools aimed at mitigating stellar activity \citep[e.g.][]{Cretignier2020a, CollierCameron2021, DeBeurs2022} in parallel with the constant RV monitoring of the Sun \citep{CollierCameron2019, Dumusque2021}. 

In order to fulfil these scientific objectives and discover low mass exoplanets, stellar activity has to be treated carefully. The RV amplitude induced by small planets on host main-sequence quiet stars is comparable to the contaminating effect of stellar activity. In order to mitigate the latter, we first adapt the observational strategy of the RPS sample. Following the prescriptions of \citet{Dumusque2011}, we systematically took two exposures of fifteen minutes each per night, with at least two hours between each exposure. This allows us to mitigate noise emanating from stellar oscillations and granulation on different scales \citep{Dumusque2011, Chaplin2019}. Coupling this observational strategy with intense monitoring led to the detection of four planets around HD\,219134 \citep{Motalebi2015}. It is one of the few systems where a planet was first discovered via the RV technique, and subsequently found to transit its host star (HD\,219134 b, with the use of the Spitzer space telescope). Two years later, a second planet was also observed to transit \citep{Gillon2017}.  

In addition to the above, significant efforts have been dedicated to develop tools for stellar activity mitigation and RV time-series analysis \citep[e.g.][]{Delisle2020a, Delisle2020b, CollierCameron2021, Cretignier2021, Hara2022, DeBeurs2022}. Simultaneously, the RPS program recently benefited from new Data Reduction Software (DRS) inspired from the ESPRESSO data reduction pipeline, and further augmented by the removal of some specific HARPS-N systematics via a study of the solar data \citep{Dumusque2021}. The convergence of all these efforts motivated us to revisit the RPS sample, in search for new exoplanet candidates. A first outcome of this work is the validation of HD\,79211 b \citep{DiTomasso2023}, an exoplanet in a 24.4 days orbit and with a minimum mass of 10.6 $M_{\oplus}$. More recently, the RV analysis of HD\,166620 and HD\,144579 demonstrated the potential for HARPS-N to reach sub-metre per second detectability in blind searches \citep{AnnaJ2023}. 

In this paper, we present the results of the analysis of three systems, namely HD\,99492, HD\,147379, and HD\,190007, each of which was already known to host a planet candidate. We aim to present an homogeneous procedure to analyse the data, using the new techniques introduced above. The present paper is organised as follows. Sect. \ref{Sect:Obs_and_StellarParam} reports the observations available for each star, and focuses on the estimation of the stellar parameters. Sect. \ref{Sect:Strategy} describes the various tools as part of our strategy to analyse the data. Then, Sect. \ref{Section_HD99492DataAnalysis}, \ref{Section_HD147379DataAnalysis}, and \ref{Section_HD190007DataAnalysis} present our analyses of the data and their modelling with planetary Keplerian orbits. Finally, we conclude and discuss these results in Sect. \ref{Conclusion}.

\section{Observations and stellar parameters}
\label{Sect:Obs_and_StellarParam}
\subsection{Observations}
\subsubsection{HD\,99492}
\begin{figure} 
    \centering
    \includegraphics[width=\columnwidth]{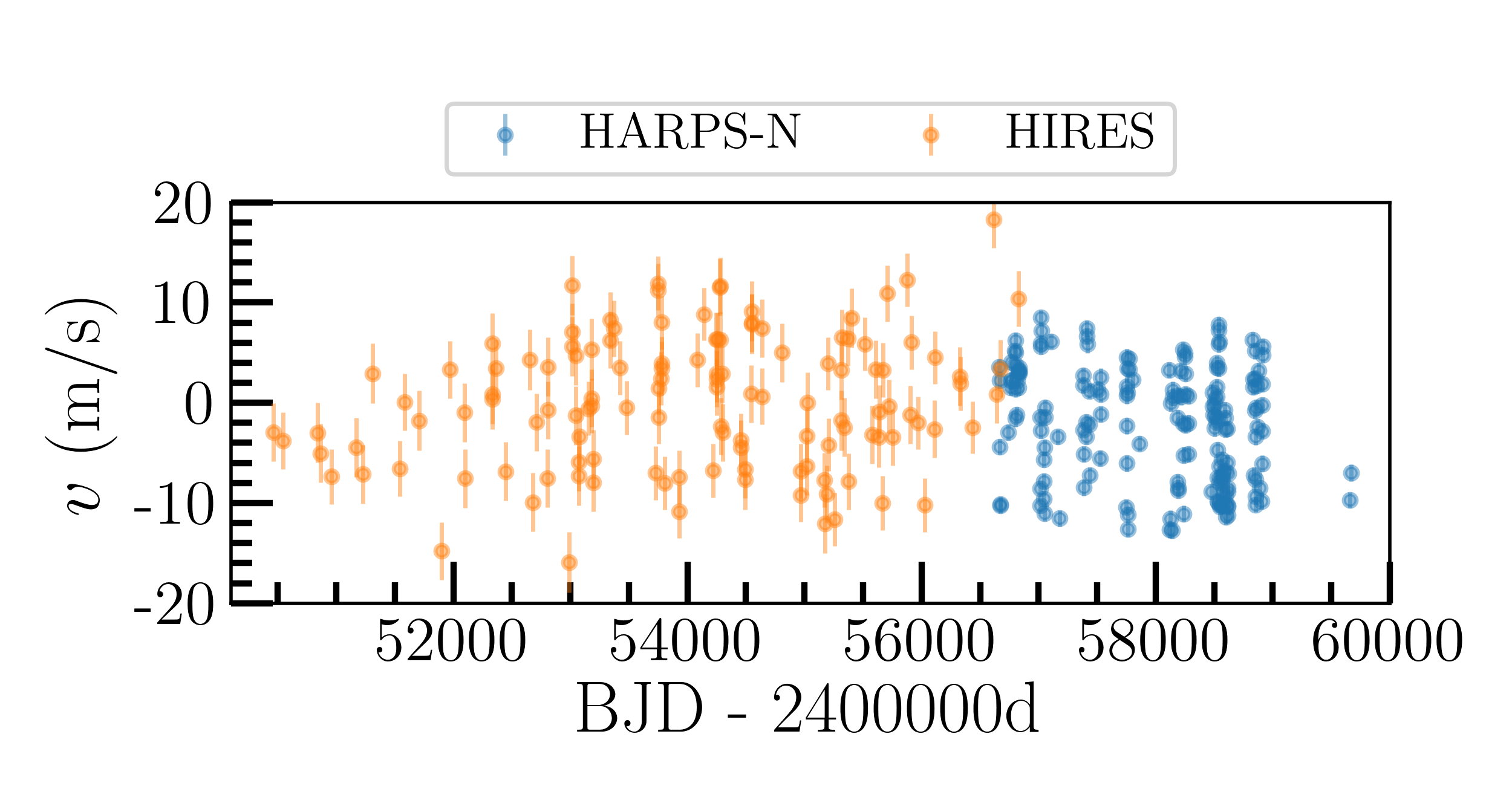}
        \includegraphics[width=\columnwidth]{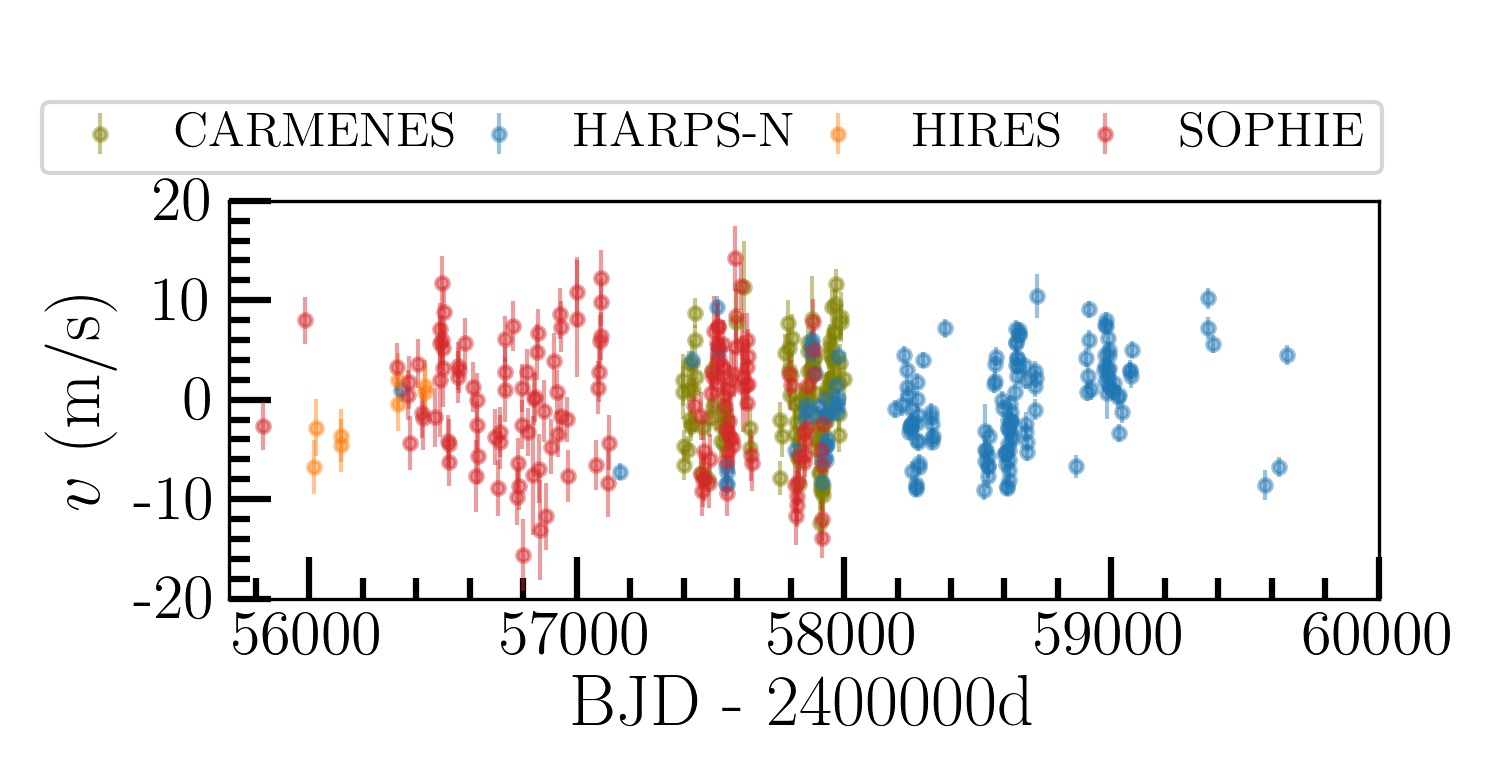}
            \includegraphics[width=\columnwidth]{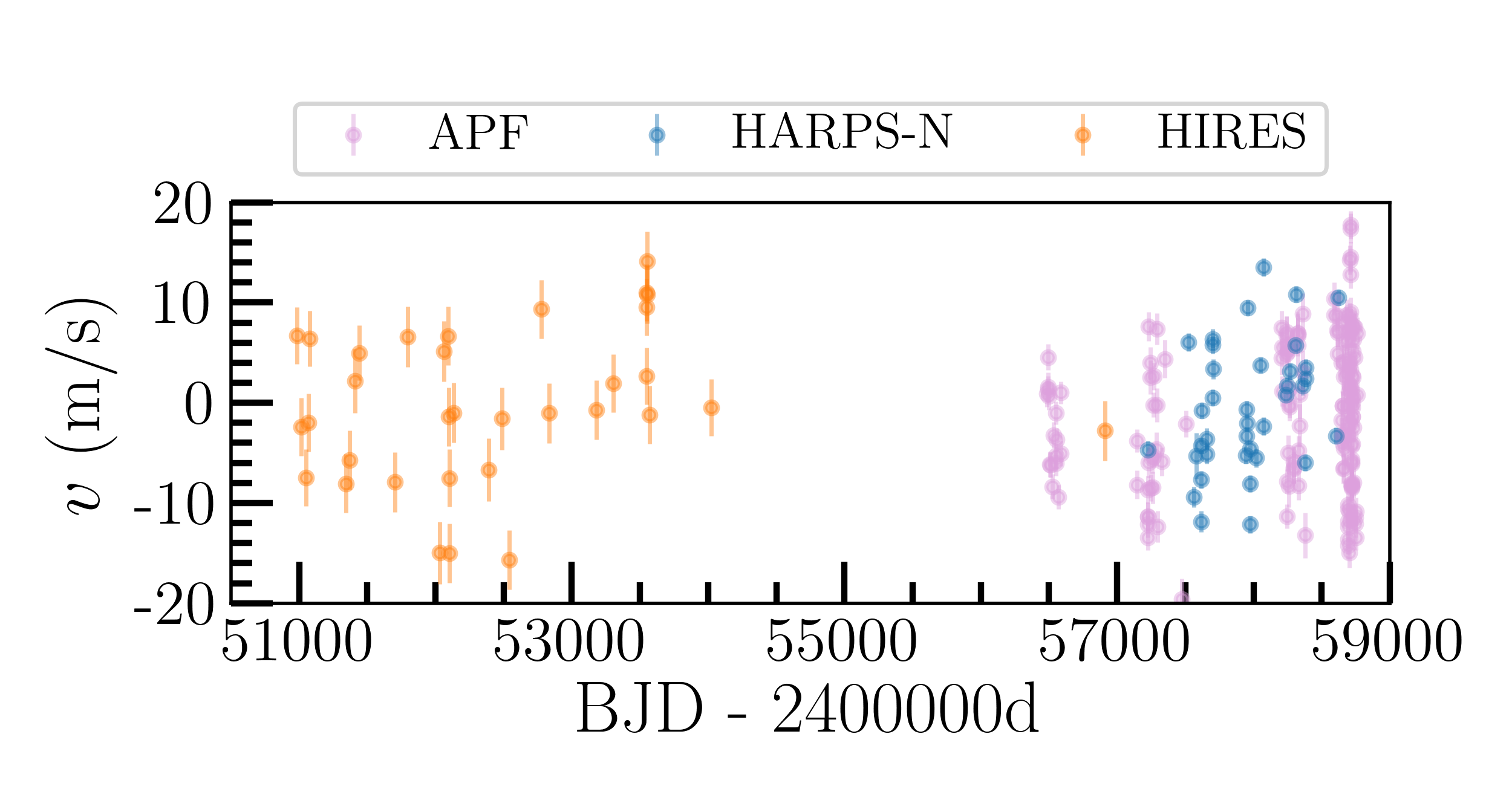}
    \caption{RV time-series of the three stars analysed in this study. \textit{Top} -- HD\,99492 (HARPS-N and HIRES data), \textit{Middle} -- HD\,147379 (HARPS-N, SOPHIE, CARMENES and HIRES data), \textit{Bottom} -- HD\,190007 (HARPS-N, HIRES and APF data).}
    \label{fig:RVtimeseries}
\end{figure}

HD\,99492, also denoted as 83\,Leo B, was observed with HARPS-N between January 15, 2014 and June 17, 2022 using the observational strategy mentioned above (per night, two 15 minutes exposures separated by at least two hours). A total of 202 nightly binned spectra were recorded. For each binned spectrum, we extracted the RV information from cross-correlation of the stellar spectra with a K2 mask. We also derived different stellar activity indicators such as the bisector span, the full width at half maximum (FWHM) of the cross-correlation function (CCF), the S-index and H$_{\alpha}$. 178 Keck/HIRES \citep{Vogt1994} publicly available RV measurements were also taken over a time-span of 6365 days, between Jan 13, 1997 and June 19, 2014. In the top panel of Fig. \ref{fig:RVtimeseries} we present the RV time-series of the combined HIRES and HARPS-N datasets, including an adjusted offset. 

HD\,99492 has 71 publicly available \textsc{Hipparcos} photometric measurements, and 368 observations from the Automatic Photoelectric Telescope \citep[APT --][]{Henry1999, Kane2009} at Fairborn observatory, Arizona (USA). This ensemble of photometric data did not allow \citet{Kane2016} to constrain the stellar rotation period, and will not be used in our analyses. Additionally, TESS has observed the star in sectors 45 and 46 (i.e. in Nov and Dec, 2021) with a 2-min cadence. 

Data from the third \textit{Gaia} Early Data Release \citep[EDR3,][]{GaiaEDR3_2020} constrain the parallax of HD\,99492 to 55.062$\pm$0.030\,masec, leading to an updated distance estimate of 18.16$\pm$0.01\,pc from the Sun. The star forms a gravitational binary with HD\,99491, also named 83\,Leo A. With a reported angular separation between HD\,99492 and HD\,99491 of 28.3\,asec, their mutual distance projected on the sky was estimated to 513.9\,AU. We note furthermore the existence of 6 CORAVEL RV measurements of HD\,99492 taken between the years 1983 and 1993, and 13 RV of HD\,99491 gathered between 1980 and 1993, with a mean RV error on the individual measurements of $\sim$300\,\ms. The precision on those data is not significant enough to detect a slope in the RV due to the gravitational influence between the two stars \citep{Halbwachs2018}. 

\subsubsection{HD\,147379}
HD\,147379, also named Gl\,617 A, was observed with HARPS-N between February 18, 2013 and August 13, 2020. A total of 165 nightly binned spectra were recorded. The same information as for HD\,99492 was extracted from the spectra, this time employing a M0 mask. The star benefits also from publicly available observations performed with three other spectrographs. HIRES gathered 30 RV measurements over about 13 years (from May 15, 2000 to May 21, 2013); the CARMENES instrument \citep{Quirrenbach2018} obtained 114 measurements over $\sim$ 600 nights (from Jan 10, 2016 to Sep 2, 2017); the SOPHIE+ spectrograph\footnote{SOPHIE+ was started in 2011, and consists in an upgraded version of SOPHIE \citep{Perruchot2008}.} \citep{Perruchot2011, Bouchy2013}, finally, brings an additional 163 high precision RV measurements obtained over nearly six years (from Sep 22, 2011 to June 17, 2017). The RV time-series of the combined instruments is presented in the middle panel of Fig. \ref{fig:RVtimeseries}. For the sake of clarity, we present a time-series centred on the CARMENES, SOPHIE and HARPS-N data. The full RV time-series is presented in Appendix -- Fig. \ref{figApp:FullRV_147379}. 

HD\,147379 was also extensively observed in 28 TESS sectors with a 2-min cadence (sectors 14-21, 23-26, 40-41, and 47-60). The total observing window covers 1279 days, from Jul 18, 2019 to Jan 18, 2023, and contains 494\,593 flux measurements. Also, the star benefits from 103 \textsc{Hipparcos} photometric measurements, obtained over a time-span of 2.5 years. Finally, it was observed by KELT \citep{Pepper2007} over slightly less than two years, for a total of 3304 measurements. 

The \textit{Gaia} EDR3 estimates the parallax of HD\,147379 to 92.877$\pm$0.015\,masec, which converts to a distance of 10.767$\pm$0.002\,pc from the Sun. The star has a nearby companion with a similar proper motion, namely Gl\,617 B. This gravitationally-bound binary presents an angular separation of 64.4 asec, which corresponds to an on-sky projected distance of 693.4\,AU. 

\subsubsection{HD\,190007}
HD\,190007 was observed with HARPS-N between July 21, 2015 and May 19, 2019, for a total of 37 nightly binned spectra. We derived the RV from cross-correlation of the stellar spectra with a K2 mask. These data complement the set of publicly available RV measurements from the HIRES and APF/Levy \citep{Vogt2014} spectrographs. The former gathered 33 nightly binned RV over more than 16 years (from Jun 19, 1998 to Sep 10, 2014), while the latter obtained 89 nightly binned high resolution spectra over more than six years (from Jul 9, 2013 to Oct 2, 2019). The RV time-series of the combined instruments is presented in Fig. \ref{fig:RVtimeseries} -- bottom panel. 

HD\,190007 was observed with the APT at Fairborn observatory for over twenty years, gathering a total of 1092 flux measurements. Furthermore, the star was followed up by TESS in sector 54 (in Jul, 2022), with a 2min cadence.

\subsection{Stellar parameters and activity}
\label{Sect:StellarParam} 
Table \ref{tab:stellar-param} reports on the main stellar parameters gathered from the literature for the three stars. We stress that the uncertainties on the effective temperature $T_{\rm eff}$ and the stellar mass $M_{\star}$ are intriguingly small. As \citet{Tayar2022} showed, important factors such as the systematic uncertainties on the fundamental stellar properties, and the scatter in the results from different stellar evolution models are often ignored. Due to these factors, they estimate a minimum uncertainty of 2.0$\% \pm$ 0.5$\%$ on $T_{\rm eff}$, to be added in quadrature to the reported uncertainties. Regarding the stellar mass estimation, the authors provide maximal fractional offsets between different model grids in the space of luminosity and effective temperature (cf. their Fig. 5). For the three stars of this work, the models variance adds uncertainties of between 5$\%$ and 10$\%$ on $M_{\star}$. This additional error budget transposes directly to the mass estimates of exoplanets (a similar uncertainty increase is also applicable on the radius, but this is not relevant for this work). Finally, we also note that the distances reported in Table\,\ref{tab:stellar-param} were obtained from the parallax measurements reported in the \textit{Gaia} EDR3 catalogue. 

\subsubsection{HD\,99492}
\label{Sect:StellarParam_99492}
HD\,99492 is a K-type main-sequence bright star (V = 7.6). Its stellar companion 83\,Leo A is a late G-type star (V = 6.5). HD\,99492 shows a low activity level, notably expressed by its \rhk = -4.93 \citep{Marcy2005}. 

\citet{Kane2016} reported the detection of a magnetic cycle on HD\,99492, with a periodicity between 3000 and 5000 days\footnote{A planet was originally proposed as the origin of this long-term periodicity, namely HD\,99492 c \citep{Meschiari2011}. The existence of this planetary companion was rejected by \citet{Kane2016}.}. In the time-span of the HARPS-N observations, which is 2270 days, we see a quadratic-like trend in the data. A similar trend is observed in the time-series of our various spectroscopic activity indicators, and we therefore attribute this trend to stellar activity (cf. Appendix -- Fig. \ref{figApp:99492_AllSpectro}). 

From the estimation \rhk = -4.93, \citet{Marcy2005} estimate the stellar rotation period to be $\sim$45 days. They used the empirical relations from \citet{Wright2004} for this estimate, but stressed that HD\,99492 has a $B-V$ index that lies outside of the verified calibration domain. \citet{Kane2016} analysed both the APT and \textsc{Hipparcos} photometry, with no conclusive result. From our 202 nightly binned HARPS-N spectra, we detect significant periodic signals at around 40 - 45 days in the periodograms of different stellar activity indicators, such as the FWHM and S-index (cf. Appendix -- Fig. \ref{figApp:99492_AllSpectro}). They strengthen the previous analyses, and indicate a stellar rotation period between 40 and 45 days. 

\begin{figure} 
    \centering
    \includegraphics[width=\columnwidth]{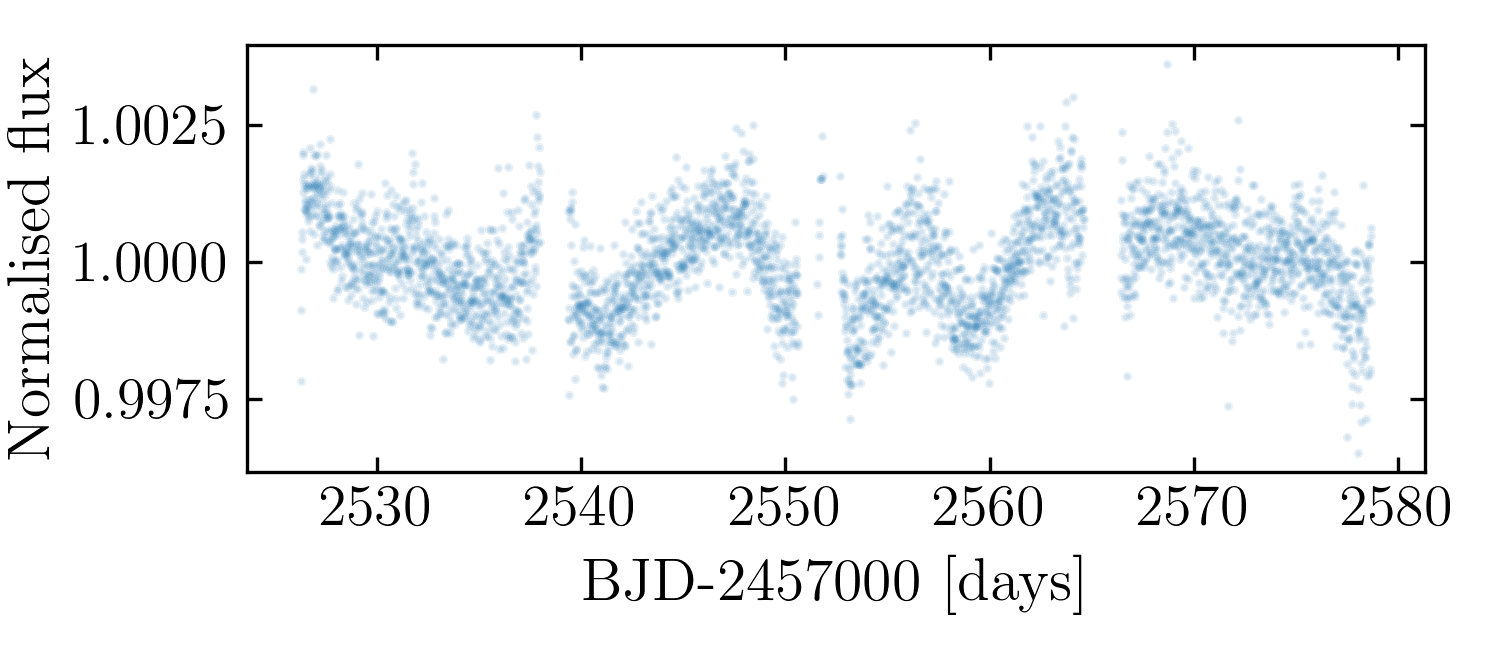}
    \includegraphics[width=\columnwidth]{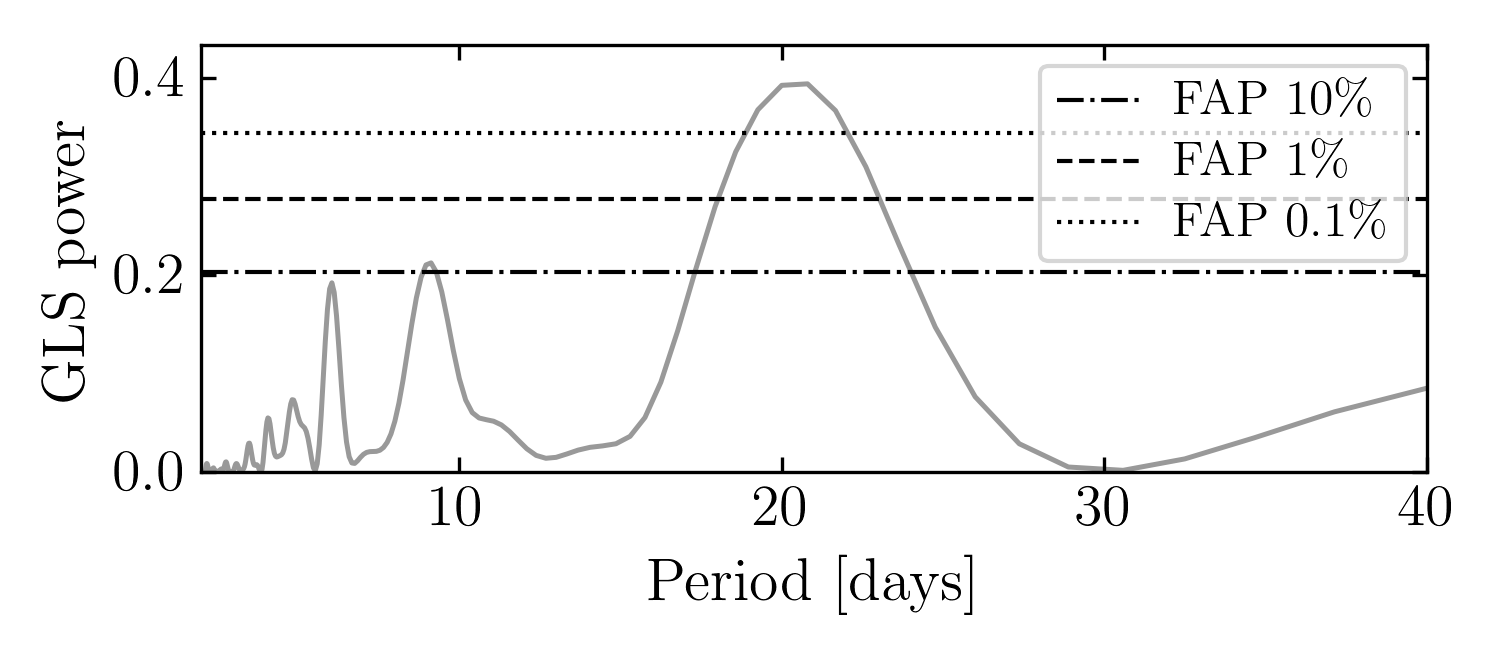}
    \caption{HD\,99492: Periodic signal search in TESS sectors 45 and 46. \textit{Top}: TESS light-curve extracted using \texttt{lightkurve}, and corrected from systematics (2 sectors). \textit{Bottom}: GLS periodogram computed between 2 and 40 days of the 1-day binned light-curve.} 
    \label{fig:HD99492_TESS_GLS}
\end{figure}

Finally, we considered the TESS photometric measurements in search for periodic signals. However, we found that the 2-min cadence Simple Aperture Photometry (SAP) fluxes of the two available sectors (45 and 46), which are derived by the TESS Science Processing Operations Center \citep[SPOC,][]{Jenkins2016}, are strongly contaminated by the Moon. Therefore, to treat the systematics we extracted our custom light-curve from the calibrated 2-min cadence target pixel files using \texttt{lightkurve}\footnote{Lightkurve is an open-source Python package for Kepler, K2 and TESS data analysis \citep{Lightkurve2018}.}. For each sector, the extraction of the light-curve was performed via aperture photometry employing the TESS SPOC pipeline mask. The latter is shifted from the photometric centre of the source, so to limit the contamination from the nearby companion HD\,99491, which lies 28$\arcsec$ away from HD\,99492 and is hence blended with the latter on the TESS detector. During the light-curve extraction, we corrected the fluxes from the background systematics, and notably we partly accounted for the contamination of HD\,99491. This was done by building a matrix of the out-of-aperture pixels (also called a design matrix), and performing a principal component analysis to retrieve the five main trends that we then removed from the pixels inside the aperture. We also retrieved the cotrending basis vectors (CBV) computed by the Pre-search Data Conditioning \citep[PDC,][]{Smith2012, Stumpe2014} unit of the TESS SPOC pipeline. These are vectors that contain the most common systematic trends found for each TESS CCD, and which are produced for every sector. They are divided into three categories: the spike CBV contain the short impulsive systematics, the single-scale CBV contain all systematics in a single set of CBV, and multi-scale CBV are spread into three different band-passes. We found that the combination of single-scale and spike CBV is best suited to correct our light-curve from systematics while preserving stellar signals with time-scales of a few tens of days. We detrended the light-curve with these CBV, jointly with the background subtraction mentioned above. We then removed outlier fluxes standing beyond five standard deviations (5$\sigma$) of the smoothed light-curve, we normalised the flux data in each sector, and merged the two sectors. Finally, we undertook a periodic signal search in the resulting light-curve, via a generalised Lomb-Scargle (GLS) periodogram \citep{Zechmeister2009}. We found a significant broad periodic signal peaking at $\sim$21 days, which is consistent with half of the expected stellar rotation period. We present those results in Fig. \ref{fig:HD99492_TESS_GLS}. 

\subsubsection{HD\,147379}
HD\,147379, or equivalently Gl\,617 A, is a M-type main-sequence star with high proper motion (V = 8.9 mag). It is also part of a gravitationally bound binary with Gl\,617 B, which is another M-type main-sequence star fainter than the former (V = 10.6 mag). HD\,147379 presents a reasonably low activity index too, with \rhk = -4.75. 

A correlation is visible in the spectroscopic data of HD\,147379 between the RV and FWHM or S-index. We observe a long-period variation in those time-series, which is attributed to the stellar magnetic activity. The HARPS-N data spans 2734 days, and like HD\,99492, does not cover a full magnetic cycle. Only a quadratic trend is observed (cf. Appendix -- Fig. \ref{figApp:147379_AllSpectro}). 

Based on an empirical relation between X-ray activity and stellar rotation, \citet{Reiners2018} estimate a stellar rotation period of $P_{rot}=31\pm20$ days. Furthermore, with the use of the empirical relation of \citet{Astudillo2017a} between \rhk and $P_{rot}$ for M dwarfs, \citet{Hobson2018} evaluate the rotation period of HD\,147379 to be $28.8\pm6.1$ days. These authors also analysed the \textsc{Hipparcos} photometry - 103 measurements over 2.5 years - but without conclusive results except a hint of signal at $\sim$ 21 days. \citet{Pepper2018} performed a photometric analysis of HD\,147379 with KELT data, and found $P_{rot}=22$ days. All the previous studies are therefore consistent with each other. We analysed the stellar activity indicators time-series of HARPS-N to search for any hint of stellar rotation. The most significant periodic signals were seen in the S-index, which reveal a cycle of 21.5 days. This strongly supports the results of previous analyses on this star.  

\begin{figure} 
    \centering
    \includegraphics[width=\columnwidth]{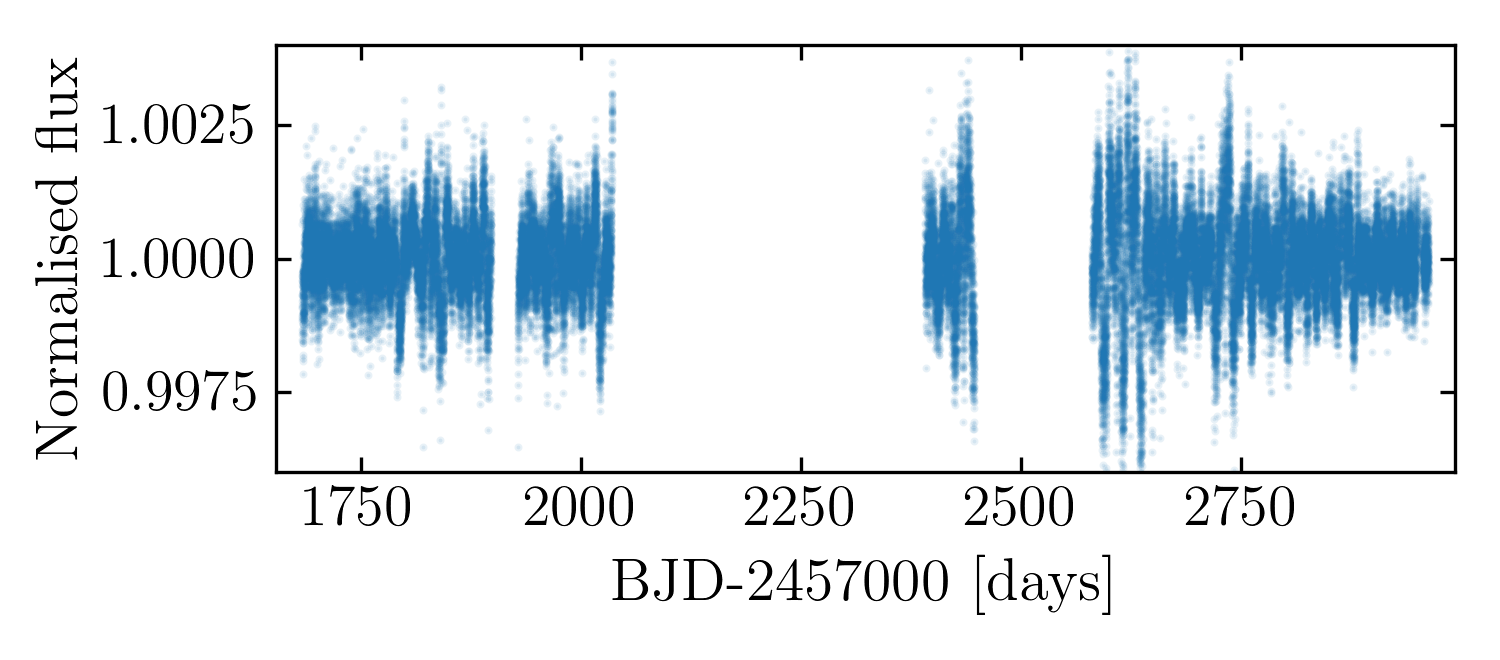}
    \includegraphics[width=\columnwidth]{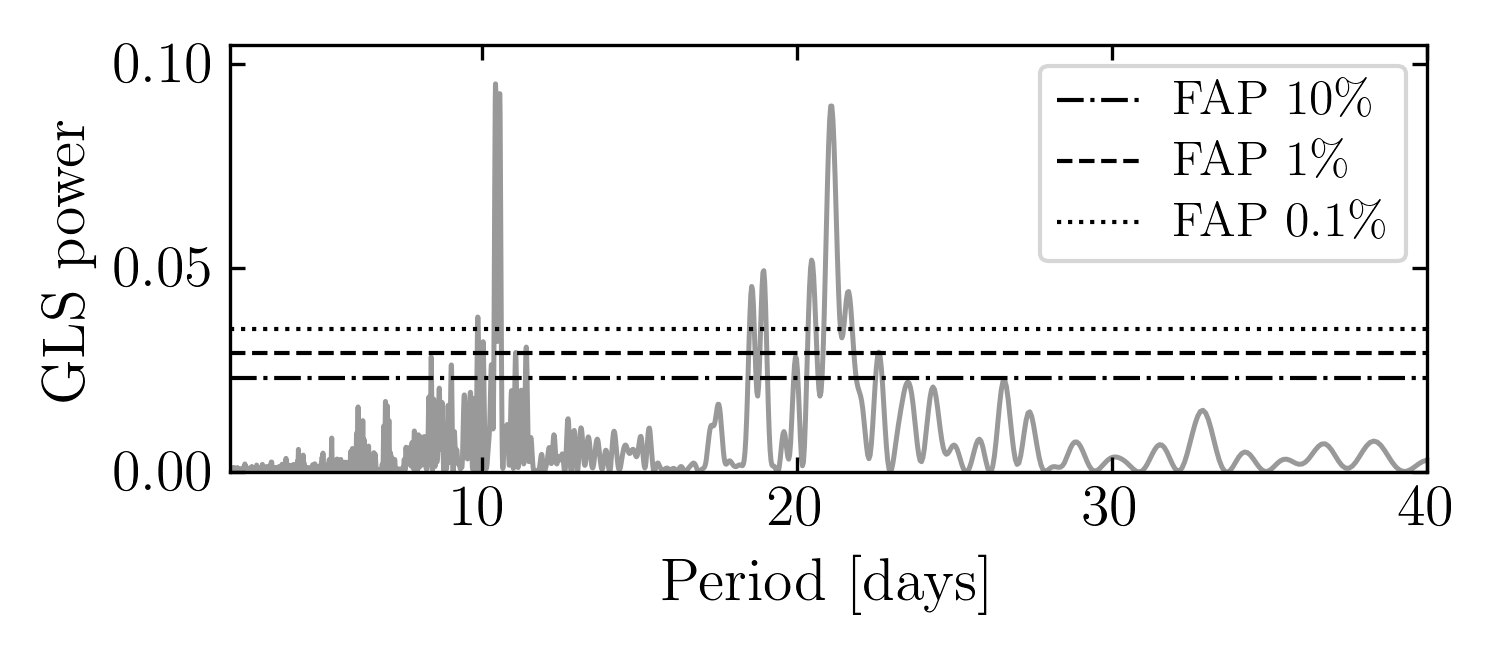}
    \caption{HD\,147379 TESS photometry. \textit{Top}: TESS light-curve extracted using \texttt{lightkurve}, and corrected from systematics (28 sectors). \textit{Bottom}: GLS periodogram computed between 2 and 40 days of the 1-day binned light curve.}
    \label{fig:HD147379_TESS_GLS}
\end{figure}

We analysed the extensive TESS 2-min cadence photometric observations (28 sectors) in search for any photometric periodic variation. To proceed, we extracted the light-curve of each of the 28 sectors following the same procedure as described in Sect. \ref{Sect:StellarParam_99492}. Notably, we note again that the SPOC pipeline mask conveniently avoids a nearby source on the detector, and we opt for this mask to perform aperture photometry. We performed a periodic signal search in our custom light-curve corrected from systematics, by computing a GLS periodogram. Both the TESS time-series and the periodogram are presented in Fig. \ref{fig:HD147379_TESS_GLS}. We found two significant signals at 21 and 10.4 days, and we associate them with the stellar rotation period and its half, respectively. This analysis is consistent with the period found in the spectroscopic observations and in \citet{Pepper2018}. We therefore conclude that the rotation period is highly likely to be around 21 days.

\subsubsection{HD\,190007}
HD\,190007 is a bright K-type main-sequence star (V = 7.46). The star belongs to the family of BY Draconis variables \citep{Kazarovets2003}. It is indeed moderately active (\rhk = -4.65) and shows a clear photometric quasi-periodic variation. It is slightly metal-rich, with an age that remains uncertain. However, \citet{Burt2021} suggest that the star is at least 1 Gyr old, based on data from the Kepler mission. 

The radial velocities combined from the different instruments, which span more than twenty years, present a linear trend of $\sim$1.4 m s$^{-1}$ yr$^{-1}$. This is in line with the measurement of a statistically significant (S/N>3) proper motion anomaly between \textsc{Hipparcos} and \textit{Gaia} DR3 mean epoch \citep{Kervella2022}. Both the RV and the \textsc{Hipparcos}-\textit{Gaia} absolute astrometry hint at the existence of a long-period companion, and future observations will help shed light onto this hypothesis. 

Both from the spectroscopic stellar activity indicators and the analysis of the available photometry, \citet{Burt2021} report a rotation period of $P_{rot}\sim$29 days. This is in agreement with the previous estimation of \citet{Olspert2018} using Gaussian Process (GP) modelling, which found $P_{rot}=27.68$ days. We observe a similar periodicity in the spectroscopic indicators when we add our HARPS-N measurements to the existing HIRES and APF data (cf. Appendix -- Fig. \ref{figApp:190007_AllSpectro}). 

\begin{figure} 
    \centering
    \includegraphics[width=\columnwidth]{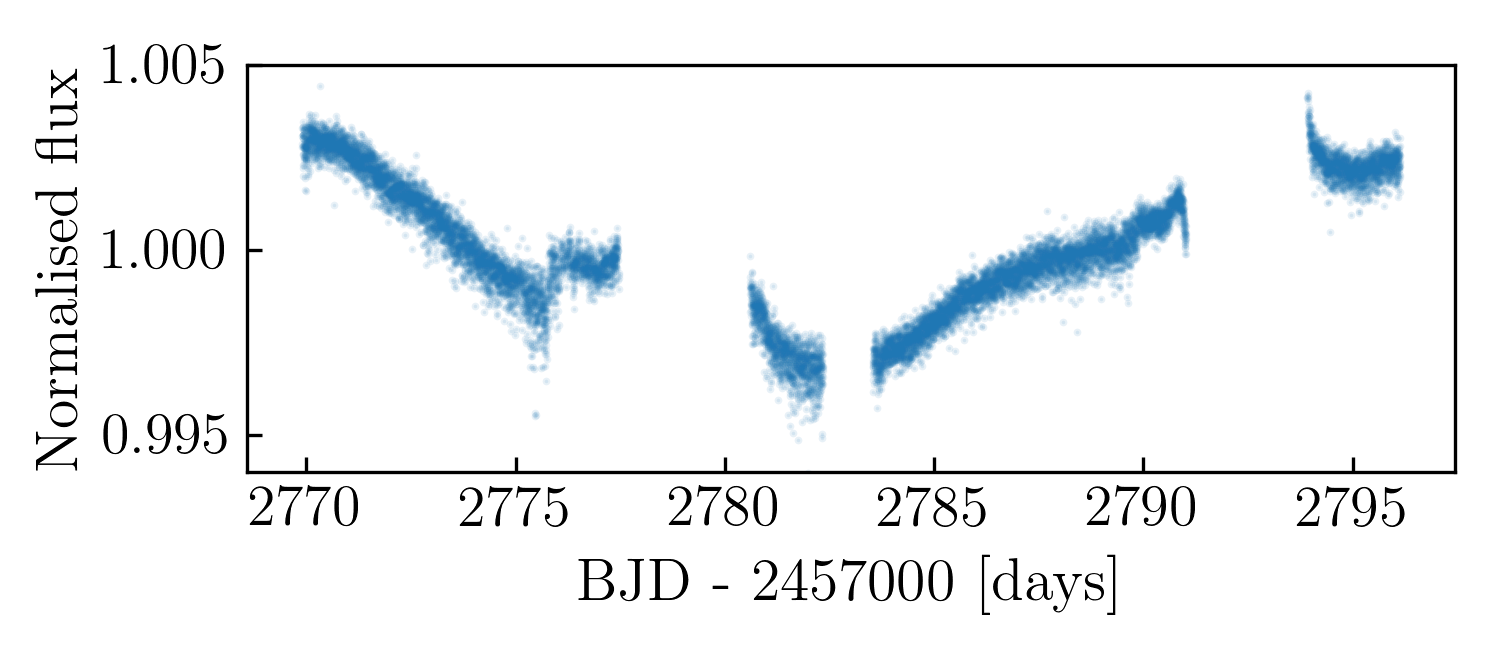}
    \caption{HD\,190007 TESS SAP normalised time-series (sector 54).}
    \label{fig:HD190007_TESS_timeseries}
\end{figure}

We also analysed the TESS photometric data of this star (sector 54) in search for potential periodicity. We were limited by the small time-span of the observations ($\sim$26 days). A simple inspection of the normalised SAP light curve, consisting of 13\,238 flux measurements, reveals a long-term variation covering nearly a full cycle over the sector. We present the normalised SAP light curve in Fig. \ref{fig:HD190007_TESS_timeseries}. This is in agreement with previous analyses, suggesting a stellar rotation period around 30 days. Yet, the time span of the TESS observations is too short to firmly confirm it.

\begin{table}
\begin{adjustbox}{width=1\columnwidth}
\footnotesize
\centering
\begin{threeparttable}
\caption{Stellar parameters for HD\,99492, HD\,147379, and HD\,190007.}
\label{tab:stellar-param}
\begin{tabular}{@{}lcccc@{}}
\toprule
Parameters [Units]  & HD\,99492     & HD\,147379    & HD\,190007    \\ \midrule
Spectral Types    & K2V$^1$         & M0.0V$^{6}$      & K4V$^{10}$          \\
$V$  {[}mag]   & $7.58^2$       & $8.9^7$  &   $7.46^{10}$ \\
$B-V$  {[}mag]   & $1.02^3$       & $1.11^7$  & $1.128^{10}$   \\
Distance  {[}pc]  & $18.161\pm0.010^{4}$         & $10.767\pm0.002^{4}$   & $12.715\pm0.003^{4}$ \\
\teff {[}K]     & $4929\pm44^2$       & $4090\pm50^7$   &  $4610\pm20^{10}$  \\
\logg  {[}cgs]   & $4.57^2$        & $4.609\pm0.012^8$     & $4.58\pm0.02^{10}$  \\
\feh {[}dex]   & $0.3\pm0.03^2$       & $0.16\pm0.16^7$     & $0.16\pm0.05^{10}$  \\
\rhk  & $-4.93^5$      &   $-4.75\pm0.14^9$   & $-4.65^{11}$ \\ 
\vsini  {[}\kms]  & $0.41\pm0.5^2$   & $2.7\pm1.5^7$   &  $2.55^{10}$ \\
$M_\star$ {[}\Msun] & $0.85\pm0.02^2$   & $0.58\pm0.08^7$     &  $0.77\pm0.02^{10}$  \\ \bottomrule 
\end{tabular}
\end{threeparttable}
\end{adjustbox}
\tablebib{\footnotesize (1)~\citet{Meschiari2011} ; (2)~\citet{Kane2016} ; (3)~\citet{Koen2010} ; (4)~\citet{GaiaEDR3_2020} ; (5)~\citet{Marcy2005} ; (6)~\citet{Alonso-Floriano2015} ; (7)~\citet{Reiners2018} ; (8)~\citet{Stassun2019} ; (9)~\cite{Hobson2018} ; (10)~\citet{Burt2021} ; (11)~\citet{Olspert2018}} 
\end{table}

\section{Data analysis tools and general strategy} 
\label{Sect:Strategy} 
In this work, we employ a systematic approach to analyse the observations, making use of a number of versatile tools. We aim to demonstrate the ability of these new tools to correct for stellar activity effects and to provide unbiased planetary parameters. Initially, we explore the spectroscopic data -- RV and stellar activity indicator time-series -- with the Data $\&$ Analysis Centre for Exoplanets (DACE). This web-platform, hosted at the University of Geneva, is dedicated to extra-solar planet data visualisation, exchange and analysis\footnote{DACE can be accessed via \url{https://dace.unige.ch}.}. Among many other functionalities, it shows the various spectroscopic time-series, but also the correlation plots between the RV and indicators. Additionally, it offers the possibility to perform interactive fits of the time-series, optionally with Keplerians. Therefore, we systematically use DACE to get a first valuable view of the data, and identify potential stellar activity signatures. As a second step, we initiate in-depth data analysis. We first correct the HARPS-N data from instrumental systematics and some stellar activity features thanks to the YARARA software \citep{Cretignier2021}. Then, the datasets are investigated in light of the correlated noise model SPLEAF \citep{Delisle2020b}. Below, we briefly summarise these two tools. Finally, we explore the model parameters via a Monte Carlo Markov Chain (MCMC) algorithm. For the latter, we employ \texttt{samsam}, a scaled adaptive metropolis algorithm \citep{Delisle2022MCMC}.

\subsection{YARARA} 
\label{Sect:YARARA}
YARARA \citep{Cretignier2021} is a post-processing method that aims to improve the spectra extracted by the classical DRS \citep{Dumusque2021} by removing the extra signatures introduced by the instrument and the Earth atmosphere, leading ultimately to an improved RV precision. The method involves correcting a residual spectral time-series with respect to a master spectrum (built to be free from the systematics), where the corrections are made of multi-linear regressions in the wavelength domain. The spectra are corrected for: cosmics, tellurics, ghosts, fringing, instrumental defocus, activity and residuals outliers. The cleaned RV are extracted with a CCF using a tailored line selection based on the master spectrum itself \citep{Cretignier2020a} in order to increase the extraction of the Doppler information content, whereas the merged-order 1d spectra are continuum normalised using RASSINE \citep{Cretignier2020b}. A few adjustments were implemented in the code in order to increase the orthogonality between the stellar activity and instrumental signatures. The main reason for this update was to be able to introduce back the stellar activity component at later stage (see Sect. \ref{Sect:S+LEAF} for a justification). In the old version of the code, both effects were simultaneously fitted using the S-index, the CCF FWHM and the CCF contrast. Those two moments contain redundant stellar activity information. The new version of the pipeline uses an improved version of the S-index using CaIIH\&K lines better corrected for ghosts, whereas the fitted CCF moments are now orthogonal to that component thanks to a Gram-Schmidt orthogonalisation algorithm (see Appendix \ref{appendix:a}). 

YARARA itself is not able to simultaneously fit in the wavelength domain the planetary signals and the systematics. However, since most systematics are fixed in the terrestrial rest-frame, the code assumes that the planetary signals are destroyed (or strongly mixed) in the wavelength domain once such a change of rest-frame is performed. In \citet{Cretignier2021}, the authors showed that injected planetary signals were not absorbed more than 20\% (and only for a planet close to a 1-year harmonic). In order to further avoid such absorption, large RV amplitude signals can be pre-fitted by shifting the spectra according to a determined Keplerian solution. In this work, that preliminary Keplerian solution will be obtained via a RV fit with DACE which matches the orbital elements of the known planets. This Keplerian solution will then be re-injected in the RV obtained on the residual spectra, before that an improved Keplerian solution could be obtained. This process can be called in an iterative way until convergence is reached, which usually occurs after a single iteration.     

Once spectra are corrected for systematics, new proxies orthogonal to a pure Doppler shift can be extracted in the time-domain \citep{Cretignier(2022)}. Under the assumption that the deformations of the line profiles are mainly driven by the line profile itself, the residual spectra $\delta f(\lambda) = f(\lambda) - f_0(\lambda)$ can be expressed rather as $\delta f(\partial f_0 / \partial\lambda, f_0)$, with $f_0$ the master spectrum of the star -- an object that the authors called a 'shell'. An advantage of that space is that a pure Doppler shift possesses an unique signature that can be fitted out analytically with a first order approximation. Other distortions, orthogonal to the pure Doppler shift can then be extracted with a principal component analysis and their associated time-coefficients are used in a multi-linear model to correct for the RV time-series. A cross-validation algorithm is used to determine the number of components to select by randomly rejecting 10\% of the observations. Such a strategy was shown successful to remove the rotational period of stars moderately active and/or potential remaining instrumental systematics uncorrected by YARARA.  

This additional correction process \citep{Cretignier(2022)} will be subsequently designated as YARARA-Shell, as opposed to YARARA-V1 which consists of YARARA \citep{Cretignier2021} further augmented with the few adjustments described above to increase the orthogonality between the stellar activity and instrumental signatures.

\subsection{SPLEAF}
\label{Sect:S+LEAF}
 Correlated noise, or Gaussian Process (GP) models have been the object of increasing research among the exoplanet community over the past ten years. They account for physical processes that might be poorly understood or poorly constrained, such as stellar activity \citep{Haywood2014}. Indeed, instead of a deterministic model, the GP regression technique aims to parametrise the covariance between the measurements. There exist different types of GP, which differ from each other by their kernels, that is by the functional form of the covariance matrices. SPLEAF \citep{Delisle2020b} refers to a class of semi-separable covariance matrices, which builds on the benefits of the celerite kernel \citep{Ambikasaran2015, Foreman-Mackey2017}. SPLEAF can take into account the calibration noise, and due to the semi-separable form of the covariance matrix, it provides low computation costs. Once the GP regression is properly performed, it should account for anything that is not due to the deterministic model nor the measurements noise. Yet in practice, there might be risks of over-fitting, and the planetary signals may be (partly) absorbed by the GP. To limit the risk, \citet{Rajpaul2015} proposed to train the GP simultaneously on the RV and on activity indicators time-series, by modelling the activity-induced RV signals as linear combinations of the GP ($G(t)$) and its derivative $\dot{G}(t)$ \citep{Aigrain2012} 
\begin{equation} \label{eq:GP_RV}
\Delta RV ~ = ~ V_c G(t) ~ + ~ V_r \dot{G}(t) 
\end{equation}
\begin{equation} \label{eq:GP_ind1}
\Delta \alpha ~ = ~ L_c G(t) 
\end{equation}
\begin{equation} \label{eq:GP_ind2}
\Delta \beta ~ = ~ B_c G(t) ~ + ~ B_r \dot{G}(t) 
\end{equation}
for $V_c$, $V_r$, $L_c$, $B_c$, $B_r$ free parameters, and $\alpha$ and $\beta$ referring to some activity indicators. 
Eq. \ref{eq:GP_RV} was first proposed by \citet{Aigrain2012} to account for the effect of stellar spots on the RV variations, where G(t) captures the suppression of the convective blueshift in the spots, and $\dot{G}(t)$ provides the effect of stellar rotation. \citet{Rajpaul2015} extended this framework to activity indicators (Eq. \ref{eq:GP_ind1} and \ref{eq:GP_ind2}), where the use of one equation or another depends on the nature of the indicator. For instance, the S-index is a proxy for the proportion of the visible stellar disk covered by active regions, and is therefore accounted for by Eq. (\ref{eq:GP_ind1}). On the other hand, the bisector span indicator, which informs about the asymmetry of the spectral lines, also depends on the radial velocity of the stellar surface at the location of the spots, and is hence described by Eq. (\ref{eq:GP_ind2}). The planetary signal being only in the RV time-series, it is less likely to be absorbed by the GP if the latter is simultaneously trained on one or several activity indicators. As a drawback though, the computation cost is significantly larger, with a covariance matrix of size $2n\times2n$ where $n$ is the number of measurements and in the case where only one activity indicator time-series is used in parallel of the RV. 
 
\citet{Delisle2022} generalised the SPLEAF model to account for different - (potentially taken at different times) as modelled by Eq. (\ref{eq:GP_RV}, \ref{eq:GP_ind1}, \ref{eq:GP_ind2}), while insuring a computation cost that scales linearly with the total number of measurements. In this work we use this generalised SPLEAF correlated noise model, which is publicly available\footnote{\url{https://gitlab.unige.ch/Jean-Baptiste.Delisle/spleaf}}. The kernel we employ for the quasi-periodic part is an approximation of the squared-exponential periodic (SEP) kernel, where the covariance has the following expression: 
\begin{equation}
k(\Delta t) ~ = ~ \sigma^2 ~ \exp \left(-\dfrac{\Delta t^2}{2\rho^2} - \dfrac{\sin^2\left(\frac{\pi \Delta t}{P}\right)}{2\eta^2}\right). 
\end{equation}
It consists of a sinusoid on top of a decreasing exponential, with $\Delta t$ being the time interval between two measurements. Therefore, this kernel correlates one measurement with the others according to this functional form. The hyperparameters $\sigma$, $\rho$, $P$ and $\eta$ describe respectively the amplitude of the correlated noise, the rate of exponential decay of the correlation, the period of variation of the sinusoid and the length-scale of the periodic component, that is its ability to capture rapidly varying features (smaller values pointing towards sharp variations). These hyperparameters are constrained from the various time-series. SPLEAF develops the periodic component of the SEP kernel in series, and keeps the two strongest harmonics. In practice, we also add a white noise term to each time-series, which consists in an additional term on the diagonal of the correlation matrix. 

In this work, we systematically apply the generalised SPLEAF model -- denoted as the SPLEAF model below -- on the RV and S-index time-series. As was described above, we find in the S-index time-series of the three stars traces of the stellar magnetic cycles and stellar rotation. Therefore, the hyper-parameters of the correlated noise model, which are estimated from the analysis of these different datasets, are expected to converge to values informing us about the stellar activity. We mostly set uninformative priors on the model parameters, that is priors following a uniform law with wide boundaries. In Table \ref{tab:priors}, we synthesise the priors we use in most of our analyses. We explicitly note in the text when certain priors are further constrained. 

We express some caution about using the YARARA dataset with a correlated noise model. Indeed, we noticed a convergence issue of the fit with any dataset that would already be (partially) corrected for the stellar activity features. The reason is that while the activity indicator time-series are still left intact, the RV time-series are corrected for the trends observed in the former. Therefore, the correlated noise model trains successfully on the activity indicators, but struggles to transpose the results to the RV time-series. In conclusion, the correlated noise model does not provide satisfactory results with YARARA-V1 and YARARA-Shell. Nevertheless, YARARA also corrects the various time-series for the instrumental systematics. The use of those datasets is preferred, so to avoid observing spurious periodic signals. Therefore, in the following analyses making use of SPLEAF, we use a modified version of YARARA-V1 where the stellar activity has been re-injected into the RV time-series, but not the instrumental systematics (see Appendix \ref{appendix:a} for an explanation of the process to separate the various components). In other words, we use an alternative version of YARARA where only the instrumental systematics have been corrected for. That allows for the convergence of the fit with the correlated noise model. 

\begin{table*}[t]
\centering
\begin{threeparttable}
\caption{List of priors used for each parameter, unless stated otherwise in the text. RV$_{\mathrm{min}}$ and RV$_{\mathrm{max}}$ are the minimum and maximum measured RV, respectively, for each star.}
\label{tab:priors}
\begin{tabular}{@{}llll@{}}
\toprule
\textbf{Parameter} & \textbf{Units}  & \textbf{Prior Distribution} & \textbf{Description}  \\ \midrule

\underline{Offsets and noise} \\ 
 \\ 

Epoch    & BJD & Fixed at 2\,455\,500    & Reference epoch \\
$\gamma_{inst}$    & \ms   & U [RV$_{\mathrm{min}}$, RV$_{\mathrm{max}}$] & Constant velocity offset    \\
$\sigma_{RV}$ & \ms         & logU [0.001, 100]       & Additional white noise (Jitter)  \\ 
\\ 
$P_{GP}$  &  days  &  U [1, obs timespan]  &  Period of correlated noise \\ 
$\rho_{GP}$  &  days  &  U [1, obs timespan]  &  Decay timescale \\ 
$\eta_{GP}$  &    &  U [0, 100]  & Smoothing parameter \\ 
$\sigma_{GP}$ & \ms & U [0, 100]  &  Amplitude of correlated noise  \\
 \\ 
\underline{Keplerians} \\ 
\\ 
$P$  &  days  & U [1.5, 5000]  &  Orbital period \\ 
$K$  &  \ms  &  logU [0.1, 10$^{5}$]  &  RV semi-amplitude \\ 
$e$  &    &  U [0, 0.8]  &  Orbital eccentricity \\ 
$\omega$  &  radians  &  U [0, 2$\pi$]  &  Argument of periastron \\ 
$\lambda_0$  &  radians &  U [0, 2$\pi$]  &  Mean longitude at epoch \\ \bottomrule
\bottomrule
\end{tabular}
\begin{tablenotes}
\item U denotes a Uniform prior and logU denotes a logarithmic Uniform prior.
\end{tablenotes}
\end{threeparttable}
\end{table*}

\section{HD\,99492 data analysis} 
\label{Section_HD99492DataAnalysis} 
HD\,99492 is known to be orbited by a planet. \citet{Marcy2005} first announced the detection with 35 HIRES RV measurements, and reported an orbital period of 17 days and a minimum mass of $\sim$25 $M_{\oplus}$. The solution was refined by \citet{Meschiari2011} with an additional 58 HIRES velocities, and the authors proposed another planet candidate at an orbital period of $\sim$5000 days. A few years later and with 130 HIRES RV, the latter signal was shown to be related to stellar magnetic activity \citep{Kane2016}.  

\begin{figure} 
    \centering
    \includegraphics[width=\columnwidth]{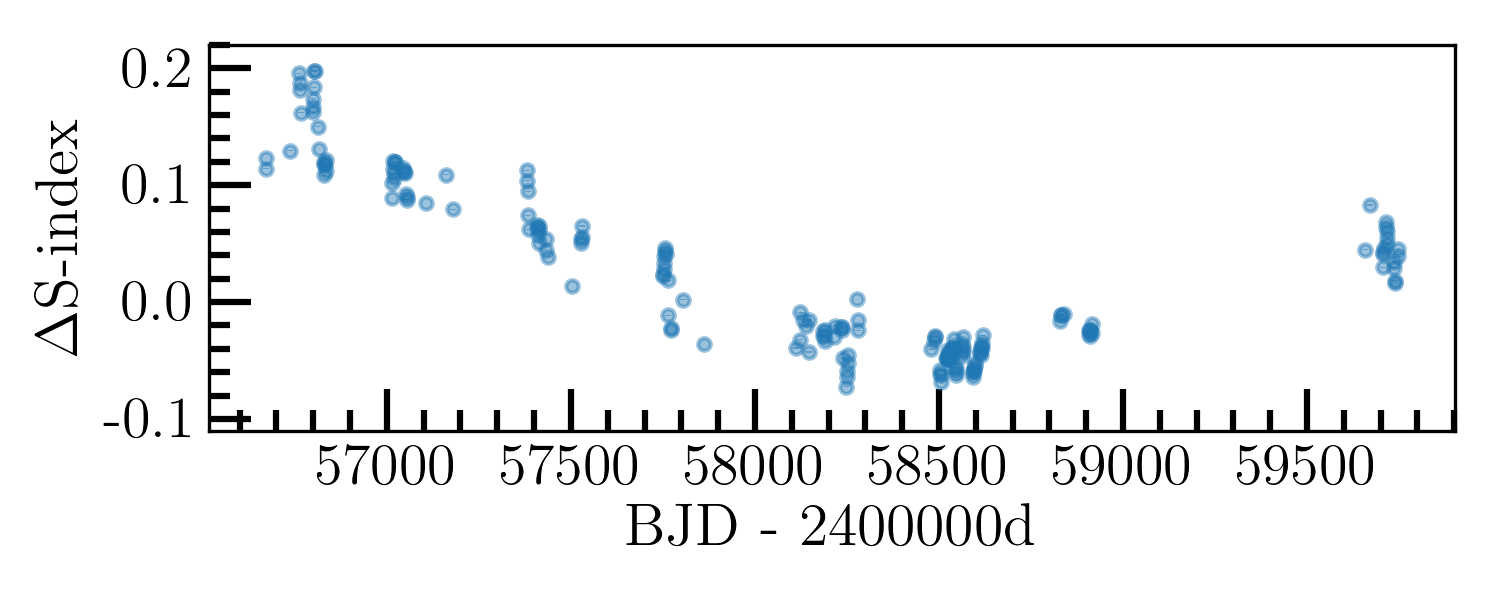}
        \includegraphics[width=\columnwidth]{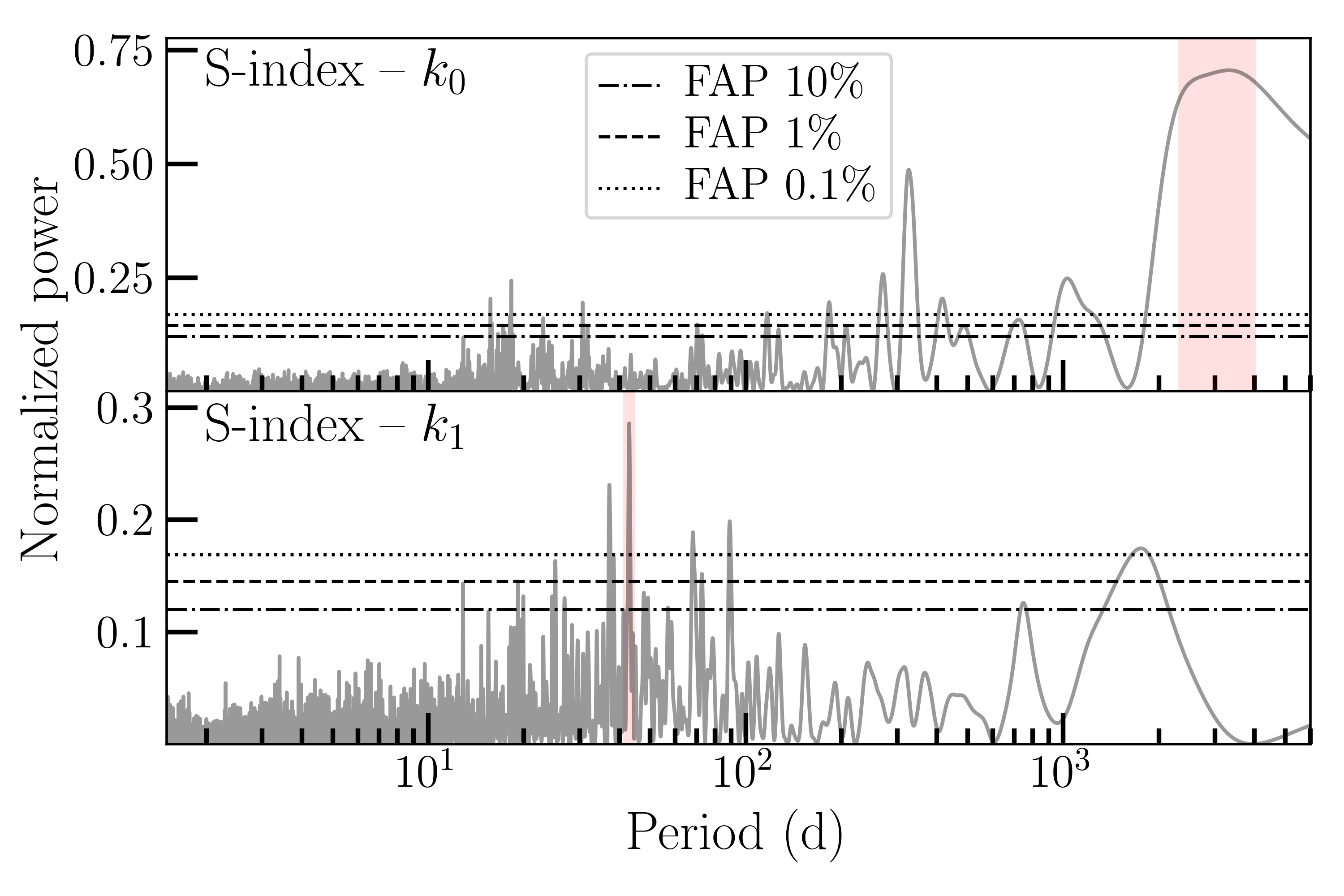}
    \caption{HD\,99492: YARARA-V1 S-index dataset. \textit{Top} -- S-index time-series. \textit{Middle} -- GLS periodogram of that time-series ($k_0$ stands for no Keplerian in the model). \textit{Bottom} -- GLS periodogram of the residual S-index -, after subtracting a Keplerian model with $P \sim 3000~$days ($k_1$). The red bands locate the periodicities seen in this activity indicator.}
    \label{fig:HD99492_S-index}
\end{figure}

We added our 202 nightly binned HARPS-N spectra processed with YARARA-V1, and performed an exploratory analysis on DACE. First, we focused on stellar activity signatures. In this dataset, the S-index time-series displays a significant long-term trend that peaks at a periodicity of $\sim$3000 days in a GLS periodogram. The latter signal is associated with the magnetic cycle described in \citet{Kane2016}. Fig. \ref{fig:HD99492_S-index} presents this time-series (top plot) and the corresponding periodogram (middle plot). At short periods, we observe a strong signal at 42.9 days, which appears once we fit the long-period variation with a Keplerian. This is illustrated in the bottom plot of Fig. \ref{fig:HD99492_S-index}. Furthermore, we also detect its 1 year aliases at 47.3 and 38.4 days. The S-index indicator reveals a notable correlation with the RV, with a correlation coefficient of $r = 0.50$. We detected a similar 42.9 days periodicity in other activity indicator time-series such as the CCF contrast and CCF FWHM. We also systematically identified the 1 year aliases of this signal. Therefore, all these observations constitute strong evidence for the 42.9 days signal to be associated with the stellar rotation period. This estimate is consistent with previous works (cf. Sect. \ref{Sect:StellarParam}). 

The HIRES publicly available data also contain the S-index activity indicator. In this dataset, we observe a clear quasi-periodic variation covering several magnetic cycles of the star, whose correlation with the RV is weaker than in the HARPS-N dataset. After fitting the most significant periodic signal in the S-index with a Keplerian (which converges to $\sim$3000 days), we do not find any significant residual periodicity at $\sim$43 days. In fact, there is no remaining significant signal below 800 days. We searched for the presence of the stellar rotation in the other activity indicator available (i.e. the H$_{\alpha}$ index). Again, we did not find any significant signal. As a result, we note that the rotation of the star does not present a measurable signature in the HIRES data. A potential cause of this non detection is the larger scatter observed in the S-index time-series of HIRES combined with a more sparse sampling. Finally, combining HARPS-N and HIRES datasets, we set tighter constraints on the period of the stellar magnetic cycle. With now an extended baseline of 9286 days, we constrain the main period of the S-index time-series to around 3300 days. 

Taking the above information into account, we then undertook a search for planets. We performed the latter using different models and datasets, so as to compare several approaches. In the first, we used the HARPS-N data as derived from the DRS together with the HIRES dataset. In a second approach, we analysed the YARARA-V1 dataset of HARPS-N together with HIRES, and included a correlated noise model. As a third approach, we repeated this procedure on the YARARA-V1 data only. Finally, in a fourth approach we analysed the YARARA-Shell data without a correlated noise model. We describe our planet search analyses below. 

\subsection{Approach 1: DACE -- HARPS-N DRS + HIRES} \label{subsect:Approach1} 
As a first preliminary analysis, we looked for periodic signals in the nightly binned RV time-series of the combined HARPS-N + HIRES datasets, using the DRS version of the HARPS-N data. While we noticed inconsistencies in the H$_{\alpha}$ time-series of the HIRES data, the S-index is consistent throughout the entire time-span. We selected this activity indicator to detrend the RV via the inclusion of a scaling parameter. Additionally, we included a quadratic drift to remove what remains of long-term variations. After adding these components in the model, we searched for periodic signals in the periodogram. The signal of the planet at 17d is very strong, with a FAP of 7.1\,10$^{-77}$. We included a Keplerian at that period into our model, and computed the periodogram of the residuals from the fit. This periodogram shows a very significant second signal. It has a period of 95.5 days and FAP of 2.1\,10$^{-21}$, and is not associated with any of the activity indicators. This favours the planet hypothesis, and we fitted that signal with another Keplerian. A third signal is revealed in the new periodogram of the residuals, with a period of 13.9 days and FAP of 0.23$\%$. Not only its FAP is above our defined detection threshold of 0.1$\%$, but also, this period is about a third of the expected stellar rotation period. Hence, it has to be interpreted carefully. 

We explored the two Keplerians model via a MCMC algorithm, which allows us to constrain the planet parameters. The parameters of HD\,99492 b are in accordance with previous publications, with results converging to $P_b$=17.0492$\pm$0.0007 days and $K_b$=7.30$\pm$0.23 \ms. Concerning the new planet candidate, we found a moderate eccentricity $e_c$=0.237$\pm$0.080, while the orbital period and the RV semi-amplitude are estimated to $P_c$=95.373$\pm$0.050 days and $K_c$=3.08$\pm$0.27 \ms. 

\begin{figure} 
    \centering
    \includegraphics[width=\columnwidth]{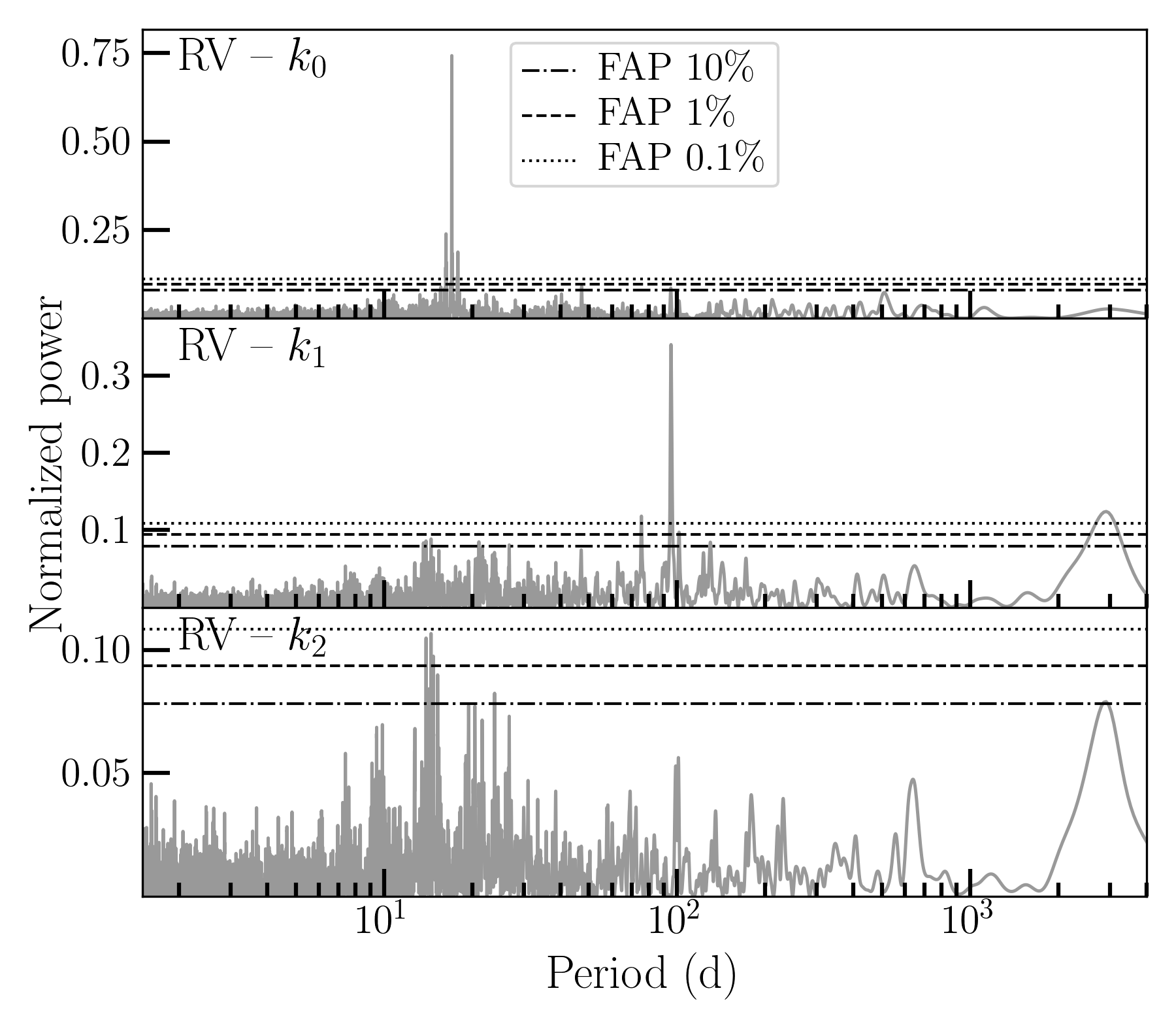}
    \caption{HD\,99492: Periodograms of the combined YARARA-V1 and HIRES RV time-series. The periodograms are computed from a fit of the time-series with a detrend + drift model. \textit{Top} -- No Keplerian. \textit{Middle} -- Periodogram of the residual time-series, after removing a Keplerian at $\sim$17 days. \textit{Bottom} -- Periodogram of the residuals after removing Keplerians at $\sim$17 and 95 days.}
    \label{fig:HD99492_RVperiodograms}
\end{figure}

\subsection{Approach 2: SPLEAF -- HARPS-N YARARA-V1 + HIRES} 
We repeated the process highlighted in Sect. \ref{subsect:Approach1} with HIRES and the YARARA-V1 data reduction of HARPS-N. Again, the signals at 17 and 95.5 days are unambiguous. Fig. \ref{fig:HD99492_RVperiodograms} illustrates our successive fitting process. The top plot presents the periodogram of the RV time-series (the highest peak has a FAP of 1.1\,10$^{-80}$), while the middle plot shows the periodogram of the residuals after fitting the RV with a Keplerian at 17 days. The highest peak in this middle plot, located at P=95.4 days, has a FAP of 3.2\,10$^{-22}$. The bottom plot displays the periodogram of the residuals after performing a two Keplerian fit, including one at 17 days and one at 95.4 days. Now, the remaining signal at 13.9 days is no longer detected in the latter plot, but we observe another signal at 14.5 days with a FAP of 0.13$\%$. Its significance is still below our threshold of 0.1$\%$ in FAP. Both the low significance and the different period detected for this third signal, in addition to its periodicity of about a third of the expected stellar rotation period, cast doubt on its planetary origin. 

In order to shed light on the origin of this third signal, we scrutinised the HIRES and HARPS-N datasets in a correlated noise model via SPLEAF (cf. Sect. \ref{Sect:S+LEAF}), using both the RV and S-index time-series. At this stage, we used the alternative YARARA-V1 dataset where the stellar activity was injected back into the RV. We did not set any constraining prior on the period of the correlated noise, nor on any other parameter (cf. Table \ref{tab:priors}). However, we initialised the correlation period to 43 days, and the decay time-scale to 500 days. The large value for the latter is aimed at mitigating the magnetic cycle, which takes place on a long time-scale. In order to facilitate the fit convergence, we fitted the correlated noise model parameters on a step-by-step strategy, successively adding an additional parameter. Then, we progressively incorporated Keplerians until no more significant signal was observed. In addition to these parameters, we also included an offset for each instrument and a white noise term, both included for each time-series -- namely the RV and S-index time-series. Finally, we added a RV linear drift in the model, which provided the best results in terms of signal significance and fit convergence compared to a quadratic drift or no drift. As a result of our fit, we found a first significant periodic signal at 17.05 days, with a negligible FAP of 4.5$\times$10$^{-52}$. After adding a Keplerian in the model and proceeding with a new fit, we detected another significant signal with a period of 95.3 days and FAP=2.5$\times$10$^{-18}$. Therefore, we incorporated a second Keplerian in our model, and performed a new fit. In the periodogram of the residuals, there remains a significant power around 3500 days which we attribute to a remnant of the stellar magnetic cycle, while the decay time-scale of the correlated noise converged to 194 days. Concerning the correlation period, it peaked at 44.8 days, which is compatible with the stellar rotation period. Then, we explored the parameter space of the two-Keplerian model with a MCMC algorithm. We performed 1M iterations, and obtained an effective sample of 4288 independent solutions. The period of the correlated noise is estimated to be 44.9$\pm$0.5 days. The outermost planet, at an orbital period of 95.4 days, has an estimated RV semi-amplitude of 2.7$\pm$0.2 m/s, which corresponds to a planet minimum mass of 17.4 $M_{\oplus}$. Its eccentricity is moderate, with $e_c$=0.112$\pm$0.086.

\subsection{Approach 3: SPLEAF -- HARPS-N YARARA-V1} 
In a third approach, we employed again SPLEAF for a correlated noise model but analysed only the HARPS-N data -- YARARA-V1 with activity injected back into the time-series. We undertook the same process as described above for the second approach. Once again, we found two significant signals at 17 and 95.5 days. Their FAP are 3.0$\times$10$^{-35}$ and 2.5$\times$10$^{-20}$ respectively. A third residual signal with a period of 14.7 days just reaches our detection threshold, and has a FAP of 0.05$\%$. It has a small semi-amplitude of 1.0 m/s. This period is different from the previously found signals at 13.9 and 14.5 days, but is also compatible with a third of the stellar rotation period. We note that the period of the correlated noise is estimated to 43.1 days. We remind the reader that SPLEAF approximates the periodic component of the SEP kernel by a development to the second order. To investigate further the stellar rotation origin for the 14.7d signal, we developed the periodic component to a higher order, so as to account for the third harmonic of the stellar rotation period. We applied this refined kernel to the same dataset. After the removal of two Keplerians at 17 and 95.5 days, the residual signal at 14.7d is still significant with a FAP of 0.07$\%$. Developing the kernel up to one additional order so as to entail the fourth harmonic yielded similar results. The residual 14.7d signal has a FAP of 0.08$\%$ with this refined development. However, in all cases, the period of the correlated noise unambiguously converges towards the stellar rotation period. Therefore, developing the periodic component of the kernel to a higher order should absorb the harmonics of the stellar rotation. Consequently, the survival of the 14.7d signal irrespective of the kernel development suggests this signal is not a direct outcome of the third harmonic of stellar rotation. 
Instead, its origin is likely different. 

We undertook an MCMC exploration of the model with three Keplerians. As the RV semi-amplitude of the third signal is small and in order to facilitate the MCMC convergence, we fixed the eccentricity and argument of periastron of the third Keplerian to 0. We performed 1M iterations, and obtained a total of 4856 independent solutions. Out of this exploration, we found RV semi-amplitudes of $K_b$=6.99$\pm$0.17, $K_c$=2.83$\pm$0.20 and $K_d$=0.94$\pm$0.17 \ms. The orbital eccentricity of planet c is now smaller: $e_c$=0.052 with a 68.27$\%$ confidence interval of [0.019, 0.102]. 

We also explored the parameter space of a model with only two Keplerians, given the doubt that we cast on the origin of the 14.7d signal. We performed a new MCMC exploration of the model with two Keplerians, again with 1M iterations. This time, we obtained a sample of 5440 independent solutions. The correlated noise parameters display similar values at the end of the MCMC exploration, compared to the model with three Keplerians. The same is observed for planets b and c RV semi-amplitudes, and the orbital eccentricity of planet c is once again constrained to be small. For this two Keplerian model, we estimate a Bayesian information criterion (BIC) of -327.1, while the BIC of the three Keplerian model amounts to -349.5. Therefore, the BIC indicates a slight preference for the model with three Keplerians. As such, we further tested the planetary nature of the 14.7 days signal, by investigating its resistance to the removal of RV measurements. If the signal was due to a planet, its power would monotonously decrease with the number of removed observations. Instead, a signal of stellar origin would react irregularly to the suppression of RV measurements. We performed different tests, following various patterns of data removal (random, lowest S/N, highest S/N, data in quadrature of the 14.7 days signal) and removing up to 10$\%$ of the time-series. We found that while the signals at 17 and 95 days successfully passed our tests, the significance of the 14.7 days signal did not decrease monotonously with the number of removed observations. Instead, its power in the periodogram evolved irregularly with the number of removed observations. From this analysis, we conclude that the current observations do not support a third planetary signal. 
On the other hand, the resilience of the 14.7 days signal to the degree of development of SPLEAF indicates that it does not emanate from the third harmonic of stellar rotation. Instead, it could be the result of a combination between stellar rotation (or its second harmonic) and the spectral window of the observations.

\begin{figure*} 
    \centering
    \includegraphics[width=\textwidth]{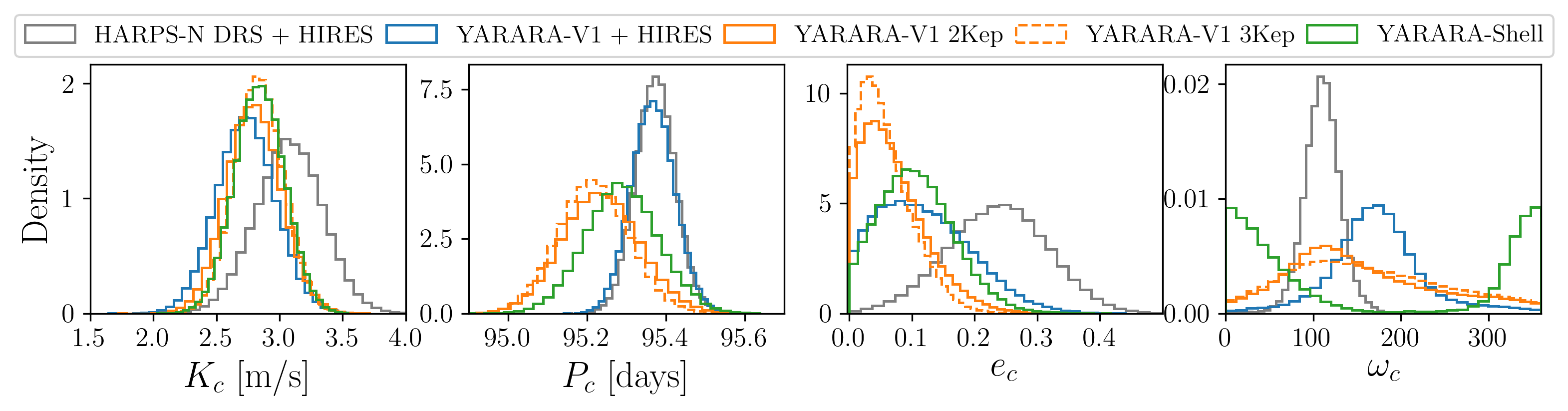}
    \caption{HD\,99492 c posterior distributions for our different models and datasets. The grey, blue, orange and green distributions refer to our first, second, third, and fourth approaches to derive the posteriors, respectively. In the case of the third approach, we further show the distributions resulting from two different models. The plain line distribution was obtained from a model with two Keplerians, while the dashed line corresponds to a model with three Keplerians.}
    \label{fig:HD99492-Results_AllApproaches}
\end{figure*}

\begin{figure*}
    \centering
    \includegraphics[width=\textwidth]{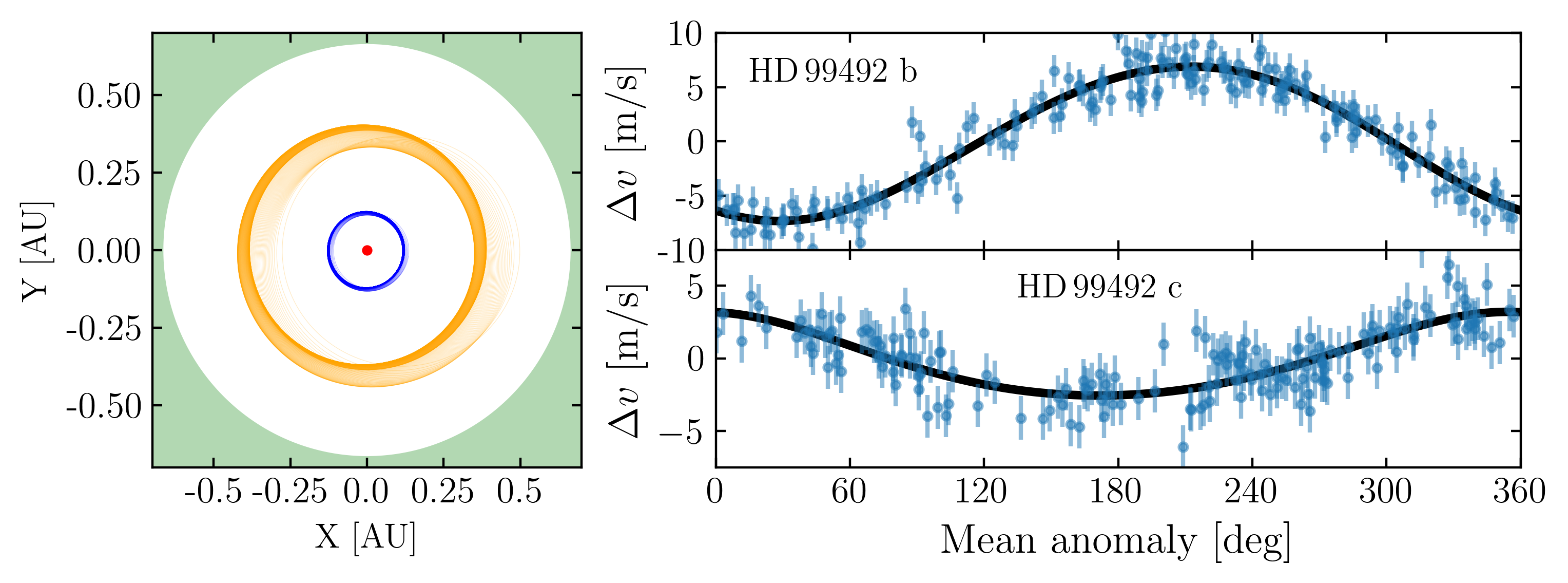}
    \caption{HD\,99492 orbital solutions for planets b and c derived from the HARPS-N YARARA-V1 dataset. \textit{Left} -- Cartesian representation of the planetary system for a sample of 205 MCMC solutions. The star is in red, the orbit of HD\,99492 b in blue and the orbit of HD\,99492 c in orange. Additionally, the green area marks the beginning of the conservative habitable zone, which continues outwards. This was calculated from the prescription of \citet{Kopparapu2013}. \textit{Right} -- RV measurements phase-folded on the best fit Keplerian solutions, together with the model curves.}
    \label{fig:HD99492-KeplerianSolutions}
\end{figure*} 

\subsection{Approach 4: HARPS-N YARARA-Shell} 
Finally, we investigated the YARARA-Shell dataset which, as compared to YARARA-V1, further corrects from stellar activity features. We designed a model that includes a white noise term, but does not include correlated noise. After an iterative search, we found two significant signals at the periods already mentioned above. The first signal, at 17.05 days, has a FAP of 5.1 10$^{-62}$, while the second signal at 95.3 days has a FAP of 3.5 10$^{-28}$. In the residuals of this two planet model, we did not find any significant period. The 14.7 days signal has now a FAP of 9.7$\%$. This observation further comforts us in the non detection of a third Keplerian signal with the current dataset. The analysis of YARARA-Shell, following the analysis of YARARA-V1 with SPLEAF, hereby demonstrates the power of using complementary analysis techniques to shed light onto the nature of periodic signals. 

We explored the two-planet model via the MCMC algorithm introduced above. We performed 1M iterations, and obtained a set of 4095 independent solutions that constitute our posterior distribution. While the RV semi-amplitude of the outer planet $K_c$ is very similar to its estimate obtained using approach 3, the orbital eccentricity of that planet converges towards a larger value of about 0.1. 

In Fig. \ref{fig:HD99492-Results_AllApproaches}, we synthesise the results from the four approaches described above. It shows the posterior distributions obtained with the different approaches, and projected on the orbital parameters of planet c. Additionally, we also distinguish between the two and three planets fits performed in the third approach. We conclude that the inclusion of the HIRES RV measurements increases the estimation of the eccentricity of planet c. This is probably due to a less efficient mitigation of stellar activity effects on the HIRES measurements. Indeed, as was already mentioned,we did not find a clear periodicity of $\sim$43 days in the S-index of HIRES. We note that the analysis without the HIRES data provides the best estimations on the planet parameters. Furthermore, using the YARARA-V1 data alone, the estimates we obtain for the parameters of planet c are very similar in the models with two or three Keplerians. The third periodic signal at 14.7 days is absent from the YARARA-Shell data. This latter dataset, however, leads to a larger estimate of the eccentricity $e_c$. As pointed out in \citet{Hara2019}, complex noise patterns -- if not modelled -- can boost the eccentricity estimations. This additional eccentricity induced by stellar activity likely also drives the shift observed in the distribution of $\omega_c$ from YARARA-Shell. Indeed, small orbital eccentricities do not constrain well the argument of periastron, as illustrated with the orange distributions. Any additional eccentricity would hence carry a significant influence on the distribution of $\omega$. In conclusion, we favour the approach leading to the smallest eccentricities. We opt for the approach using the YARARA-V1 dataset only together with a correlated noise analysis, namely approach 3 with a two-planet model. These results are synthesised in Table \ref{tab:FinalSolutions}. They suggest that planet c has a minimal mass of 17.9$\pm$1.3\,$M_{\oplus}$, which is equivalent to the mass of Neptune. So far, few planets in that mass range have been found with large (>50 days) orbital periods. We further discuss this result in Sect. \ref{Conclusion}. We present in Fig. \ref{fig:HD99492-KeplerianSolutions} the plot of the orbits in the cartesian space, and the RV phase-folded on the period of the two planets.

\subsection{Independent confirmation using {\sc tweaks}}
The HARPS-N CCF were independently analysed for planetary signals using {\sc tweaks}. This pipeline was particularly designed for attaining a sub-m/s detection threshold at long orbital periods, by combining the wavelength-domain and time-domain stellar activity mitigation \citep{AnnaJ2022, AnnaJ2023}. We first conducted a blind search of the radial velocities, using a model with up to five unidentified Keplerian signals. For this, we used the {\sc kima} nested-sampling package \citep{Faria2018}. Using {\sc scalpels} \citep{CollierCameron2021}, which involves doing a principal-component analysis on the autocorrelation function of the CCF, time-domain activity-decorrelation vectors were produced. These basis vectors were then used for the spectral line-shape decorrelation \citep{CollierCameron2021} in {\sc kima}, as \citet{AnnaJ2022} reported that de-trending the RV for line shape variations using the SCALPELS basis vectors yields a model that is significantly better than a model that does not account for these stellar activity signatures.

The joint posteriors showed clear detection of two Keplerian signals at orbital periods 17.05 days and 95.2 days. We conducted a False Inclusion Probability (FIP) analysis \citep{Hara2022} in frequency space, with the bin size set to the Nyquist frequency resolution over the entire data duration to search for multiple planet signals simultaneously. While a strong FIP of 0.13 was found for the 17.05 day signal, the 95.2 day signal was detected even more strongly with a FIP of 0.04; in other words, 96\% of the mutually independent {\sc kima} posterior solutions favoured a planet detection at an orbital period of 95.2 days. We present these results in the Appendix -- Fig.\,\ref{figApp:99492_Tweaks}. The RV semi-amplitudes obtained from this analysis were  combined with the stellar mass of 0.85 $M_{\odot}$, leading to a minimum mass determination of 27.13 ± 1.18 $M_{\oplus}$ and 19.85 ± 2.19 $M_{\oplus}$ respectively. We also validated these detections by confirming the coherency of the signals across different data subsets. Unlike the signals occurring from sampling patterns, stellar activity or aliases, both the 17.05 and 95 d signals were strongly detected in all the subsets with sigma $\geq$9$\sigma$. Finally, as opposed to these two planetary signals, the signal at 14d is not consistent in time. It is only detected in the second half of the data. This further motivates us to discard this signal as a potential additional planet.

\subsection{Transit search} 
\label{Sect:99492_TransitSearch} 
We analysed the TESS photometry in search for transit signals. To treat the instrumental systematics, we extracted our custom light-curve from the 2-min cadence pixel files with \texttt{lightkurve}, following the process explained in Sect. \ref{Sect:StellarParam_99492}. To aim at the transit search, we detrended our light curve from a combination of multi-scale and spike CBV. Indeed, we found that this model performs better at minimising the Combined Differential Photometric Precision (CDPP) metric, which is a measure of the remaining scatter in the light curve expressed in parts-per-million (ppm). Our corrected light curve presents a CDPP of 123 ppm.  

As a second step, we undertook the removal of the remaining stellar systematics on our corrected light-curve. We applied a detrending via splines fitting  using \texttt{keplersplinev2}\footnote{\url{https://github.com/avanderburg/keplersplinev2}}, and letting free the parameter describing the time-scale of variation of the spline, only imposing a lower boundary of 0.5 day. The fit converged to a time-scale of 0.74 day, and the modelled trend was removed from our light-curve. The resulting detrended light-curve is presented in Appendix, Fig. \ref{figApp:HD99492_TransitSearch} (top panel). 

We undertook a multi-transits search in this detrended light-curve. It was carried out with the \texttt{transitleastsquares} software\footnote{\url{https://transitleastsquares.readthedocs.io/en/latest}}. This search was done via the computation of a box least square (BLS) periodogram, with an orbital period search comprised between 2 and 20 days. We did not find any significant transit signal in this period range. From the estimated minimum mass of planet b, the mass-radius relationship for rocky planets\footnote{While the minimum mass of planet b points towards a gas-rich composition, here we explored the unfavourable case of rocky composition so to estimate the transit detectability in an unfavourable scenario. This comment also applies for the transit searches presented in Sect. \ref{Sect:147379_TransitSearch} and \ref{Sect:190007_TransitSearch}.} from \citet{Otegi2020} and the stellar radius reported on exofop\footnote{\url{https://exofop.ipac.caltech.edu/tess/}}, we estimated a transit depth of $\sim$1000ppm. We do not observe any hint of a transit signal at the period and phase of planet b estimated from our RV analysis. Concerning planet c, the expected transit time occurred at the beginning of sector 45. Under the hypothesis of a rocky composition, the planet would induce a transit depth of 850ppm on its parent star, which we do not see in the data. For both planets, the hypothetical transit depths are large enough compared to the light-curve noise, and they would be detected. Therefore, we rule out the transiting configuration of HD\,99492 b. Concerning the outer planet, the TESS observations do not cover all the orbital phases, and given the uncertainty on the transit time it is possible that a potential transit was missed. However, it is unlikely given the low transit probability of such a long-period planet ($\sim$1$\%$). Details of the transit searches can be found in Appendix \ref{App:Photometry} and Fig. \ref{figApp:HD99492_TransitSearch}. 

We also analysed the TESS sectors 45 and 46 in search for potential single transit features. To proceed, we fitted our detrended light curve based on two distinct models: no planet and one planet models. We compared the posterior probabilities of the models using the true inclusion probability (TIP) detection criterion \citep{Hara2022}, which provides us with a rigorous transit detection threshold. A TIP of 1 would favour the one planet model with a probability of 100$\%$ (TIP=1-FIP). This framework was first designed for RV data, and was later on applied to transit analyses \citep[][Wilson et al. in prep.]{Hoyer2022, Ehrenreich2023}. To aid convergence we performed the TIP analyses on one-day slices of the TESS sectors. We ran two sets of analyses employing different priors to first search for any transit signal and subsequently to probe the transiting nature of HD\,99492 b: wide planetary priors in one case, priors constrained on the HD\,99492 b planet parameters as reported in this study in the other case. We did not find any preference for the one-planet model as the TIP never exceeds 0.5 across the entire light curve (cf. Appendix -- Fig. \ref{figApp:HD99492_TIP}). Therefore, we conclude on the absence of transit signals in this data set.

\section{HD\,147379 data analysis}
\label{Section_HD147379DataAnalysis} 
HD\,147379 was first found to harbour a planet by the CARMENES team \citep{Reiners2018}. With the help of 114 CARMENES RV measurements and an additional 30 HIRES RV, they detected a planet with a minimum mass of $\sim$25 $M_{\oplus}$ in a 86.5 day orbit. This detection was independently confirmed by the SOPHIE team, with a dataset of 163 SOPHIE RV measurements \citep{Hobson2018}. A noticeable fact about HD\,147379 b is that it lies in the conservative habitable zone of its host star, as defined by \citet{Kopparapu2013}. 

\subsection{RV analysis} 
We analysed this system using exclusively the 165 nightly binned HARPS-N RV measurements. As with HD\,99492, we first looked in the modified YARARA-V1 dataset for activity signatures -- that is YARARA-V1 where the stellar activity was injected back in the RV. Besides the long-term trend due to the stellar magnetic cycle, the most significant signal is detected in the S-index indicator and peaks at 21.6 days. Furthermore, we observe a weak correlation between the RV and the S-index measurements, with a correlation coefficient of R=0.39 (cf. Appendix -- Fig. \ref{figApp:147379_AllSpectro}). 

\begin{figure}
    \centering
            \includegraphics[width=\columnwidth]{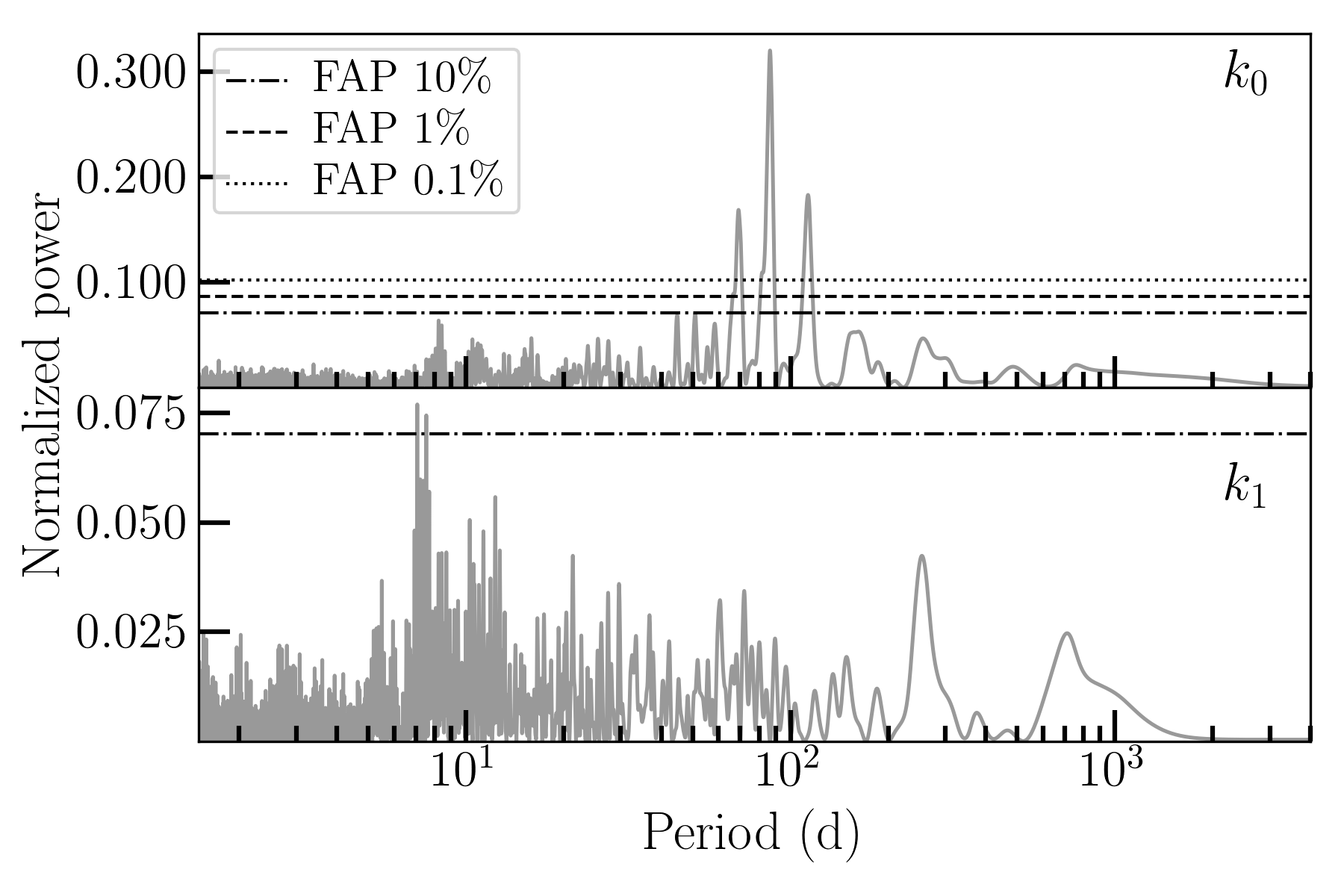}
            \includegraphics[width=\columnwidth]{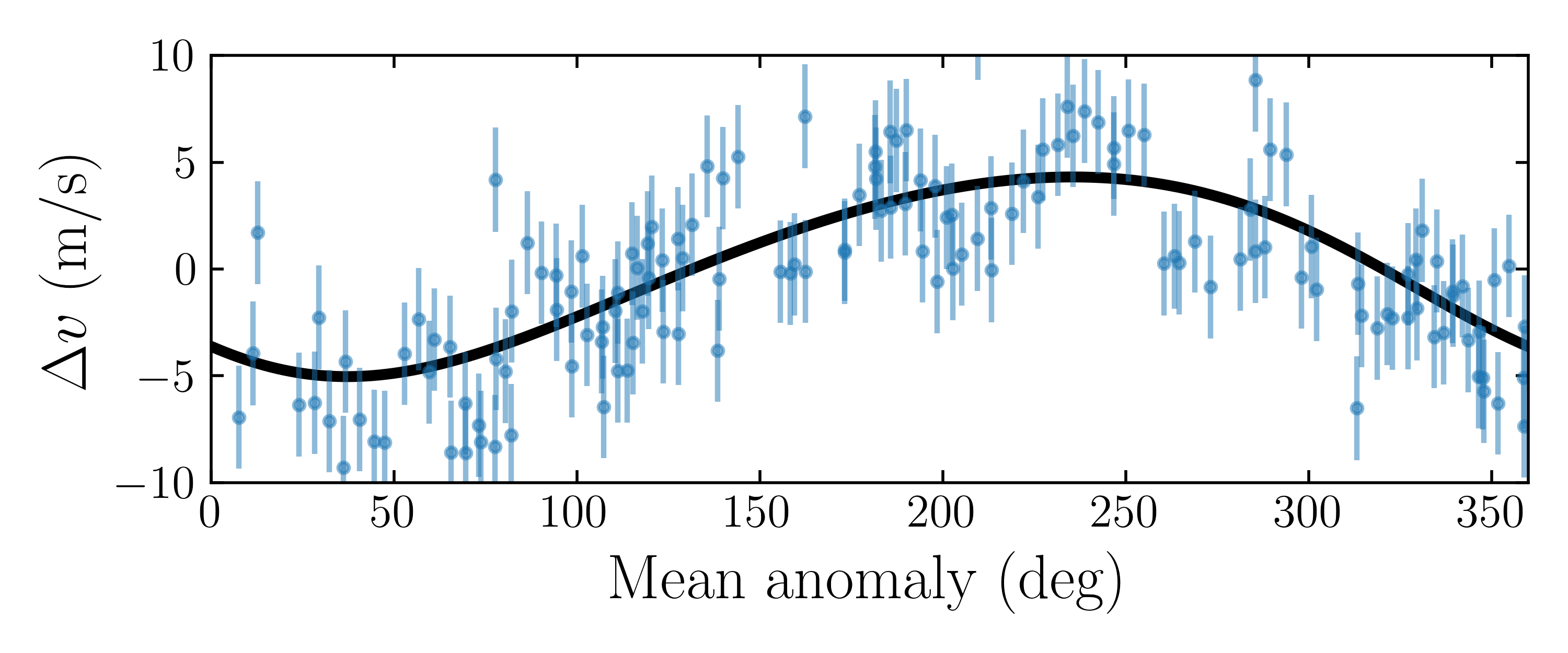}
    \caption{HD\,147379 planet search. \textit{Top and Middle panels} -- GLS periodograms of the HARPS-N YARARA-V1 RV time-series, in a model with correlated noise and with no Keplerian in the model ($k_0$) and one Keplerian ($k_1$) at a period of 86.5 days. \textit{Bottom panel} -- RV measurements phase-folded on the 86.5 days signal.}
\label{fig:HD147379_RV_PlanetSearch}
\end{figure}

In a second step, we undertook an analysis of the RV in search for planetary signals. We employed a correlated noise model with SPLEAF, using both the RV and S-index time-series. We also included in the model an offset, a white noise term and a linear drift term\footnote{The inclusion of a linear drift in our model yielded the largest likelihood.}. Wide uniform priors were injected, and we initialised the correlated noise with a period of 22 days and a large correlation decay time-scale of 500 days. After performing a first fit -- again through a successive addition in the model of the different correlated noise parameters -- we find a significant signal at 86.5 days in the periodogram of the residuals (Fig. \ref{fig:HD147379_RV_PlanetSearch}). We include the latter in our model as a Keplerian, and fit again the RV time-series with the new model. In the periodogram of the residuals, we do not find additional significant signals and the correlated noise period is fitted to 21.8 days, which is the expected stellar rotation period. The most prominent signal that we find in the residuals has a period of 7.1 days, which is about a third of the estimated stellar rotation period, and FAP=3.8$\%$. This signal is pushed down to a FAP of 23$\%$ when we use the augmented version of SPLEAF, which develops further the periodic component of the kernel to account for the third harmonic. This result further suggests the stellar rotation origin of the 7.1d signal. Furthermore, we do not find any signal around 500 days, which was reported by \citet{Hobson2018} using the SOPHIE data alone. Therefore, we stop our planet search and consider a model composed of one Keplerian only. The exploration of this model was performed with a MCMC, using 500k iterations leading to a final sample of 3298 independent solutions. The results are presented in Table \ref{tab:FinalSolutions}. The stellar rotation period that we estimate ($P_{GP}$=21.9$\pm$0.4 days) is consistent with previous investigations, and our correlated noise model converges to a correlation decay time-scale of 28.3$\pm$4.4 days. The minimum mass of HD\,147379 b is found to be smaller than both the estimates of \citet{Reiners2018} with 24.7$\substack{+1.8 \\ -2.4}$ $M_{\oplus}$ and \citet{Hobson2018} with 28.6$\pm$1.5 $M_{\oplus}$ -- our result suggests 21.6$\pm$1.1 $M_{\oplus}$, or about 1.26 times the mass of Neptune, and was derived from a careful mitigation of stellar rotation features. The bottom plot of Fig. \ref{fig:HD147379_RV_PlanetSearch} presents the RV measurements folded on the orbital period and phase of HD\,147379 b.

As a comparison, we searched for planetary signals with HARPS-N using the YARARA-Shell post-process. After deriving the cleaned RV, we computed a periodogram which unambiguously revealed the 86.5d planet (FAP=3.3\,10$^{-33}$). After fitting the latter with a Keplerian model and computing the periodogram of the residual RV, we did not find any signal that reaches our detection threshold of FAP=0.1$\%$. The strongest peak stands at a period of 12.3d, and has FAP=0.5$\%$. Because of this too large FAP and the weakness of the signal in V1-SPLEAF (FAP>10$\%$), we did not retain this signal as a planetary candidate. Further observations are needed to shed light on this signal. 

Finally, we also analysed the entire dataset composed of the HARPS-N (DRS), CARMENES, SOPHIE and HIRES RV. This ensemble consists of 458 RV measurements\footnote{Nine measurements from the SOPHIE spectrograph were neglected following \citet{Hobson2018}, because of an uncertainty larger than 35 m/s on the RV measurement.}. However, we note that the CARMENES and SOPHIE public data do not contain spectroscopic activity indicators, but only the RV measurements. Therefore, we could not carry a reliable modelling of stellar activity. We only applied a linear drift fit, in order to account for an observable trend. We computed the GLS periodogram of the residuals on this combined dataset, and found the 86.5d planet as the most significant peak. After its subtraction from the time-series and the computation of the updated periodogram, we found several periodic signals with a FAP below 0.1$\%$ at 10.6, 12.3 and 21.4 days. While the signals at 10.6 and 21.4 days are attributed to stellar rotation (they correspond to 0.5 and 1 time the stellar rotation period, respectively), the signal at 12.3 days is more unclear. Nevertheless, we note that its fit with a Keplerian yields a large eccentricity of 0.25. With the HARPS-N data alone, this remaining signal not only is less significant, but also leads to a large orbital eccentricity of 0.62 if we fit it with a Keplerian. As a result, we put strong caution about the planetary nature of this 12.3d signal, and presently interpret it as an artefact of stellar activity. The Keplerian orbital parameters of HD\,147379 b that we derived from the combined dataset suggest a minimum mass of msini=25.2M$_{\oplus}$ (K=5.25\ms). In Fig. \ref{figApp:FullFoldedRV_147379b}, we present the RV folded on this Keplerian solution. Again, we stress that these results do not include a careful modelling of stellar activity. Therefore, our analysis of the HARPS-N YARARA-V1 data alone together with SPLEAF is favoured, and its results defines our final solution for the orbital parameters of HD\,147379 b.

\subsection{Transit search} 
\label{Sect:147379_TransitSearch}
We scrutinised the full TESS light curve in search of any transiting signal. To proceed, we extracted the light curve of each of the 28 sectors following the same procedure as described in Sect. \ref{Sect:StellarParam_99492}. After downloading the CBV of those sectors, we found that the combination of multi-scale and spike CBV provides the best corrections in terms of minimising the CDPP metric, with an average CDPP of $\sim$90 ppm over the sectors. We further detrended the resulting light curve from any remaining stellar systematics with \texttt{keplersplinev2}. Our detrending fit converged to a spline variation time-scale of 0.9\,d.  

We then undertook several successive transit searches, for orbital periods between 2 and 20 days, between 20 and 70 days, and between 70 and 100 days. In none of the BLS periodograms did we find significant peaks. Concerning HD\,147379 b, using the mass-radius relationship of the rocky population from \citet{Otegi2020} -- considering the unfavourable case of a high bulk density -- and the stellar radius provided on exofop, we estimated the transit depth to $\sim$1200ppm. The planet would hence be detected given the large time-span of the TESS observations, and we conclude that it does not transit its parent star. We refer to the Appendix -- Fig. \ref{figApp:HD147379_TransitSearch} for more details.

\section{HD\,190007 data analysis}
\label{Section_HD190007DataAnalysis}
A planet candidate was recently suggested to orbit HD\,190007 \citep{Burt2021}. The authors used the combined RV dataset of APF and HIRES in order to determine the planetary characteristics. They found a body that orbits its star with a period of 11.7 days and a minimum mass of 16.46 $M_{\oplus}$. The stellar rotation is visible in the data, and they modelled the latter using a Keplerian with a period consistent with the photometric variability ($\sim$ 29 days). 

\begin{figure*}[t!]
    \centering
            \includegraphics[width=\textwidth]{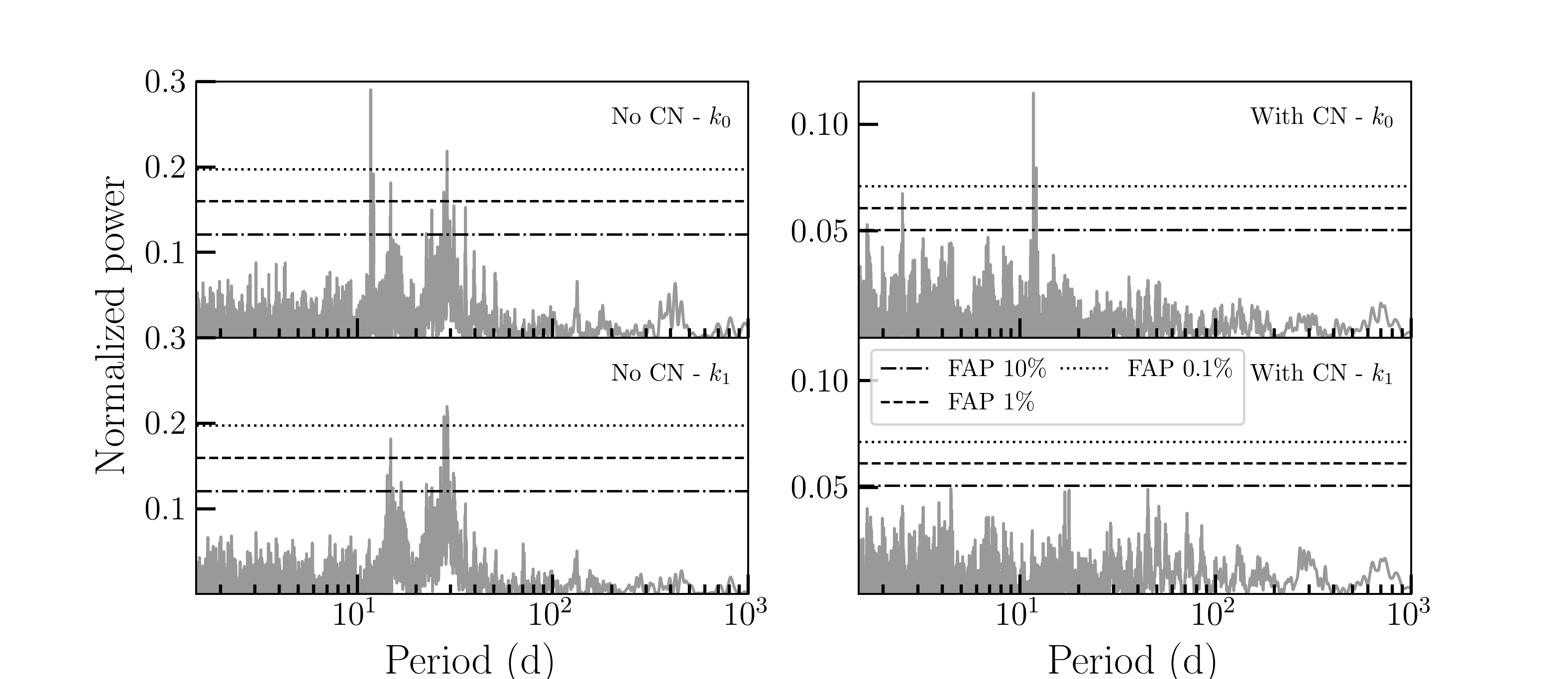}
    \caption{HD\,190007: Periodograms of the full RV time-series. \textit{Top left} -- No Keplerian ($k_0$), and no correlated noise  (CN) in the model. \textit{Bottom left} -- One Keplerian ($k_1$) in the model, at a period of 11.7 days. \textit{Top right} -- Inclusion of a CN model, no Keplerian. \textit{Bottom right} -- Inclusion of both CN and one Keplerian models.}
\label{fig:HD190007-Periodograms}
\end{figure*}

\begin{figure}
    \centering
            \includegraphics[width=\columnwidth]{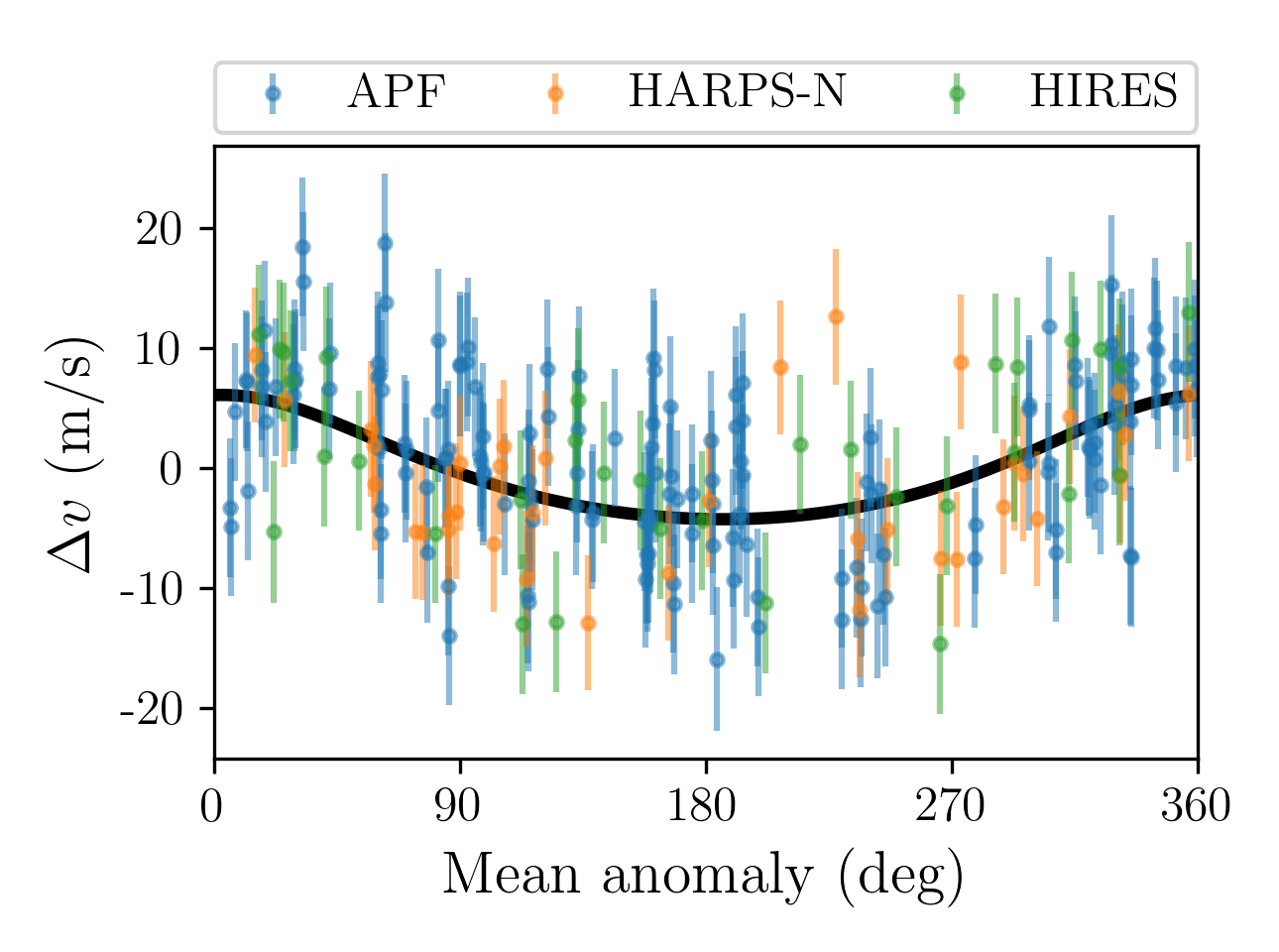}
    \caption{HD\,190007: RV measurements phase-folded on the Keplerian solution of HD\,190007 b.}
\label{fig:HD190007-Phase-fold}
\end{figure}

\subsection{RV analysis} 
We reviewed this solution by adding our set of 37 nightly binned HARPS-N measurements, and analysing the combined dataset. In total, we used 159 RV. Given the small number of HARPS-N RV measurements, we could not reliably process them with YARARA. Indeed, the algorithms behind the post-processing require several dozens of observations with different barycentric Earth RV (BERV) in order to not over-fit the data. Also, to work properly, having several observations per BERV bin element is mandatory, a condition not satisfied in this case. Instead, we used the data derived from the new DRS \citep{Dumusque2021}. After correcting for the different instrument offsets and accounting for drifts, the periodogram of the RV time-series clearly shows the presence of the 11.7 days planet. Additionally, a significant signal is visible with a period of $\sim$28 days, consistent with the stellar rotation period. This periodogram is presented in the top left plot of Fig. \ref{fig:HD190007-Periodograms}. At the bottom left, we show the periodogram of the residual RV, after subtraction from a Keplerian model with a period of 11.7 days. Significant power is still observed at the rotation period and half of it. \citet{Burt2021} fitted the 28 days signal with another Keplerian.  

Our three datasets -- HARPS-N, HIRES and APF -- contain the S-index indicator. We used this information to train a correlated noise model, and apply it simultaneously on the RV time-series. Again, we employed SPLEAF to model the correlated noise. Given the tight constraints on the stellar rotation period, we set a uniform prior on the correlated noise period between 25 and 30 days. We set priors on the other parameters of the correlated noise model according to Table \ref{tab:priors}. We initialised the model with a correlated noise period of 28 days and a decay time-scale of 200 days. Beside the correlated noise parameters, we also included in the model the instrumental offsets and a white noise term. We fitted the data with all the parameters of the model, and present the resulting periodogram of the RV residuals at the top right plot of Fig. \ref{fig:HD190007-Periodograms}. The planetary signal at 11.7 days is more significant. Its has a FAP of 3$~$10$^{-10}$. Naturally, we further note that the signal at $\sim$28 days disappeared. As a second step, we included a Keplerian with a period of 11.7 days in the model, and performed a new fit. The bottom right plot of Fig. \ref{fig:HD190007-Periodograms} shows the periodogram of the new residuals. Without ambiguity, no significant signal remains. Therefore, the correlated noise modelling provides a solution cleaned from the stellar rotation effects. It is able to model the main stellar rotation period and its harmonic better than a single Keplerian as previously done by \citet{Burt2021}. The minimum mass of the planet and the orbital eccentricity are estimated to 15.5$\pm$1.3 $M_{\oplus}$ and 0.14$\pm$0.08, respectively. This is in agreement with \citet{Burt2021}, which estimated a minimum mass of 16.5$\pm$1.7 $M_{\oplus}$ and an eccentricity of 0.14$\pm$0.07. We present the updated planetary parameters in Table \ref{tab:FinalSolutions}, and the phase-folded RV time-series in Fig. \ref{fig:HD190007-Phase-fold}. 

\subsection{Transit search}
\label{Sect:190007_TransitSearch} 
We analysed the photometry of TESS sector 54 in search for transit signals. Again, we retrieved the 2-min cadence target pixel files and downloaded the CBV of sector 54. We applied aperture photometry and corrected the light-curve for instrumental systematics via a joint fit of multi-scale and spike CBV and background subtraction at the pixels level (as explained in Sect. \ref{Sect:StellarParam_99492}). A special care with the detrending was needed due to the rapidly-varying variability of the star. When let free, the parameter describing the time-scale of variation of the spline converged to the lower limit that we imposed (i.e. 0.5d). We undertook several BLS transit searches between orbital periods of 0.5 and 13 days, exploring various detrend time-scales in the interval [0.5, 1.5] days with the parameter \texttt{bkspace} of the function \texttt{keplersplinev2}. Irrespective of the detrend time-scale, we did not find any significant signal in the BLS periodogram. 

We computed the expected transit times, depth and duration of planet b. To proceed, we used the results of our RV analysis, the stellar radius reported on exofop, and the mass-radius relationship for rocky planets taken from \citet{Otegi2020} -- again, we tested the transit detectability in the unfavourable case of a high bulk density, noting however that HD\,190007 b is more likely gas-rich. We estimated the transit depth to 920ppm, which is significant compared to the scatter observed in our detrended light curve. However, we did not find hints of transits that match with the expected ephemerides of planet b. We refer to the Appendix -- Fig. \ref{figApp:HD190007_TransitSearch} for more details. 

In a second approach, we scrutinised the TESS sector 54 considering potential single transit features. To proceed, we applied the same single-transit search process as the one employed for HD\,99492 (cf. Sect. \ref{Sect:99492_TransitSearch}) and based on the TIP framework. Again, we ran two distinct analyses with wide priors on one side, and priors constrained on the ephemerides of HD\,190007 b as derived from our RV modelling on the other side. In both cases, the absence of planetary transits was unambiguously favoured. Therefore, we rule out the presence of transits in the data.

\begin{table*}[]
\centering
\caption{Final results of the fits on HD\,99492, HD\,147379, and HD\,190007. We report the median values from the posterior distribution of the MCMC exploration. The uncertainties in the parameters are 68.27$\%$ confidence intervals.}
\label{tab:FinalSolutions}
\begin{tabular}{@{}lccccc@{}}
\hline\hline \\
Parameter  & Units  & HD\,99492 & & HD\,147379 & HD\,190007  \\ \hline 
\underline{Offsets ($\gamma$) and jitter ($\sigma$)} \vspace{0.1cm} \\ 
 \\ 
$\gamma_{HARPS-N}$   &  m s$^{-1}$  &  -2.18$\pm$0.69  &  &   -113.2$\pm$22.8   &   0.94$\pm$0.97   \vspace{0.15cm} \\ 
$\gamma_{APF}$   &  m s$^{-1}$  &  /  &  &  /  &  -1.44$\pm$0.44    \vspace{0.15cm} \\ 
$\gamma_{HIRES}$   &  m s$^{-1}$  &  /  &  &  /  &   -1.37$\substack{+1.37 \\ -1.40}$    \vspace{0.15cm} \\ 
$\sigma_{RV}$   &  m s$^{-1}$  &  1.56$\pm$0.10  &  &  0.98$\substack{+0.13 \\ -0.12}$  & 2.57$\substack{+0.30 \\ -0.27}$     \\ 
 \\ 
\underline{Correlated noise} \vspace{0.1cm} \\ 
$P_{GP}$   &  days  &  43.65$\substack{+0.47 \\ -0.56}$  &  &  21.94$\substack{+0.42 \\ -0.44}$   &  30.71$\substack{+0.89 \\ -0.72}$   \vspace{0.15cm} \\ 
$\rho_{GP}$   &   days  &   152.7$\substack{+22.5 \\ -20.7}$  &  & 28.31$\substack{+4.43 \\ -4.12}$  &  39.0$\substack{+7.1 \\ -5.7}$   \vspace{0.15cm} \\ 
$\eta_{GP}$   &  &   0.94$\substack{+0.16 \\ -0.13}$  &  &  0.61$\substack{+0.10 \\ -0.09}$  &  0.97$\substack{+0.17 \\ -0.15}$   \\ 
 \\ 
\underline{Planets}  \\ 
Parameter  & Units & HD\,99492 b  &  HD\,99492 c  &  HD147379 b  &  HD\,190007 b  \\ \hline  
 \\ 
$P$  &  days  &  17.0503$\pm$0.0016  &  95.233$\substack{+0.098 \\ -0.096}$  &  86.58$\pm$0.14  &  11.724128(99)  \vspace{0.15cm}  \\ 
$K$  &  m s$^{-1}$  &  7.05$\pm$0.18  &  2.79$\substack{+0.22 \\ -0.21}$  &  4.49$\pm$0.22     &  4.91$\pm$0.45 \vspace{0.15cm} \\ 
$e$  &    &  0.034$\substack{+0.025 \\ -0.021}$  &  0.063$\substack{+0.060 \\ -0.040}$  &  0.063$\substack{+0.047 \\ -0.038}$  &  0.136$\substack{+0.085 \\ -0.080}$  \vspace{0.15cm} \\ 
$\omega$  &  deg  &  154.3$\substack{+48.6 \\ -48.0}$  &  137.5$\substack{+107.7 \\ -62.9}$    &   130.1$\substack{+58.3 \\ -48.1}$  &  2.3$\substack{+50.5 \\ -46.3}$  \vspace{0.15cm} \\ 
$\lambda_0$  &  deg  &  276.4$\pm$5.2   &  351.2$\pm$10.3  &  204.6$\substack{+20.1 \\ -19.7}$  &  78.1$\substack{+8.2 \\ -7.9}$ \vspace{0.15cm} \\ 
$m \sini$   &  $M_{\oplus}$  &  25.5$\pm$0.6  &  17.9$\pm$1.3   &   21.6$\pm$1.1  &  15.5$\substack{+1.2 \\ -1.3}$   \\ \bottomrule
\end{tabular}
\end{table*}

\section{Discussion and conclusions}
\label{Conclusion}
In this work, we presented the HARPS-N spectroscopic data acquired on three stars: HD\,99492, HD\,147379 and HD\,190007, all of them already known to harbour planets. We reviewed those planetary systems with the help of advanced data analysis tools. YARARA-V1 provided us with HARPS-N datasets cleaned from any instrumental effects. Additionally, we used the generalised SPLEAF model to carefully and efficiently mitigate stellar activity. The combination of these two tools was shown to provide the best results with respect to other strategies, as discussed for HD\,99492 in Sect. \ref{Section_HD99492DataAnalysis}. We updated the orbit of the known planet HD\,99492 b ($P$=17.05 days), and unambiguously detected a second planetary companion, namely HD\,99492 c, that we confirmed independently using {\sc tweaks}. This second planet orbits its parent star in 95.2 days, and has an estimated minimum mass of 17.9$\pm$1.3 $M_{\oplus}$. The analysis of the 165 nightly binned HARPS-N measurements of HD\,147379 did not lead to a new planet detection. We also did not find any transit signal in the extensive 28 TESS sectors. However, we updated the parameters of HD\,147379 b, and notably found a minimum mass smaller than previous studies \citep{Reiners2018, Hobson2018} -- our new estimate, peaking at 21.6$\pm$1.1 $M_{\oplus}$, is 2.6$\sigma$ away from the nearest value provided by \citet{Reiners2018}. Finally, we reviewed the system HD\,190007 with the addition of our 37 nightly binned HARPS-N measurements to the publicly available HIRES and APF data. We performed a correlated noise analysis to account for the strong stellar rotation signal. We updated the solution of the known planet at $P = 11.7$~days, and obtained a minimum mass of 15.5$\pm$1.3 $M_{\oplus}$. The results from our review of the three planetary systems are presented in Table \ref{tab:FinalSolutions}. We also carried out a systematic transit search in the available TESS sectors, and did not find transit signals.

\begin{figure*} 
    \centering
    \includegraphics[width=0.45\textwidth]{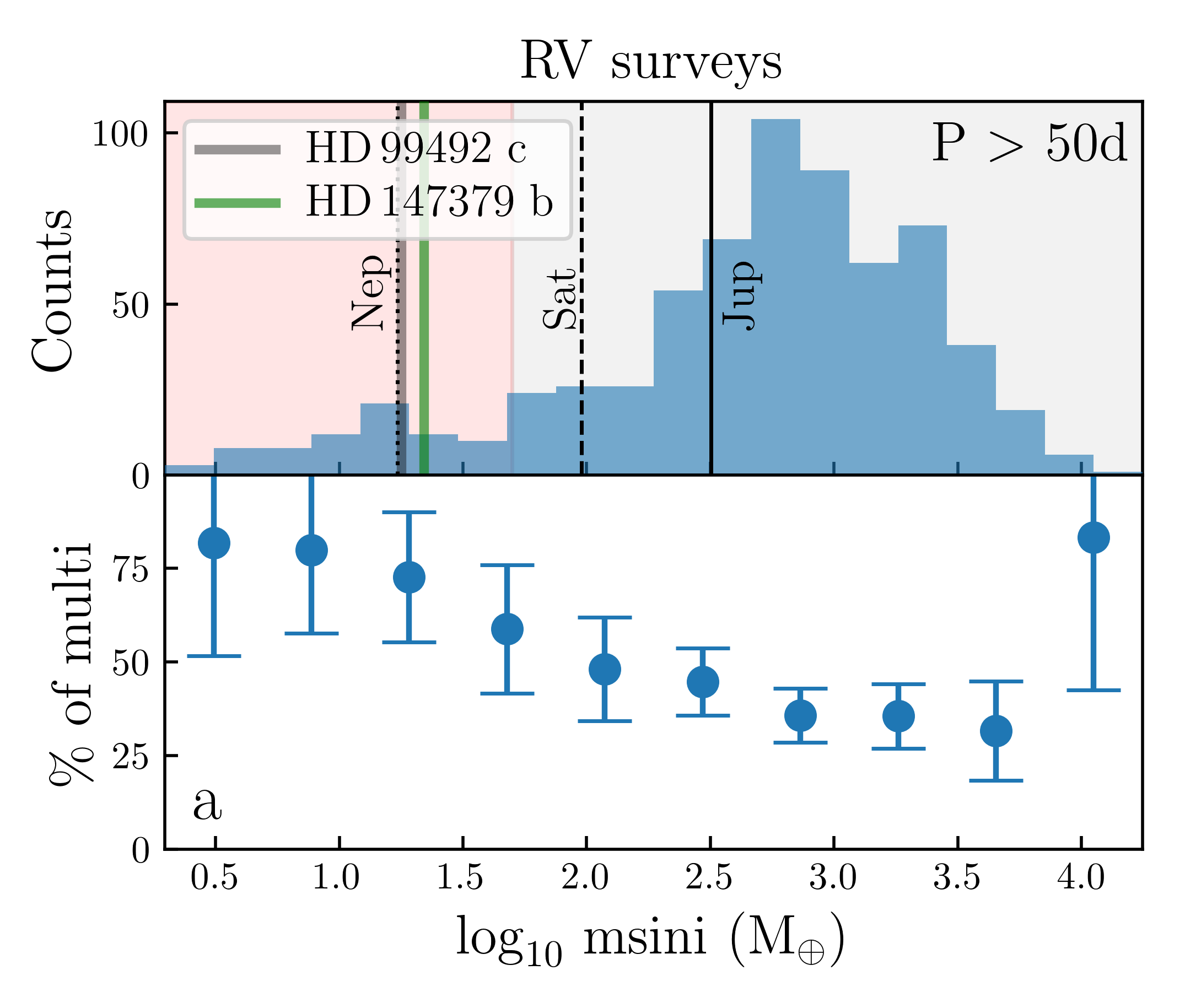}
    \includegraphics[width=0.45\textwidth]{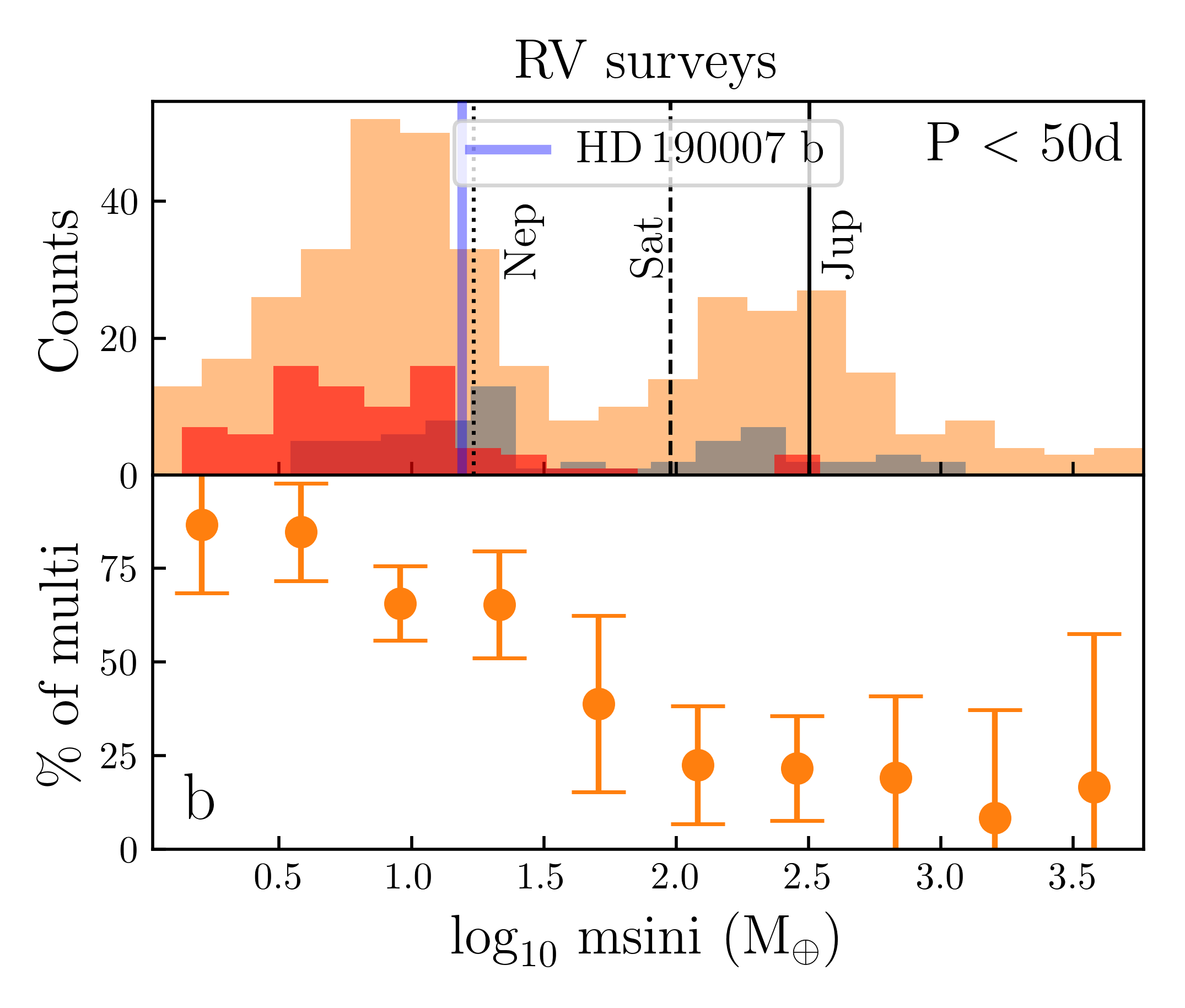}
    \caption{Exoplanet population with a minimum mass measurement msini. Panel a: Sub-population with large orbital periods (P>50 days). This sample contains 428 exoplanets. \textit{Top} -- Histogram of that population, with respect to the planet minimum mass. The vertical grey and green lines identify the position of HD\,99492 c and HD\,147379 b, respectively. \textit{Bottom} -- Proportion of planets that are part of multi-planet systems, among the considered sub-population. Panel b: Sub-population with small orbital periods (P<50 days). \textit{Top} -- Distribution of that sub-population according to msini (in orange). The red histogram indicates companions to the low-mass long-period planets (red zone in Panel a). The grey histogram focuses on companions to the large-mass long-period planets (grey zone in Panel a). \textit{Bottom} -- Proportion of planets part of multi-planet systems.} 
\label{fig:PlanetPopmsini}
\end{figure*}

\begin{figure} 
    \centering
    \includegraphics[width=0.45\textwidth]{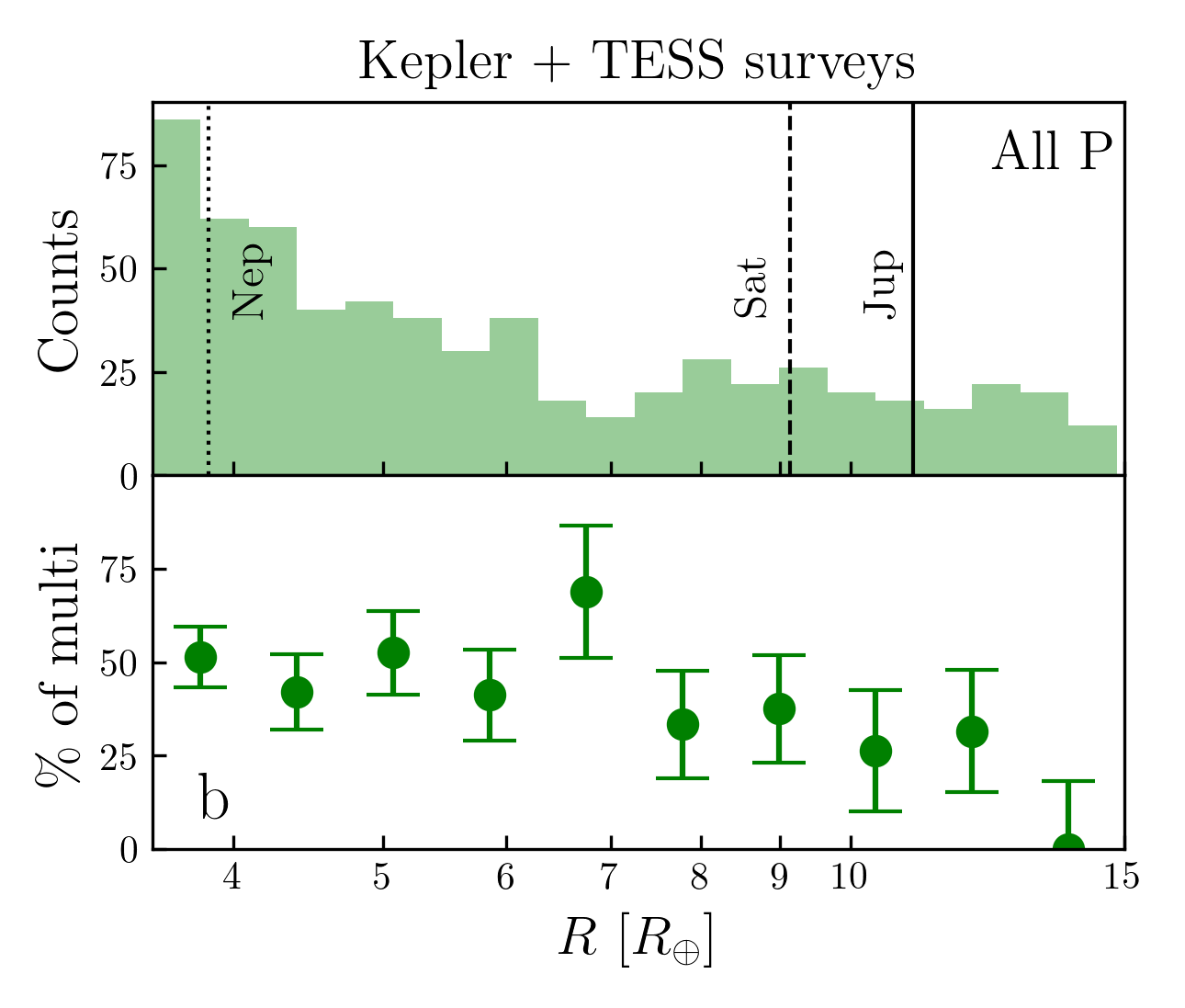}
    \caption{Exoplanet population from the Kepler and TESS surveys, distributed in planet radius between 3.5 and 15~$R_{\oplus}$. This population is composed of 632 planets as of March 22, 2023. \textit{Top} -- Histogram of the considered populations. \textit{Bottom} -- Proportion of planets part of multi-planet systems.} 
\label{fig:PlanetPop_KeplerTess}
\end{figure}

HD 99492 c and HD 147379 b have orbital periods and minimum masses that stand in an underpopulated region of the parameter space. Fig. \ref{fig:PlanetPopmsini}a presents the distribution of known exoplanets with measured minimum masses $m\sini$, and with large orbital periods $P > 50$~days. This population is composed of 962 planets, according to the Exoplanet Archive\footnote{\url{https://exoplanetarchive.ipac.caltech.edu/}} as of March 22, 2023. We note that this distribution also includes long-period transiting planets whose masses are constrained with RV measurements. The positions of HD\,99492 c and HD\,147379 b in this plot are represented by vertical grey and green lines, respectively. For comparison, we indicate the position of Neptune, Saturn and Jupiter in this histogram. This distribution unambiguously reveals a larger number of long-period giant exoplanets. Naturally, observational biases play an important role in this picture, since Neptune-mass exoplanets on long orbital periods are technically more challenging to detect. However, a closer inspection of this histogram may reveal a potential lack of planets between 30 and 50 Earth masses, giving rise to two distinct populations that the observational biases hardly explain. 

To estimate the statistical significance of this bimodality, and hence the existence of two planet populations, we tested the similarity between this distribution and a single-mode Gaussian distribution. To proceed, we defined the large-mass sub-sample by setting a mass threshold at 50~$M_{\oplus}$, and fitted a Gaussian law on this sub-sample. Then, we generated a random sample following this law, and with the same size as the full distribution. Finally, we computed the p-value between this Gaussian and the blue histogram. We repeated this procedure 10\,000 times, and derived a distribution of p-values. In those results, the maximum p-value obtained is 3~10$^{-4}$, which indicates that all of our 10\,000 tests support the hypothesis of two distinct distributions between the planet distribution and the fitted Gaussian. As a second statistical test, we performed a Z-test between the sample of planets with m\sini < 50~$M_{\oplus}$ (red shaded area), and the sample with m\sini > 50~$M_{\oplus}$ (grey shaded area) to quantify their mutual difference. This test compares the means and standard deviations of two samples, and indicates if these samples significantly differ from each other. We fitted both large-mass and small-mass sub-samples with a Gaussian, and applied the Z-test on these adjusted laws. The test returned the value of 2.94, which points towards a significant difference between the two populations. We repeated the Z-test with a new separation threshold of 30~$M_{\oplus}$ between the two populations, and obtained the value of 3.06, again indicating a significant difference between these two populations. Therefore, our analyses support the bimodality of the planet distribution. 

We bring an additional support to the bimodality in the bottom panel of Fig. \ref{fig:PlanetPopmsini}a. It presents, among the considered population, the percentage of planets found in multi-planet systems. The planets seem to be globally found in distinct system architectures. Despite the large uncertainties in the small-mass population caused by the small number of detections, we observe that these planets are most often found in multi-planet systems. For higher-mass planets, this ratio is significantly smaller. Those results support previous RV-detected planet statistics. Notably, \citet{Winn2015} pinpoint a bimodality in the distribution of minimum masses, without distinction of orbital period -- see their Fig. 4. In that same figure, the authors illustrate, at the population level, the increase in eccentricity scatter for larger minimum masses. Besides, among the exoplanet population, lower multiplicity systems present also a larger eccentricity scatter \citep{Wright2009, Limbach2015}. Through the link of orbital eccentricity, we hence deduce that large-mass planets tend to be found in systems with smaller multiplicity, according to the RV surveys. The results that we present in the lower panel of Fig. \ref{fig:PlanetPopmsini}a corroborate these observations, since larger-mass planets seem to be more often found in single-planet systems. Again, while we raise a hint of a trend, we stress that a treatment of detection biases is needed to confirm this statement. 

We compared these results with the short-period planet population. Fig. \ref{fig:PlanetPopmsini}b shows the population of exoplanets with measured minimum masses, but this time with a focus on short-period planets ($P<50$~days). The double peak distribution is more prominent than for the long-period population, and a similar trend with the system multiplicity is observed. We undertook a new Z-test for a threshold mass of 50~$M_{\oplus}$, and obtained a value of 2.78 which confirms the likely bimodality of the distribution. We note that at short orbital periods, low-mass planets appear more common than giant planets. The opposite is observed for long-period planets. This observation is in line with classical theories of planet formation, for which gas giants are formed in cooler regions of the protoplanetary disk, as opposed to smaller-mass planets. However, we stress that detection biases likely alter this picture. Furthermore, we highlighted with the red histogram the short-period planet companions to the low-mass long-period planets located in the red zone of Fig \ref{fig:PlanetPopmsini}a. Most of these planets have smaller masses too. Conversely, we highlighted with the grey histogram the short-period planet companions to the large-mass long-period planets inside the grey zone of Fig. \ref{fig:PlanetPopmsini}a, which does not indicate a strong preference for large or small mass companions. Additionally, the red histogram constitutes most of the planet companions to low-mass long-period planets with a measured minimum mass, with 80 planets out of 94 companions in total, that is 85$\%$. The remaining 15$\%$ are long-period planets. As a result, most of the companions to the small-mass long-period planets have small masses and short periods. 

Based on all those observations, we would therefore expect HD\,147379 to be orbited by at least a second planetary companion inner to HD\,147379 b, with the hypothesis that this system follows the observed trend. We stress, however, that the statistics have low numbers and that the observational biases could alter this picture. Continuing observational efforts using high-resolution spectrographs will help to augment those statistics and refine the architectures of planetary systems. Finally, we note that a similar bimodality is observed in the true mass distribution of exoplanets. 

We put these results in light of the Kepler and TESS transit surveys in Fig. \ref{fig:PlanetPop_KeplerTess}. The histogram presents the population of confirmed Kepler and TESS planets spread over radius, in the range [3.5, 15]~$R_{\oplus}$ so to entail the Neptune-size regime and above. In these data, there might also be a tentative hint of two distinct populations separated around 7~$R_{\oplus}$, which corresponds to a mass of 38$M_{\oplus}$ using the mass-radius relationship for volatile-rich planets \citep{Otegi2020}. The location of this boundary hence overlaps with the ones observed in Fig. \ref{fig:PlanetPopmsini}a and b. Nevertheless, this separation is here too shallow for any further prospects. Moreover, there is no clear trend in the proportion of planets in multi-planet systems as a function of planet radius in this considered radius range, as is illustrated in the bottom panel. Different factors could explain this discrepancy with the RV surveys, such as a non homogeneous scaling law for the planets bulk density, or probing different types of systems from the transit and RV surveys. This is briefly discussed in \citet{Leleu2023}, in the context of the mass dichotomy between RV and transit-timing variations (TTV) techniques. 

This work has once again proven the importance of long-term RV surveys with stable high-precision spectrographs for the detection and characterisation of warm planets, which constitute a yet under-explored family of exoplanets. Additionally, the use of advanced tools for the removal of instrumental systematics and for modelling correlated noise sheds light on the noise pattern, and pushes further the barriers of small-mass planet detections.

\begin{acknowledgements}  
This work is based on observations made with the Italian {\it Telescopio Nazionale
Galileo} (TNG) operated by the {\it Fundaci\'on Galileo Galilei} (FGG) of the
{\it Istituto Nazionale di Astrofisica} (INAF) at the
{\it  Observatorio del Roque de los Muchachos} (La Palma, Canary Islands, Spain). 
The HARPS-N project was funded by the Prodex Program of the Swiss Space Office (SSO), the Harvard University Origin of Life Initiative (HUOLI), the Scottish Universities Physics Alliance (SUPA), the University of Geneva, the Smithsonian Astrophysical Observatory (SAO), the Italian National Astrophysical Institute (INAF), University of St. Andrews, Queen’s University Belfast, and University of Edinburgh. This work has been carried out within the framework of the National Centre of Competence in Research PlanetS supported by the Swiss National Science Foundation under grants 51NF40\textunderscore 182901 and 51NF40\textunderscore 205606. The authors acknowledge the financial support of the SNSF. 
This work has made use of data from the European Space Agency (ESA) mission {\it Gaia} (\url{https://www.cosmos.esa.int/gaia}), processed by the {\it Gaia} Data Processing and Analysis Consortium (DPAC,
\url{https://www.cosmos.esa.int/web/gaia/dpac/consortium}). Funding for the DPAC has been provided by national institutions, in particular the institutions participating in the {\it Gaia} Multilateral Agreement. 
M.C. acknowledges the SNSF support under grant P500PT\_211024. 
R.D.H. is funded by the UK Science and Technology Facilities Council (STFC)'s Ernest Rutherford Fellowship (grant number ST/V004735/1). FPE and CLO would like to acknowledge the Swiss National Science Foundation (SNSF) for supporting research with HARPS-N through the SNSF grants nr. 140649, 152721, 166227 and 184618. The HARPS-N Instrument Project was partially funded through the Swiss ESA-PRODEX Programme. TGW acknowledges support from STFC consolidated grant number ST/V000861/1 and UKSA grant number ST/R003203/1. This project has received funding from the European Research Council (ERC) under the European Union’s Horizon 2020 research and innovation programme (grant agreement SCORE No 851555). 
A.S. acknowledges financial support from the agreement ASI-INAF n.2018-16-HH.0. 
Finally, the authors thank the referee for their insightful comments and suggestions on the paper. 
\end{acknowledgements}

\bibliographystyle{aa} 
\bibliography{bib.bib}

\begin{appendix}
\section{Complementary spectroscopic data} 
\label{App:RV}
\begin{figure}[h!] 
    \centering
    \includegraphics[width=\columnwidth]{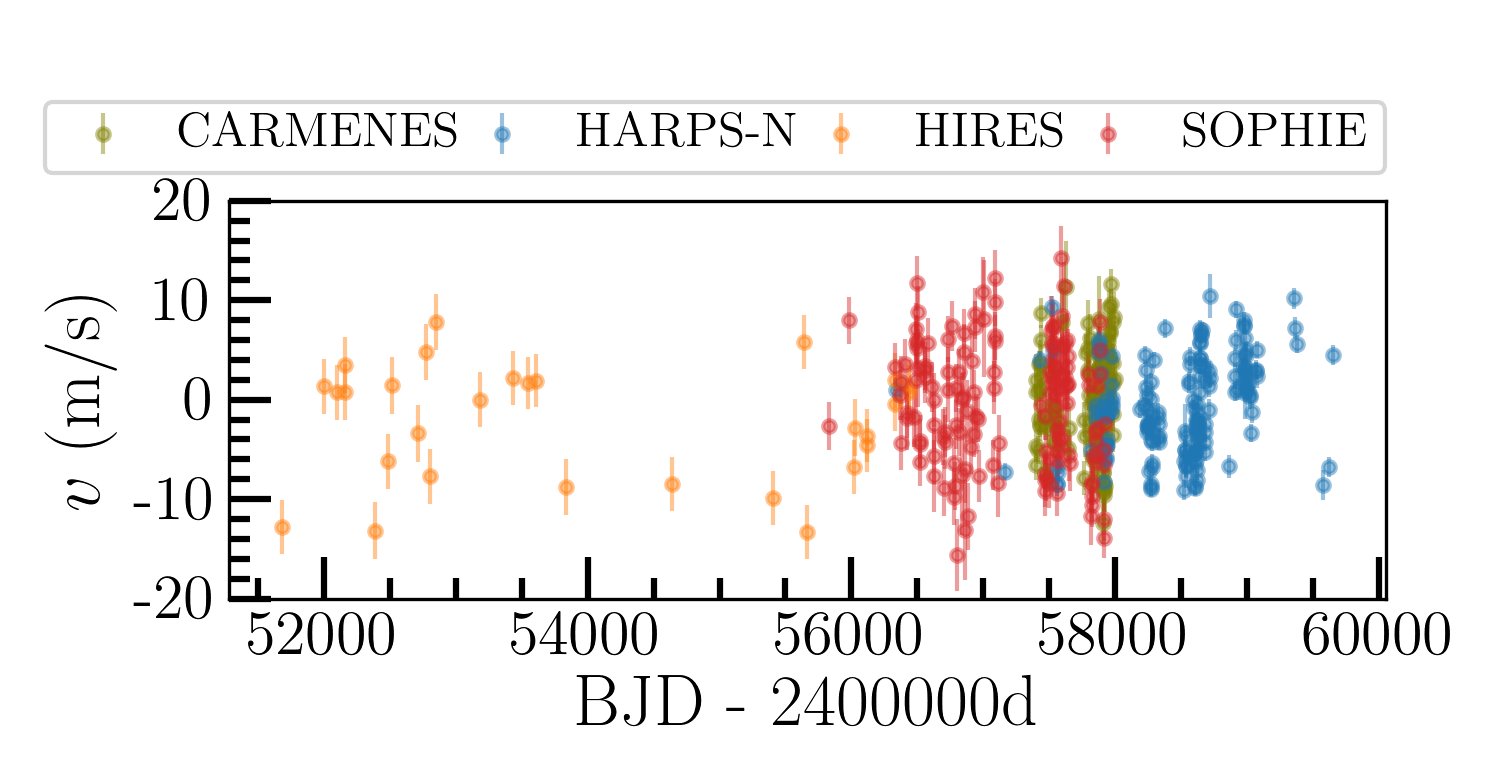}
    \caption{Entire RV time-series of HD\,147374. It combines our HARPS-N RV measurements with the public data from HIRES, CARMENES, and SOPHIE high-resolution spectrographs.}
\label{figApp:FullRV_147379}
\end{figure}

\begin{figure*}[h!] 
    \centering
    \includegraphics[width=\textwidth]{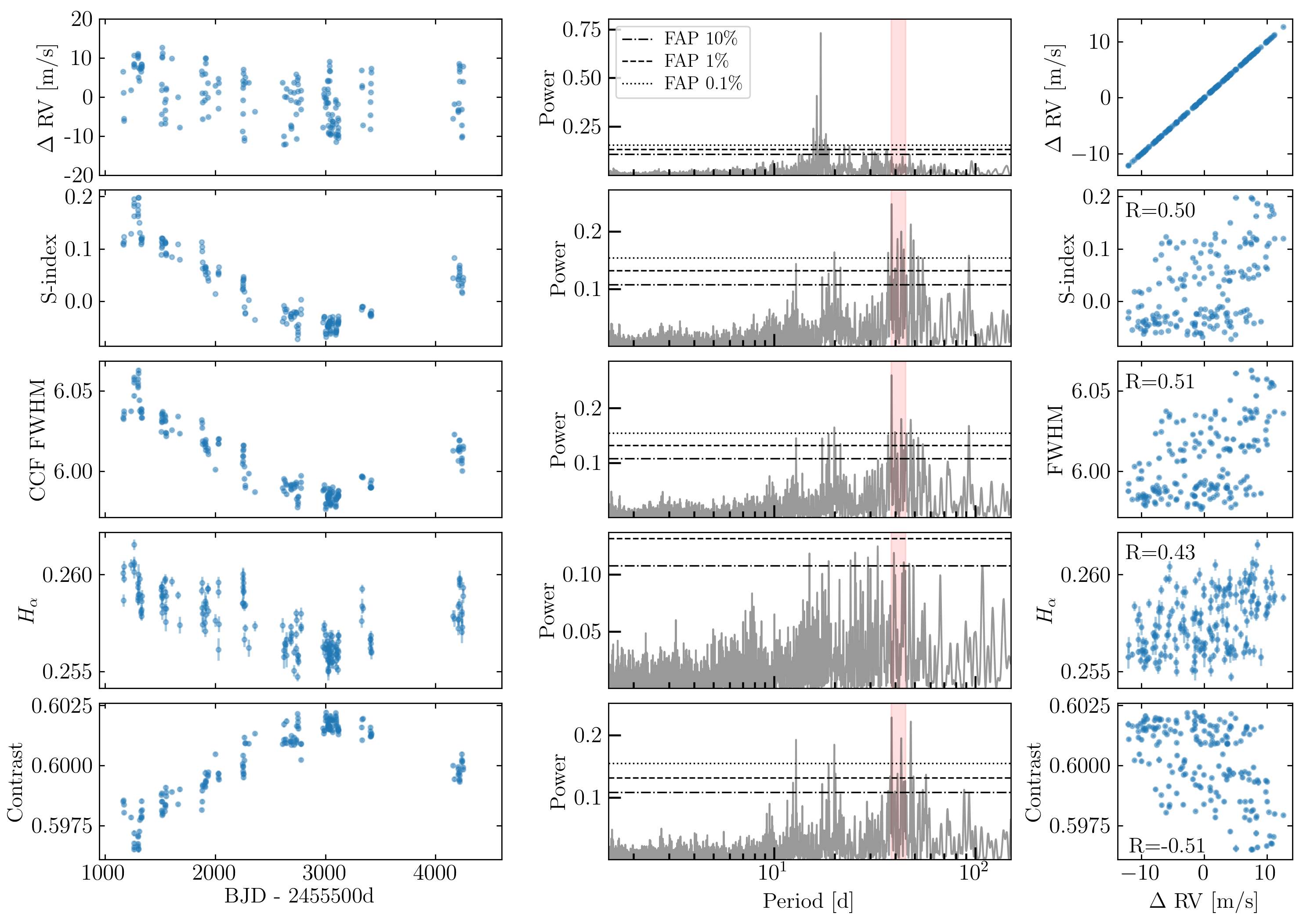}
    \caption{HARPS-N RV and spectroscopic activity indicators of HD\,99492. This dataset is cleaned from instrumental systematics but still contains the stellar activity features, as obtained with the YARARA-V1 post-process described in Sect. \ref{Sect:YARARA} and Appendix \ref{appendix:a}. The left column presents the RV and indicator time-series, while the middle column contains the periodograms of these time-series. No drift nor white noise was included prior to their computation. However concerning the activity indicators, we included a Keplerian in the model so to absorb most of the long-period power, which impacts the short-period signals. In those cases, we present the periodograms of the residuals. The red bands depict the estimated range for the stellar rotation period, obtained from both literature and a preliminary analysis of our data. The right column illustrates the correlation between the various indicators and the RV.} 
\label{figApp:99492_AllSpectro}
\end{figure*}

\begin{figure*}[h!]
    \centering
    \includegraphics[width=\textwidth]{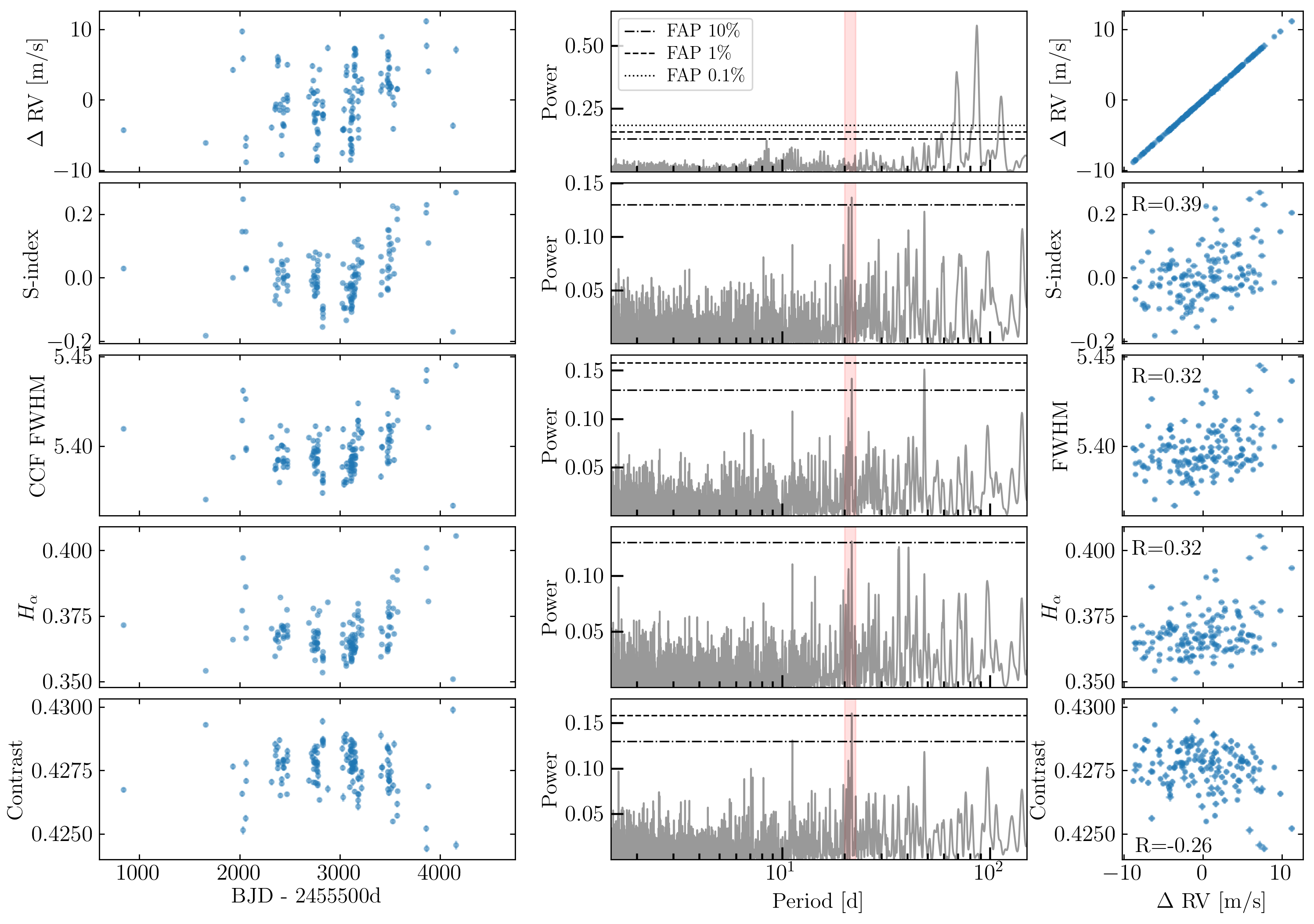}
    \caption{HARPS-N YARARA-V1 RV and spectroscopic activity indicators of HD\,147379. Similar description to Fig. \ref{figApp:99492_AllSpectro}, except with the periodogram computation. The long-period trends observed in the time-series of the spectroscopic indicators are better fitted by quadratic drifts instead of Keplerians. The corresponding periodograms were computed from the residuals of these fits.}
\label{figApp:147379_AllSpectro}
\end{figure*}

\begin{figure*}[h!]
    \centering
    \includegraphics[width=\textwidth]{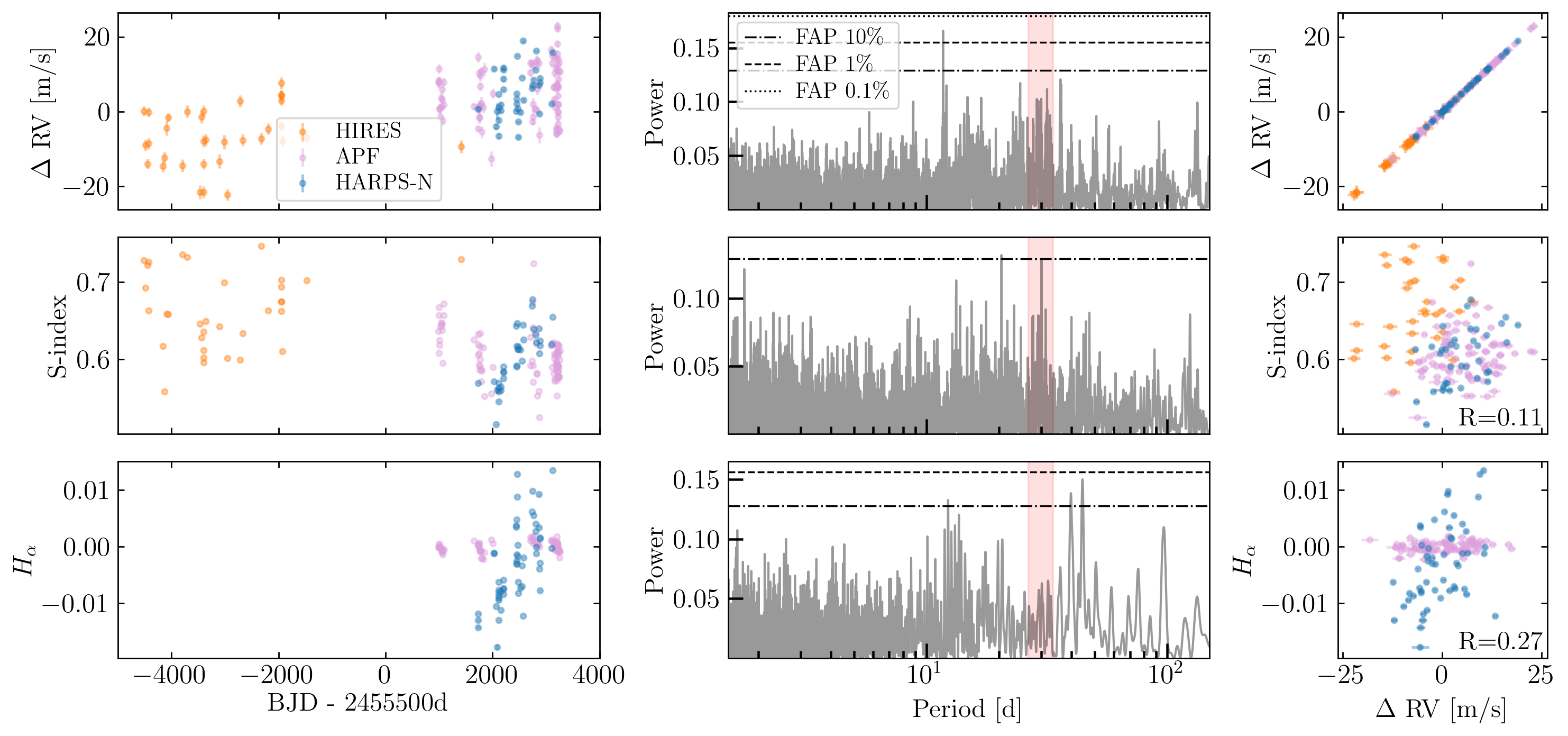}
    \caption{HARPS-N YARARA-V1, HIRES and APF RV and spectroscopic activity indicators of HD\,190007. Similar description to Fig. \ref{figApp:99492_AllSpectro}. We did not include any Keplerian nor drift prior to the computation of the periodograms. We note that only two and one chromospheric activity indicators are publicly available for the APF and HIRES data, respectively (namely the H$_{\alpha}$ and S-index indicators). H$_{\alpha}$ being absent from the HIRES data \citep{Butler2017}, the correlation coefficient between the RV and H$_{\alpha}$ only takes the HARPS-N and APF data into account. Furthermore, this correlation is stronger in HARPS-N, for which R=0.47.}
\label{figApp:190007_AllSpectro}
\end{figure*}

\begin{figure*}[h!] 
    \centering
    \includegraphics[width=\textwidth]{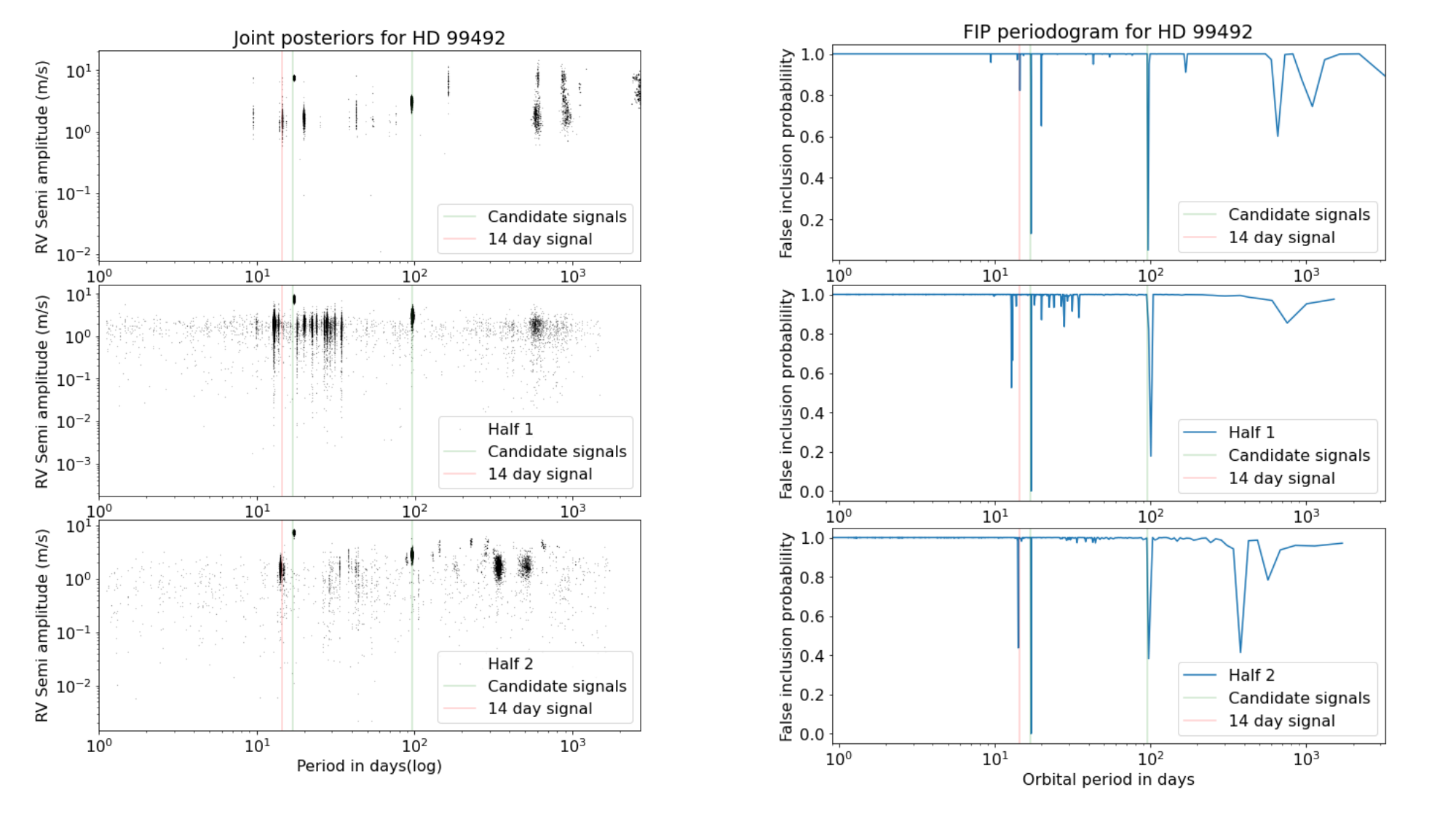}
    \caption{HD99492: Results from the \texttt{TWEAKS} pipeline. \textit{Left column}: posterior distribution. \textit{Right column}: FIP periodogram. \textit{From top to bottom rows}: Analyses on the entire RV time-series, on the first half and on the second half of the time-series, respectively. The vertical plain lines (light green) depict the two planet detections, at 17.05 and 95.2 days, respectively. The red dashed lines help to guide the eye on the 14d signal, which is inconsistently present in the data.} 
\label{figApp:99492_Tweaks}
\end{figure*}

\begin{figure}[h!] 
    \centering
    \includegraphics[width=\columnwidth]{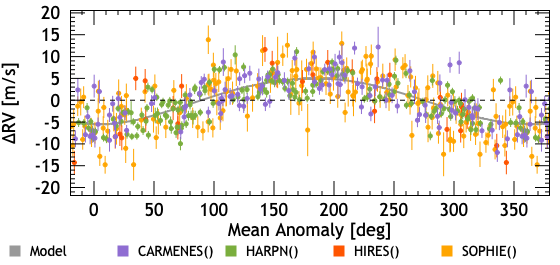}
    \caption{Entire RV time-series of HD\,147374 folded on the best-fit Keplerian solution for the 86.5d planet. This time-series combines our HARPS-N RV measurements with the public data from HIRES, CARMENES, and SOPHIE high-resolution spectrographs. The best-fit Keplerian model is illustrated with the grey curve.}
\label{figApp:FullFoldedRV_147379b}
\end{figure}

We present in this section complementary spectroscopic data, in the order with which they are mentioned in the main text. Fig. \ref{figApp:FullRV_147379} illustrates the entire high-precision RV time-series of HD\,147379, built up from the combination of HIRES, CARMENES, SOPHIE and HARPS-N data. Fig. \ref{figApp:99492_AllSpectro}, \ref{figApp:147379_AllSpectro}, and \ref{figApp:190007_AllSpectro} display the various spectroscopic activity indicator data on HD\,99492, HD\,147379 and HD\,190007, respectively. Trends and correlations with the RV are hinted. Fig. \ref{figApp:99492_Tweaks} presents the \texttt{TWEAKS} results on HD\,99492, which confirm the detection of two planet signals and help to rule out the planetary nature of the 14d signal. Finally, Fig. \ref{figApp:FullFoldedRV_147379b} depicts the HD\,147379 RV folded on our Keplerian solution for planet b obtained with all the available RV data. This solution results from an analysis of the RV only, and hence it is not favoured. We refer to Table\,\ref{tab:FinalSolutions} for the preferred updated Keplerian parameters of HD\,147379 b, based on the analysis of the HARPS-N YARARA-V1 data together with SPLEAF.

\FloatBarrier

\section{Transit searches} 
\label{App:Photometry}
\begin{figure} 
    \centering
    \includegraphics[width=\columnwidth]{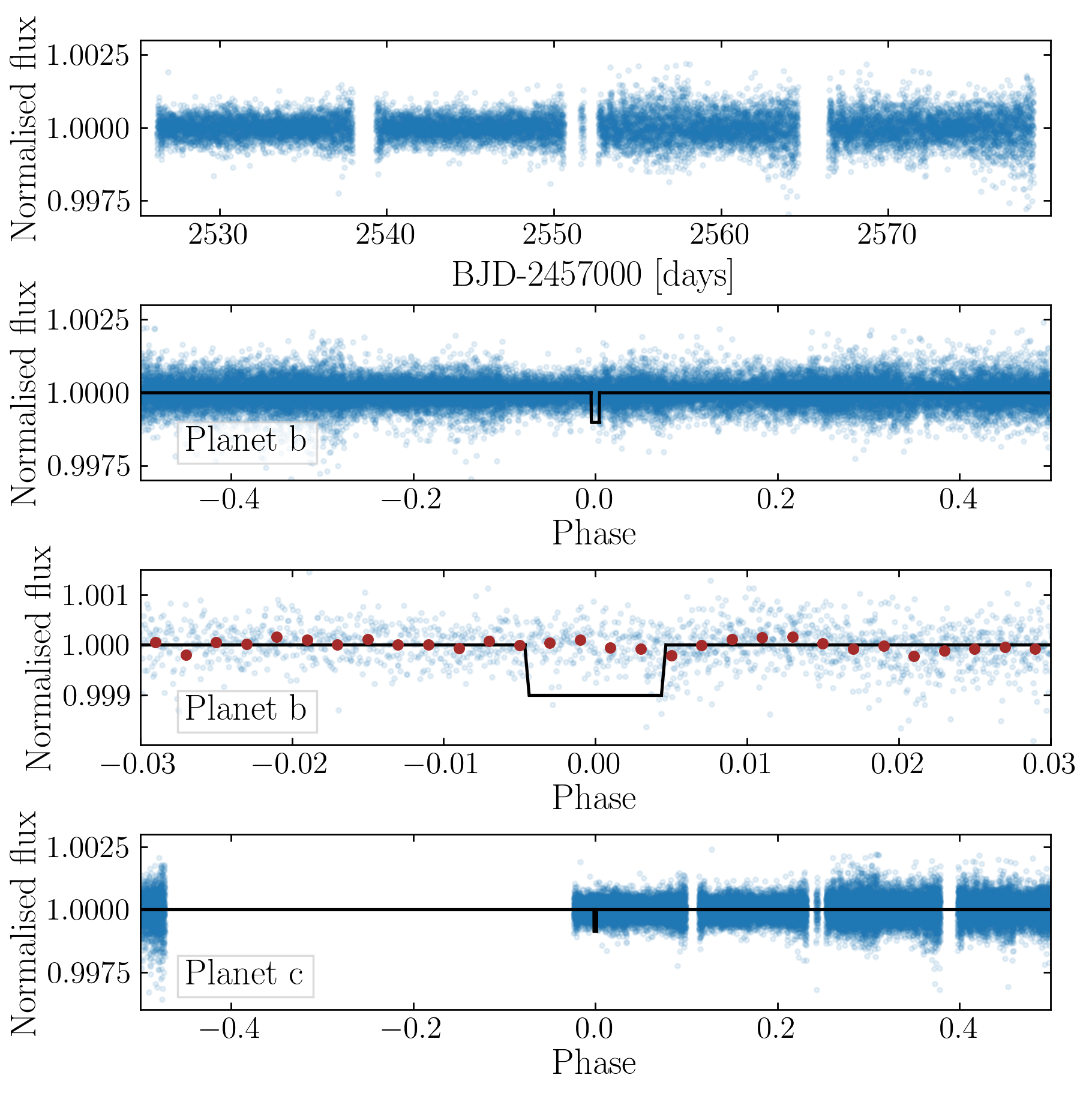}
    \caption{HD\,99492: Transit search in the TESS light-curve, sectors 45 and 46. \textit{Top panel}: cleaned light-curve following our custom extraction from the target pixel files. \textit{Second and third panels}: Light-curve folded on the orbital period and phase of planet b predicted from our RV analysis, together with the transit model curve (the modelled transit depth is 1038ppm for a rocky composition). \textit{Bottom panel}: Light-curve folded on the ephemerides of planet c, together with its theoretical transit curve for a rocky composition (the modelled transit depth is 850ppm). } 
    \label{figApp:HD99492_TransitSearch}
\end{figure}

This section provides details on our transit searches in the three stars. Fig. \ref{figApp:HD99492_TransitSearch}, \ref{figApp:HD147379_TransitSearch}, and \ref{figApp:HD190007_TransitSearch} present the corrected and detrended light-curves together with the hypothetical transit models for the four planets around HD\,99492, HD\,147379, and HD\,190007, respectively. To obtain a light-curve cleaned from systematic trends and amenable to a reliable transit search, we employed a two-stages process. First, we used \texttt{lightkurve} to treat the instrumental systematics mostly (see Sect. \ref{Sect:StellarParam_99492} for more details), and then, we detrended the light-curve from the remaining stellar activity trends with \texttt{keplersplinev2} (see Sect. \ref{Sect:99492_TransitSearch} for more details). These cleaned light-curves are presented in the top panel of each of the three figures. To illustrate the transit detectability of the planets, we generated hypothetical transit model curves with the assumptions of null impact parameter and null eccentricity. Furthermore, these curves are drawn from the unfavourable assumption of rocky composition, even though we expect those planets to stand in the volatile-rich regime. This decision aims to strengthen the transit detectability, by stating that the planets would be detected even under unexpectedly high bulk densities. The transit depths of the model curves were derived from the mass-radius relationship for rocky planets provided by \citet{Otegi2020}, utilising the results from our RV modelling, together with the stellar radius indicated on exofop. In the bottom panels of Fig. \ref{figApp:HD99492_TransitSearch}, \ref{figApp:HD147379_TransitSearch}, and \ref{figApp:HD190007_TransitSearch}, we present the cleaned light-curves phase-folded onto the ephemerides of the considered planets, together with the transit model curves. When relevant, we also provide zoomed-in views of these plots, centred around the expected transit time. The red dots are binned flux measurements, and allow for a better representation of the variations in the light curves. 

\begin{figure} 
    \centering
    \includegraphics[width=\columnwidth]{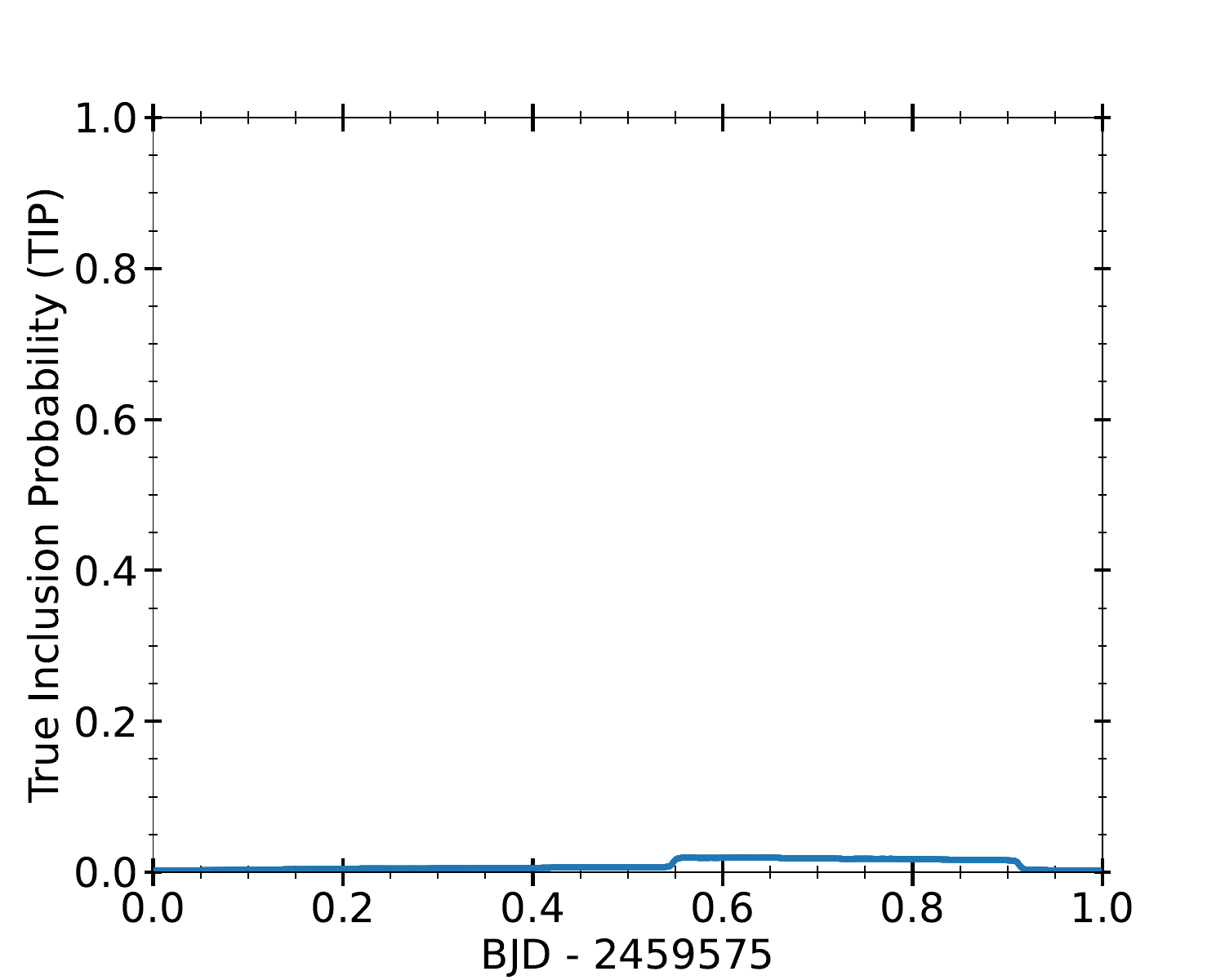}
    \caption{Illustration of the single-transit search in the HD\,99492 TESS sectors 45 and 46 data. }  
    \label{figApp:HD99492_TIP}
\end{figure}

\begin{figure} 
    \centering
    \includegraphics[width=\columnwidth]{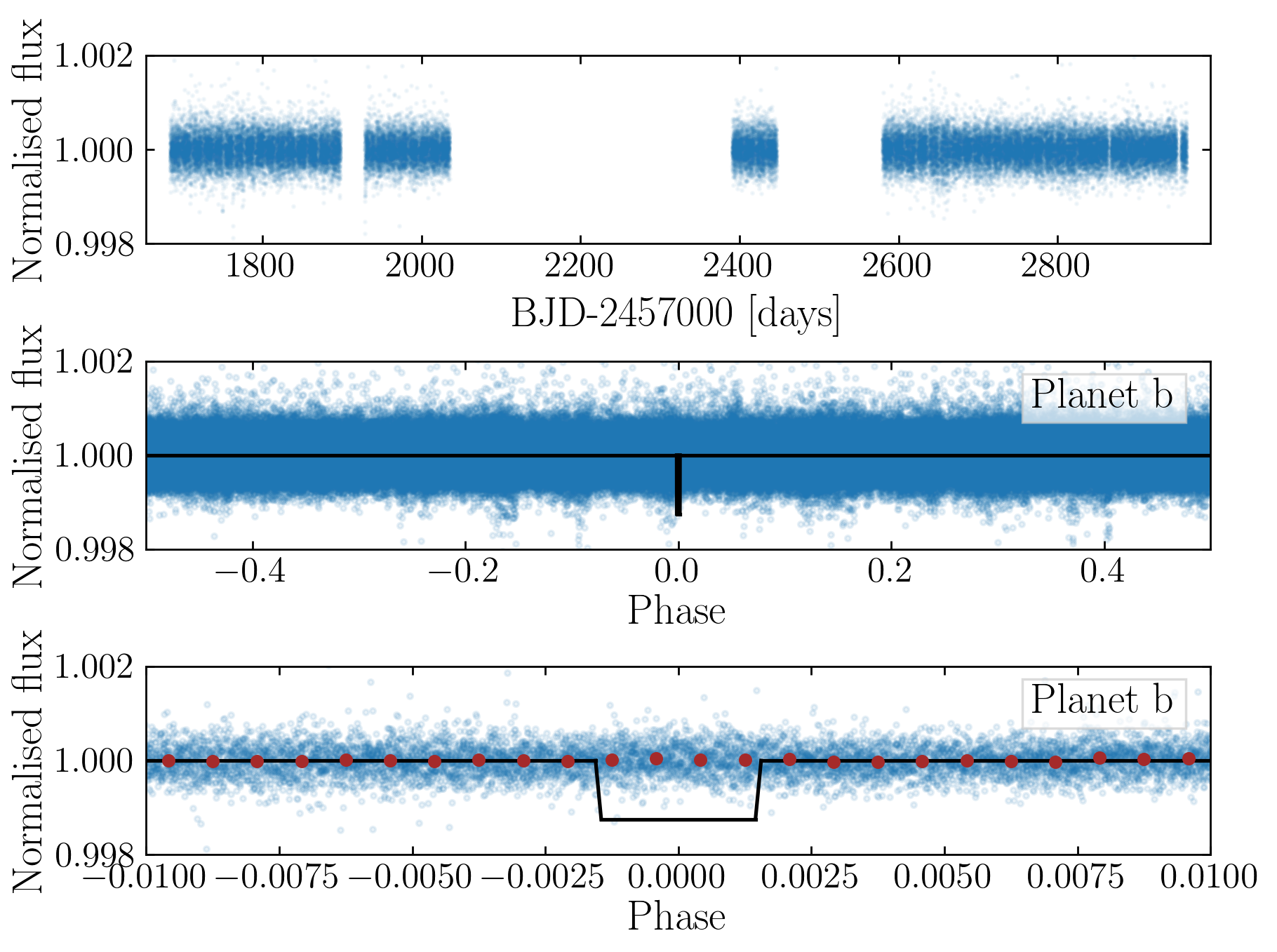}
    \caption{HD\,147379: Transit search in the TESS light-curve composed of 28 sectors (sectors 14-21, 23-26, 40-41 and 47-60). \textit{Top panel}: cleaned light-curve. \textit{Second and third panels}: Light-curve folded on the orbital period and phase of planet b predicted from our RV analysis, together with the transit model curve (the modelled transit depth for a rocky composition is 1250ppm).}  
    \label{figApp:HD147379_TransitSearch}
\end{figure}

\begin{figure} 
    \centering
    \includegraphics[width=\columnwidth]{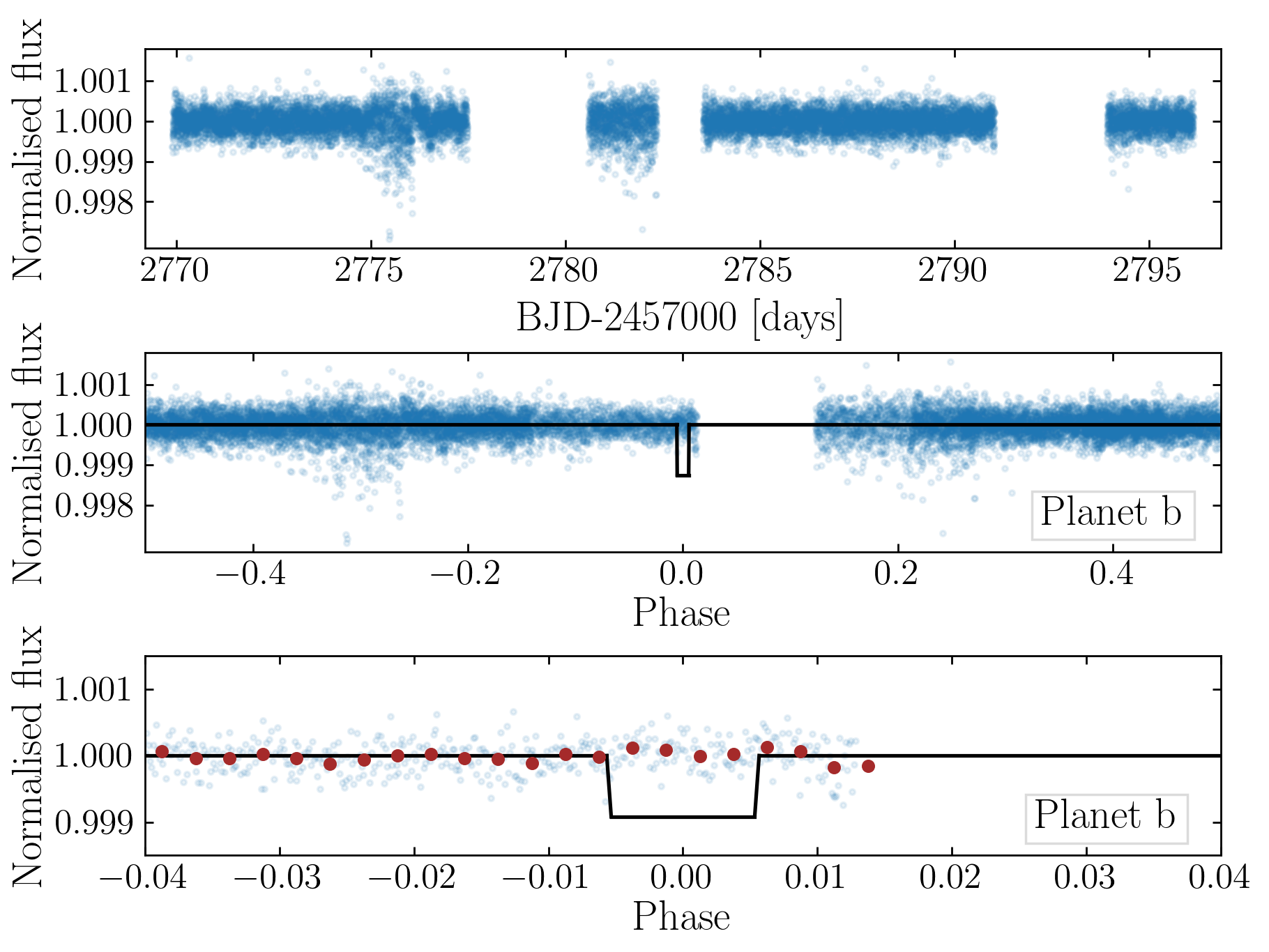}
    \caption{HD\,190007: Transit search in the TESS light-curve, sector 54. \textit{Top panel}: cleaned light-curve. \textit{Second and third panels}: Light-curve folded on the orbital period and phase of planet b predicted from our RV analysis, together with the transit model curve (the modelled transit depth and duration for a rocky composition are 920ppm and 190 minutes, respectively). } 
    \label{figApp:HD190007_TransitSearch}
\end{figure}

Finally, Fig. \ref{figApp:HD99492_TIP} illustrates our single-transit search in the TESS sectors 45 and 46 of HD\,99492, using the TIP framework. The search was performed in consecutive 1-day temporal windows, and the plot presents one of the outcomes. The single-transit search process is explained in Sect. \ref{Sect:99492_TransitSearch}. The TIP compares two models, without and with one planet. In none of the one-day windows did we find a TIP higher than 0.5, which means that the zero-planet model is favoured. A similar procedure was undertaken with the TESS sector 54 of HD\,190007, and the same conclusion was reached.

\FloatBarrier

\section{Orthogonalisation of stellar activity and instrumental systematic vectors in YARARA}
\label{appendix:a}

During the cleaning process of spectra, YARARA is fitting simultaneously for stellar activity and instrumental systematics in the wavelength domain by fitting a vector basis in a multi-linear model. In the previous version of the code \citep{Cretignier2021}, the basis was fitted using the S-index, the CCF's contrast and the CCF's FWHM. However, on HARPS-N, the CCF moments usually contain the superposition of 1) a PSF variation (induced by an instrumental defocus or technical operations on the instrument) and 2) a stellar activity component that affects the individual stellar lines \citep{Cretignier2020a}. Because of that, it was not possible to re-inject the stellar activity component fitted out by YARARA, without reintroducing as well the PSF defocus systematics. In order to remove part of the degeneracy, a better splitting of activity and instrumental systematics was mandatory, which was obtained by producing a better vector basis.

Since the S-index is a broad flux integration in the core of the strong CaIIH\&K lines, the proxy is insensitive to small PSF defocus, opposite to the thin photospheric stellar lines that dominate the moments of the CCF. On the other hand, PSF variations are mainly symmetrical. This led us to develop a new methodology to derive better time-domain vectors orthogonal to the stellar activity S-index component. We show the results of this analysis on HD4628 as well as on HD99492. HD4628 is not presented in the present article and its analysis will be described in a forthcoming paper. However, the star is one of the most consistently monitored by HARPS-N from the very beginning. The star also possesses a clear magnetic cycle and has a similar spectral type than HD99492. This is why it was selected for the present demonstration. 

\begin{figure} 
    \centering
    \includegraphics[width=\columnwidth]{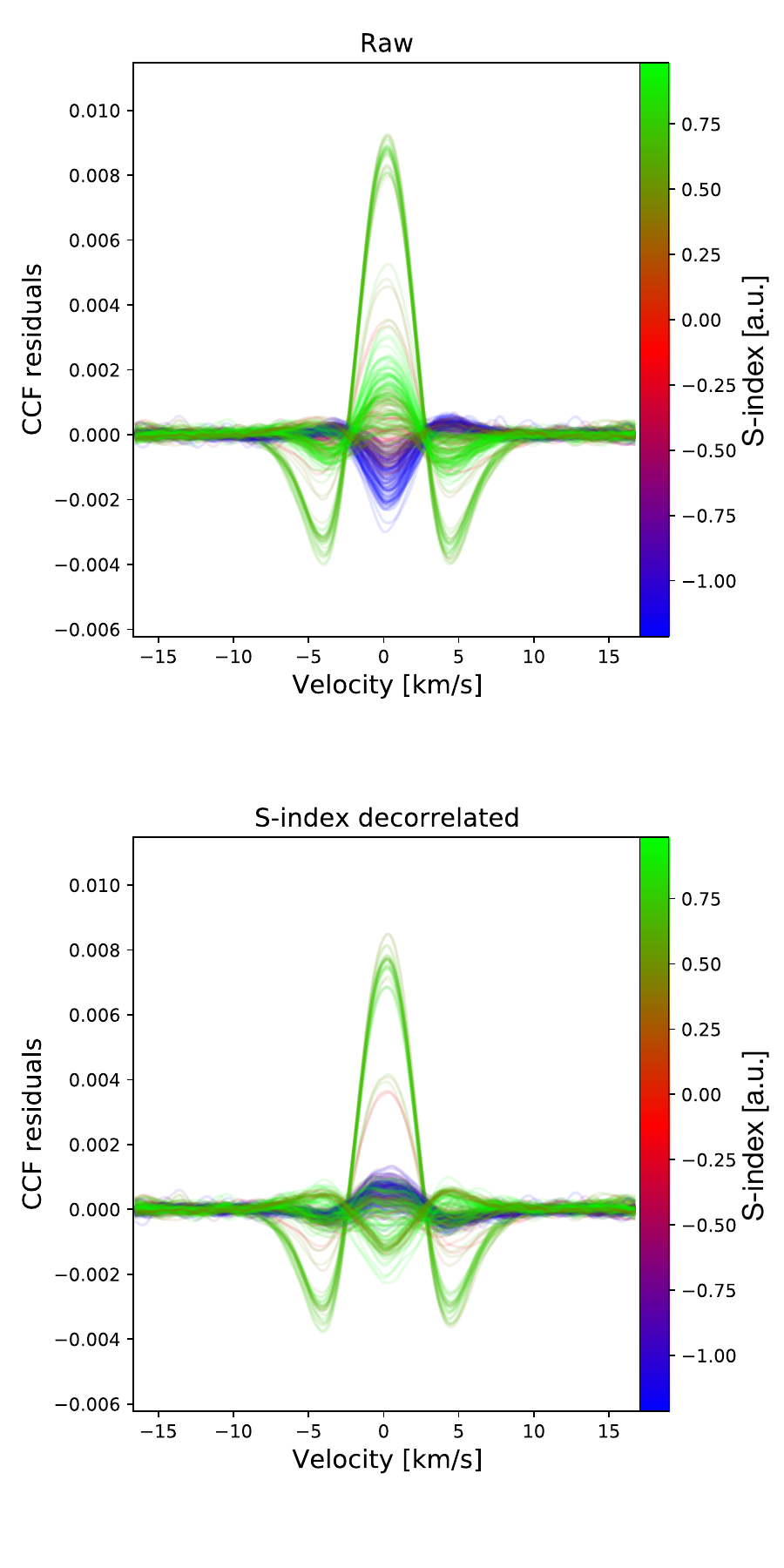}
    \caption{Visualisation of the CCF residuals profiles obtained before (top panel) and after (bottom panel) the decorrelation by the S-index for HD4628. The colour code is the Z-normalised S-index vector (in arbitrary units). The stellar activity signature, initially visible in the top panel by the blue-to-green colour gradient at the line centre ($v=0$ km/s), is removed, whereas the clear instrumental defocus is preserved. The instrumental time-domain vectors can be extracted after performing a PCA decomposition of those profiles (see Fig.\ref{fig:Appendix2}).}
\label{fig:Appendix1}
\end{figure}

Using the same  generic line lists as the DRS and for a given velocity grid $v_i$, we first derived the cross-correlation functions for all the epochs: $\text{CCF}(t,v_i)$. The CCFs were centred by an RV amount corresponding to the centre of a Gaussian fit before being interpolated on the initial velocity grid. Such an operation allows to cancel any large Doppler shift. The median CCF was then subtracted to produce residual profiles and each of these profiles was itself median centred afterwards. The resulting CCF residuals are plotted in the top panel of Fig.\ref{fig:Appendix1}. 

In order to remove the stellar activity dependency, the residual CCFs for each velocity grid element $v_i$ were linearly decorrelated from the $S_{index}(t)$ time-series: $\text{CCF}_{res}(t,v_i) = \text{CCF}(t,v_i) - \alpha_i \cdot$ $S_{index}(t)$. In \citet{Cretignier2021}, the authors showed that the S-index was strongly correlated with the line profiles distortions (see their Fig.3), hence the same correlations are expected for the CCFs. However, the S-index usually also contains long-term variations due to stellar magnetic cycles \citep{Baliunas1996}. Such signals can be easily collinear with the long-term drift defocus observed in the HARPS and HARPS-N instruments, producing incorrect scaling coefficients $\alpha_i$. In order to avoid this issue, the scaling coefficients between the S-index and the residual CCF time-series were obtained after performing a low-pass filtering. The fitted scaling coefficients between the S-index and CCF residuals were obtained on the high frequency part of the filtered vectors, which implicitly means that the correlation was done on the rotational modulation of the S-index signal and not on the magnetic cycle. The $\text{CCF}_{res}$ profiles obtained can be seen in the bottom panel of the Fig.\ref{fig:Appendix1}.

\begin{figure*} 
    \centering
    \includegraphics[width=18cm]{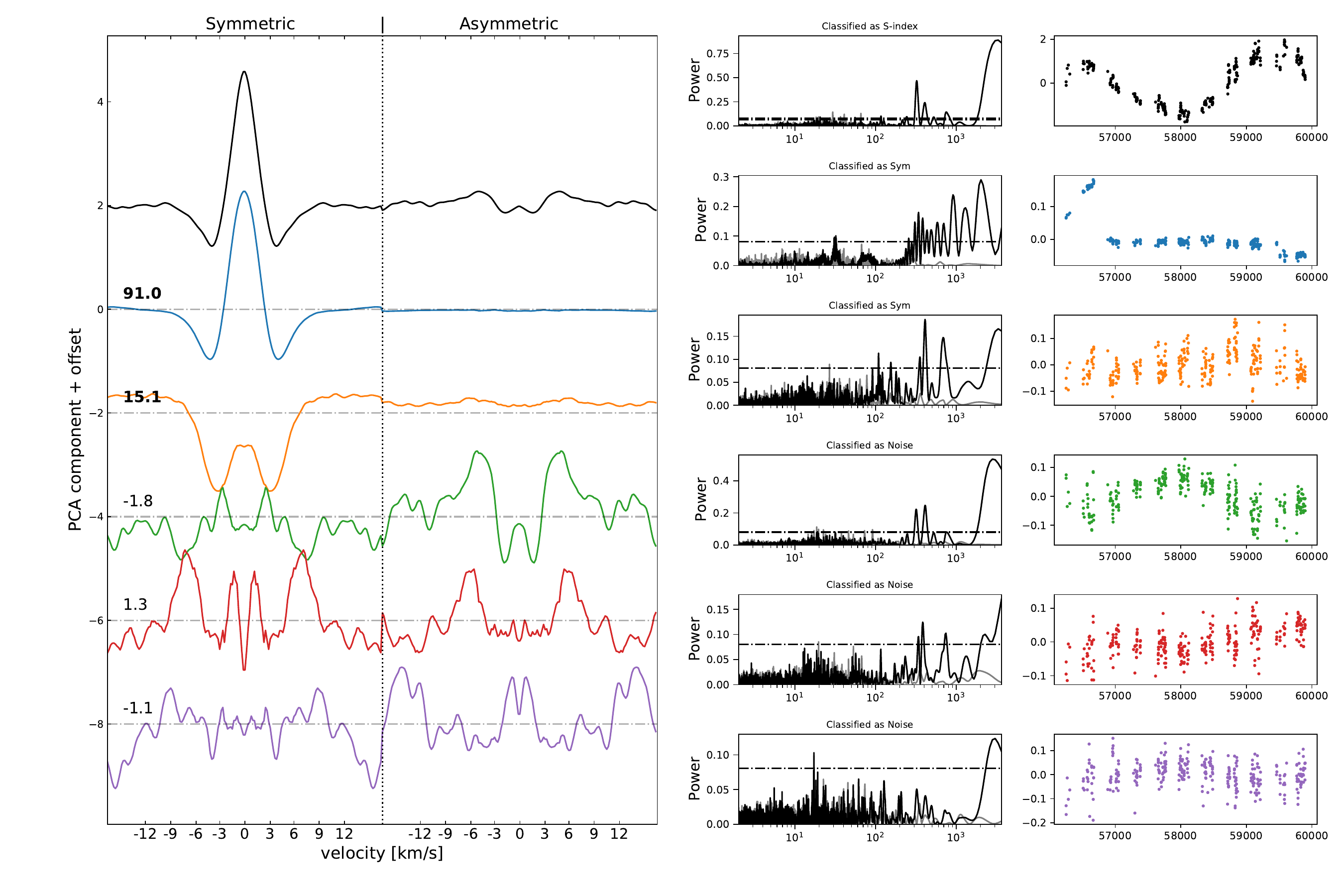}
    \caption{Extraction of the instrumental systematics basis for HD4628 by performing a PCA decomposition on the $M(t,v_i)$ matrix (as defined in the text). The first five principal components $P_j(v_i)$ are displayed (left panel) with an arbitrary offset as well as their time-domain coefficients $U_j(t)$ and their corresponding periodogram (right panel). The grey periodograms were derived after applying a high-pass filter on the time-series, while the black ones do not correspond to any filtering. The S-index (top row) is also displayed for comparison. The ratio score $R$ is displayed for each component. Only the first two components (blue and orange) are found symmetric ($R_1=91.0$ and $R_2=15.1$). The first one clearly shows the instrumental defocus during the first part of the instrument's life (1st and 2nd seasons), whereas an instrumental refocus is also visible at the middle of the penultimate season, which corresponds to a manual intervention on the instrument on the 4th of October 2021 to fix the HARPS-N cryostat (priv. communication).}
\label{fig:Appendix2}
\end{figure*}

The decorrelated residual profiles were split into a left ($L(t,v_i)=CCF_{res}(t,v_i),v_i<0$) and a right ($R(t,v_i)=CCF_{res}(t,v_i),v_i>0$) wings profile. Symmetric $S(t,v_i)$ and anti-symmetric $A(t,v_i)$ vectors were created, respectively computing $S(t,v_i)=0.5\cdot\left(L(t,-|v_i|)+R(t,|v_i|)\right)$ and $A(t,v_i)=0.5\cdot(L(t,-|v_i|)-R(t,|v_i|))$. Those vectors help categorising symmetric versus anti-symmetric CCF variations from the PCA decomposition performed here after. 

Indeed, because the PSF variation is unknown \textit{a priori}, a PCA was performed simultaneously on the symmetric and anti-symmetric profiles of the joint matrix $M(t,v_i)=[S(t,v_i);A(t,v_i)]$. The principal components $P_j(v_i)$ and time-coefficients $U_j(t)$ were recovered. The first five components are represented in Fig.\ref{fig:Appendix2} as well as the S-index coefficient on top (also converted into a symmetric and anti-symmetric component for comparison purpose). Each PCA component was then classified computing a ratio score $R$ defined as: 

\[ \left \{ 
\begin{array}{r r r}
      R = &  \sigma_{sym}/\sigma_{anti} & \text{ , if } \sigma_{sym} > \sigma_{anti} \\
      R = &  -\sigma_{anti}/\sigma_{sym} & \text{ , if } \sigma_{sym} < \sigma_{anti} \\ 
\end{array}
\right .\] \\  
with $\sigma_{sym}$ the standard deviation of $P_j(v_i)$ on the symmetric part and $\sigma_{anti}$ the standard deviation in the anti-symmetric side. We considered a component as symmetric if $R>2$, anti-symmetric if $R<-2$ and noisy otherwise. We then selected the $N$ first symmetric components until a first asymmetric or noisy component was detected. 

\begin{figure*} 
    \centering
    \includegraphics[width=18cm]{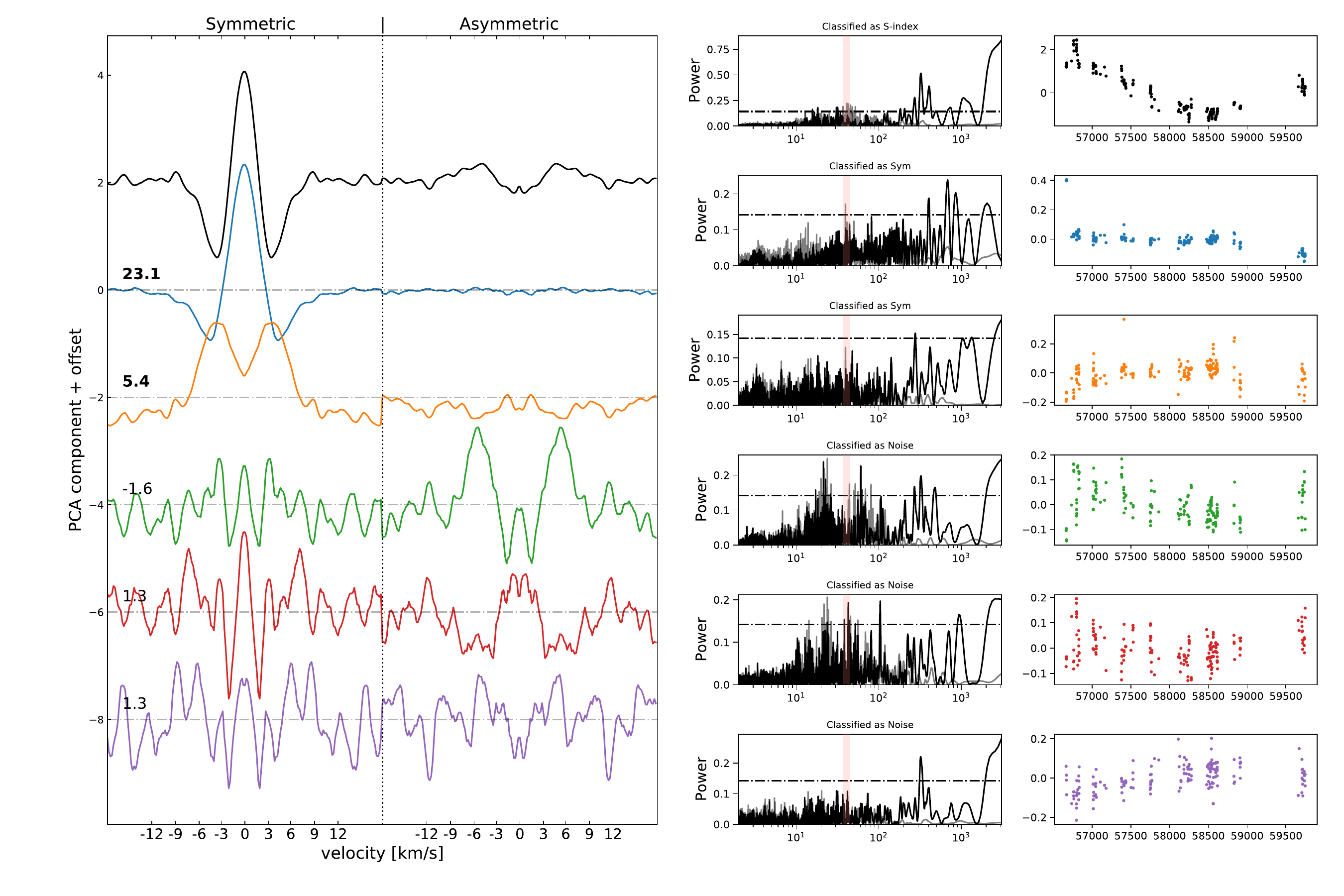}
    \caption{Same as Fig.\ref{fig:Appendix2} but applied on the HD99492 dataset.}
\label{fig:Appendix3}
\end{figure*}

Similar components were identified for both HD4628 and HD99492 (see Fig.\ref{fig:Appendix3}), even if for the latter the components are noisier as expected from the lower S/N of the observations. It can be noted how similar the activity profile and the instrumental defocus profile are, showing how challenging the decoupling of the signals would be without the S-index decorrelation. The third PCA component is anti-symmetric, even though its score is sometimes just below the threshold value (e.g $R=-1.6$ for HD99492). This component shows some magnetic cycle and rotational period (marked by red bands on the periodograms in Fig.\,\ref{fig:Appendix3}) modulations and correlates slightly with the VSPAN of the CCFs, which is a CCF's moment measuring the asymmetry of the CCF profile.  

The present algorithm was successfully tested on dozens of stars and showed to be efficient to produce time-domain vectors dominated by the instrumental defocus observed on HARPS-N and HARPS. Among those stars: the Sun \citep{Dumusque2021}, for which results will also be  presented in a forthcoming paper. Although the solar data is not in the scope of the present paper, it is tempting to use them in order to interpret the PCA components, even if it is always a dangerous and difficult task since components are often a mixture of contaminations that change from star-to-star. The first component on the Sun contains almost exclusively the change of the $v\sin i$ \citep{CollierCameron2019} that may be interpreted as a change of the PSF in some way. The second solar component seems strongly correlated with the S/N of the solar observations which are related to the ageing of the transparent dome in which the solar telescope is located \citep{Dumusque2015}. Contrarily to the Sun observations which are obtained with the solar telescope, observations of other stars are obtained with the TNG which is not covered by such a dome. While such systematics are necessarily different in origin for other stellar observations, it may still indicate that this component is related to unusual S/N perhaps due to clouds or peculiar observing conditions (seeing). 

Note that, by design, nothing prevents stellar activity to be slightly present in the derived instrumental vectors since residual symmetric activity variation uncorrelated with the S-index may be captured by the PCA. However, even if such a risk exists, such a component would be now strongly mitigated, as is depicted by the absence of rotational peak in the periodogram of the extracted components. 

\end{appendix}

\end{document}